\documentclass{aa}  
\usepackage[colorlinks=true,     linkcolor=blue, citecolor=blue, filecolor=blue, urlcolor=blue]{hyperref} 

\usepackage{color}
\usepackage{subfigure}
\usepackage{longtable}
\usepackage{multicol}

\usepackage{txfonts}
\begin{document}

   \title{Deep optical imaging of star-forming blue early-type galaxies}

   \subtitle{Color map structures and faint features indicative of recent mergers}

  \author{Koshy George
          \inst{1} }
  \institute{$^{1}$Faculty of Physics, Ludwig-Maximilians-Universit{\"a}t, Scheinerstr. 1, 81679, Munich, Germany\\
                \email{koshyastro@gmail.com}}

%   \date{Received September 15, 1996; accepted March 16, 1997}

% \abstract{}{}{}{}{} 
% 5 {} token are mandatory

  \abstract
  { 
  Blue early-type galaxies with galaxy-scale ongoing star formation are interesting targets in order to understand the stellar mass buildup in elliptical and S0 galaxies in the local Universe. We study the star-forming population of blue early-type galaxies to understand the origin of star formation in these otherwise red and dead stellar systems. The legacy survey imaging data taken with the dark energy camera in the $g$, $r$, and $z$ bands for 55 star-forming blue early-type galaxies were examined, and $g-r$ color maps were created. We identified low surface brightness features near 37 galaxies, faint-level interaction signatures near 15 galaxies, and structures indicative of recent merger activity in the optical color maps of all 55 galaxies. These features are not visible in the shallow Sloan Digital Sky Survey imaging data in which these galaxies were originally identified. Low surface brightness features found around galaxies could be remnants of recent merger events. The star-forming population of blue early-type galaxies could be post-merger systems that are expected to be the pathway for the formation of elliptical galaxies. We hypothesize that the star-forming population of blue early-type galaxies is a stage in the evolution of early-type galaxies. The merger features will eventually disappear, fuel for star formation will cease, and the galaxy will move to the passive population of normal early-type galaxies.}


%   \begin{enumerate}
%\end{enumerate}

\maketitle
%
%________________________________________________________________

\section{Introduction}

Early-type galaxies (ETGs) in the local Universe are morphologically classified as elliptical (E) and S0 galaxies. The optical spectral energy distribution of these galaxies shows features that are indicative of an old stellar population, which causes these galaxies to form a tight red sequence against a blue cloud of star-forming spiral galaxies in the optical color-magnitude diagram, and they populate the passive population in the plane of star formation rate (SFR) to stellar mass (M$\star$) \citep{Baldry_2004,Brinchmann_2004,Salim_2007,Noeske_2007,Elbaz_2007,Daddi_2007}. ETGs are normally found in denser regions, in contrast to star-forming spiral galaxies, which prefer the low-density environments of the Universe. This led to the common consensus that ETGs are formed through a short but rigorous star formation episode that then ceased, and that they are now destined to reach their final evolution stage in denser regions of the Universe. Deep-field photometric surveys of a statistically large sample of ETGs, however, revealed that the stellar mass in early-type galaxies has increased over the last 8 billion years \citep{Bell_2004,Faber_2007,Brown_2007}. This implies that there has been significant stellar mass accretion or star formation episodes that contribute to the stellar mass buildup in ETGs since redshift $(z)$ $\sim$ 1. Several processes can contribute to stellar mass growth, the most important of which are galaxy major and minor mergers. Gas-rich so-called wet mergers can bring a significant amount of neutral and molecular hydrogen that can act as fuel for star formation. Minor mergers are found to be the dominant mechanism that increases the size of ETGs from z$\sim$1 to 0 \citep{Trujillo_2011}. Major mergers between two spiral galaxies can form elliptical galaxies, which, with a significant gas content, can turn into a starburst. However, in ETGs, minor mergers are thought to be the dominant mechanism, and the major-merger rate is not preferred to account for the observed stellar mass buildup \citep{Jogee_2009,Lopez_2011,Kaviraj_2009,Kaviraj_2011,Kaviraj_2014,Weigel_2017}.\\

The existence of a class of ETGs with blue optical colors in this context changed our conventional view of ETGs as red and dead stellar systems \citep{Fukugita_2004,Schawinski_2009,Kannappan_2009,Huertas_2010,McIntosh_2014,Mahajan_2018, Moffett_2019, Dhiwar_2022, Paspaliaris_2022, Lazar_2023}. The initial report of an actively star-forming population of elliptical galaxies from a morphologically classified sample of bright galaxies at low redshift from the early release of the Sloan Digital Sky Survey (SDSS) identified galaxies with a SFR similar to that of late-type spiral galaxies \citep{Fukugita_2004}.  The star-forming ETGs reside in low-density environments compared to the quiescent ones, and a higher percentage of galaxies are members of groups. Blue ETGs are also observed in high-redshift (z$>$1) galaxy groups and clusters \citep{Mei_2006,Mei_2015,Mei_2022}. All these studies used different techniques to select galaxies and probe different mass and redshift ranges, however, but they showed in common that star-forming blue ETGs have structural properties similar to those of normal ETGs, but star formation properties in line with spiral galaxies, and that they reside in low-density environments. \citet{Schawinski_2009}(here after S09) identified 204 local Universe blue ETGs using the morphology classification from the Galaxy Zoo and $u-r$ colors that are significantly bluer than the red sequence and well within the blue cloud in the optical color-magnitude diagram occupied by star-forming galaxies. These galaxies have redshifts 0.02 $< z <$ 0.05 and luminosities greater than L$\star$. They are found to be in lower-density environtments than red-sequence ETGs and contribute $\sim$ 6$\%$ to the low-redshift ETG population. Based on an analysis using emission line diagnostic diagrams, 25 $\%$ are actively star forming, 25 $\%$ host both star formation and an active galactic nucleus (AGN), 12 $\%$ have an  AGN, and 38 $\%$ show no strong emission lines based on which they might be classified. With SFR ranging from 0.5 to 50 M$\odot$/yr, the star-forming blue ETGs are found to be hosting intense spatially extended star formation. To better understand the contribution of ongoing star formation to the origin of these systems, we used the subsample of 55 star forming blue ETGs from S09. A detailed  analysis using IFU data for a sample of blue ETGs revealed misaligned disks and kinematically decoupled cores, which is indicative of recent gas accretion \citep{Rathore_2022}. The main aim of this work is to answer the questions whether we need to consider blue ETGs as a separate category of galaxies and whether they are fading spirals, rejuvenated spirals/ellipticals, or post-merger systems. Post-merger systems are associated with significant starburst activities. Ultraluminous infrared galaxies (ULIRGs) are late-stage mergers with elliptical profiles and significant star formation \citep{Wright_1990}. We determine whether star-forming blue ETGs belong to this class of objects by answering this question, for which we need to recover any signatures of previous interaction-related events. This was challenging from the shallow SDSS imaging (2m telescope with an integration time of 54s). We ther fore turned to deeper legacy survey data to detect faint low surface brightness features around the galaxies and to construct a color map, based on which we aim to understand any underlying change in stellar population properties. The DECam Legacy Survey (DECaLS) with the 4m telescope has total exposure times of 140, 100, and 200 s in the $g,r,z$ band and can detect low surface brightness features that are indicative of recent merger events and can also help to identify the real morphology of these systems.\\

The star formation could be triggered by recent interactions, as is evident from the stellar debris seen at the galactic outskirts of a good fraction of blue early-type galaxies \citep{George_2015,George_2017}. The SFRs of blue ETGs are very high. This needs a copious amount of molecular gas that is exchanged in the recent interactions to fuel the star formation. The recently accreted gas should stabilize and settle in disks to initiate star formation. The spatially extended optical color maps of blue ETGs can be used to understand localized regions of recent star formation. Sites of ongoing star formation peak at blue optical colors and can be used to understand the triggered star formation process in blue ETGs. We used the deeper legacy survey imaging data of 55 star-forming blue ETGs to identify low surface brightness features around the galaxy and construct the $g-r$ color maps. The wet merger scenario that involves gas-rich accretion is associated with recent or ongoing star formation, in contrast to gas-free dry mergers, which are dominated by evolved stellar population.  We identify the sites of blue colors in galaxy color maps, correlate this with galaxy interactions, and determine the origin of the star formation in an otherwise red and dead elliptical galaxy. We adopt a flat Universe cosmology with $H_{\rm{o}} = 71\,\mathrm{km\,s^{-1}\,Mpc^{-1}}$, $\Omega_{\rm{M}} = 0.27$, and $\Omega_{\Lambda} = 0.73$ \citep{Komatsu_2011}.\\

\section{Data and analysis}

The optical $g,r,z$-band imaging data of 55 star-forming blue ETGs were taken from data release 9
of the legacy survey DECaLS \citep{Dey_2019}. DECaLS uses the dark energy camera, consisting of 62 2k $\times$ 4k CCDs with a pixel scale of 0.262 arcsec pix$^{-1}$ and a 3.2 deg${^2}$ field of view. It is located at the 4 m Blanco telescope of the Cerro Tololo Inter-American Observatory and is suited for wide-field imaging surveys. The observations were conducted in a dynamic observing mode in which the exposure times and target field selection can be changed on the fly, depending on the observing conditions, to ensure a uniform depth throughout the survey. The median $g,r,z$-band FWHM of the delivered image quality is $\sim$ 1.3, 1.2, and 1.1 arcsec. The  photometric calibration was made using the Pan-STARRS1 DR1 photometry through a set of color-transformation equations given in \citet{Dey_2019}. The $g,r,z$-band coadded images we used in this analysis were calibrated with pixel values stored in nannomaggies, which can be converted into magnitudes using the appropriate conversion stored in the header. The DECaLS imaging reaches $\sim$ 2 mag deeper than the SDSS and hence can detect low surface brightness features in the r band down to 28 mag arcsec$^{-2}$ (the corresponding limit for SDSS is 25 mag arcsec$^{-2}$) \citep{Driver_2016, Hood_2018}. The DECaLS covers the SDSS footprint. The 55 star-forming blue ETGs identified from previous SDSS imaging have $g,r,z$-band imaging data, except for 2 ETGs, which lack z-band images.\\

\subsection{Blue early-type galaxies: M$\star$-SFR relation}

\begin{figure}
\centering
\includegraphics[width=3.5in]{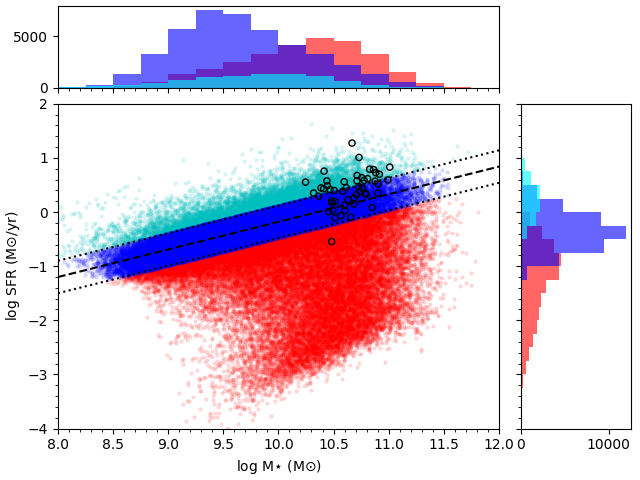}
\caption{Relation of SFR-M$\star$  for galaxies in the local Universe (0.02<$z$<0.05) using data from GSWLC. The galaxies that are on the main sequence are shown in blue, starburst galaxies are shown in cyan, and passive galaxies are shown in red. The main-sequence relation is shown as the dotted black line, and the 0.3dex width on either side of the main sequence is shown as the dashed line. The star-forming blue early-type galaxies are shown with a black circle. Fifty out of the 55 galaxies are in GSWLC, out of which 24 galaxies are on the main sequence, 25 are starburst galaxies, and one is a passive galaxy.} \label{figure:fig1}
\end{figure}

We used the spatially extended SFR and M$\star$ estimate of local Universe galaxies from the GALEX-SDSS-WISE Legacy Catalog (GSWLC; \citet{Salim_2016}). The GSWLC estimates the SFR and M$\star$ by fitting spectral energy distribution to the global ultraviolet (UV), optical, and infrared (MIR) flux values of the galaxies. The UV-MIR based SFR estimate is a better proxy for the global SFR of nearby galaxies \citep{Cortese_2020}. GSWLC covers 90 $\%$ of the SDSS area and contains galaxies within the GALEX footprint with or without a UV detection. We constructed the SFR-M$\star$ plane for 80468 galaxies in the local Universe (0.02<$z$<0.05 using the data from GSWLC \citep{Salim_2016}. We used the main-sequence relation for the local Universe galaxies described in \citet{Guo_2019}, which is shown in Fig \ref{figure:fig1}. The 42475 galaxies are on the main sequence, 8611 star burst galaxies and  29382 passive galaxies. The distribution of M$\star$ and SFR values used in the plot is shown in side and top panels. Blue early-type galaxies are with mass $>10^{10.2}$ M$\odot$. There are 24 galaxies in star burst region, 25 on the main sequence and 1 galaxy in passive population.  We note that almost all galaxies (except for one galaxy) are on the main sequence and lie in the starburst region of the plot. The galaxy with the highest SFR (ID 46) is in an ongoing interaction with a neighboring spiral galaxy, as revealed from the blue tidal feature that is visible in the color-composite, $r-$band, and color map described in following sections (see also \citet{George_2015}). \\

.

\subsection{Blue early-type galaxies: Morphological analysis}
The color-composite images for 55 galaxies were created using $grz$ imaging data and assigning blue ($g$), green ($r$), and red ($z$) to the imaging data. The morphology of the blue ETGs was evaluated using the deeper DECaLS $r$-band and color-composite images of 55 galaxies. Figure~\ref{figure:fig2} left and middle panels show the color composite and $r$-band images of 55 star forming blue early-type galaxies. Table~1 gives the morphology information from deeper imaging data analysis and Table~2 gives the details obtained from the visual inspection of the 55 star forming blue early-type galaxy color composite and $r$-band images. The $r$-band images are presented in inverted grey scale to detect faint low surface brightness features. There are 29 E and 25 S0 and 1 spiral identified from a visual morphological analysis. The galaxy with ID 47 is misclassified as a blue early-type galaxy in S09. Many galaxies show faint features at high galactic radii which we discuss in detail below.

\subsection{Merger features}
\begin{figure*}
\centering
{\includegraphics[width=0.3\textwidth]{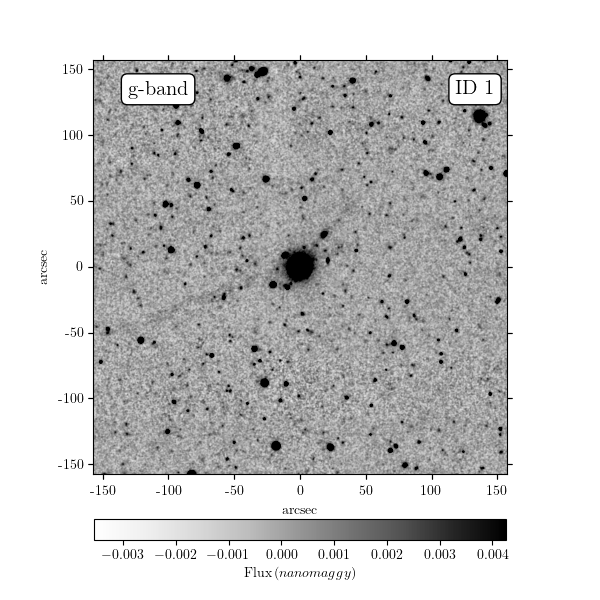}}
{\includegraphics[width=0.3\textwidth]{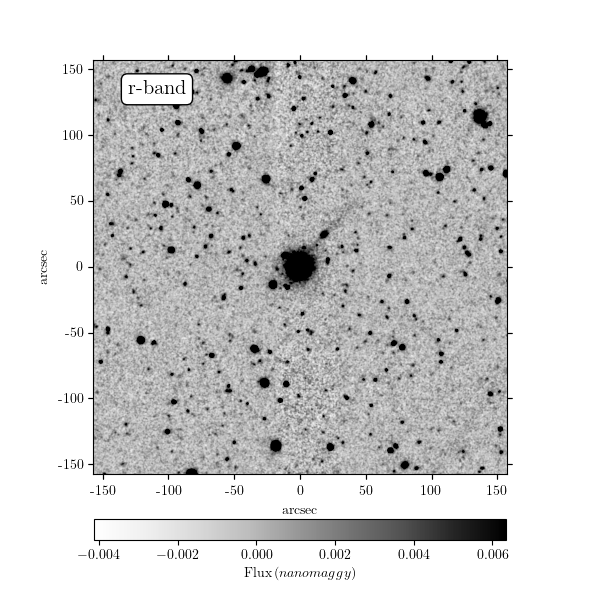}}
{\includegraphics[width=0.3\textwidth]{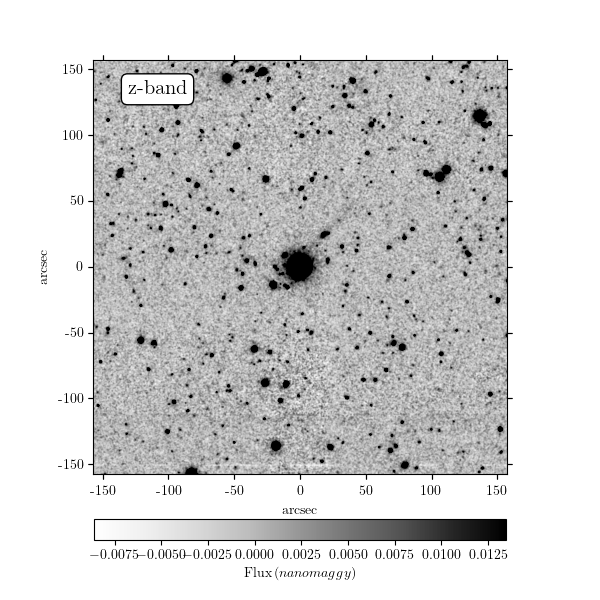}}
{\includegraphics[width=0.3\textwidth]{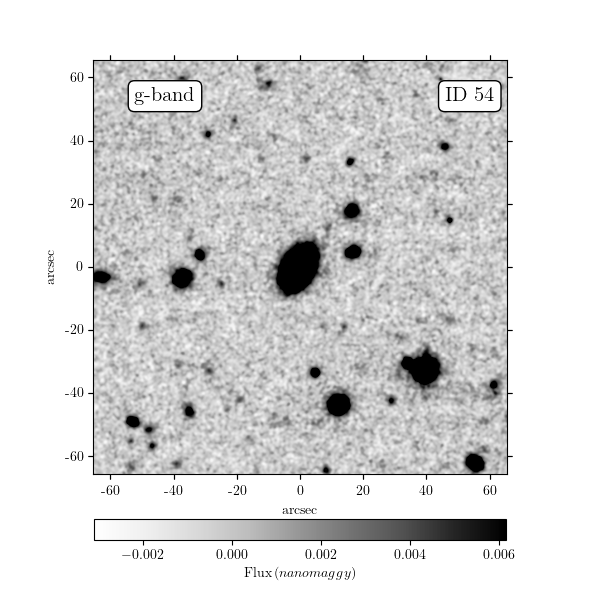}}
{\includegraphics[width=0.3\textwidth]{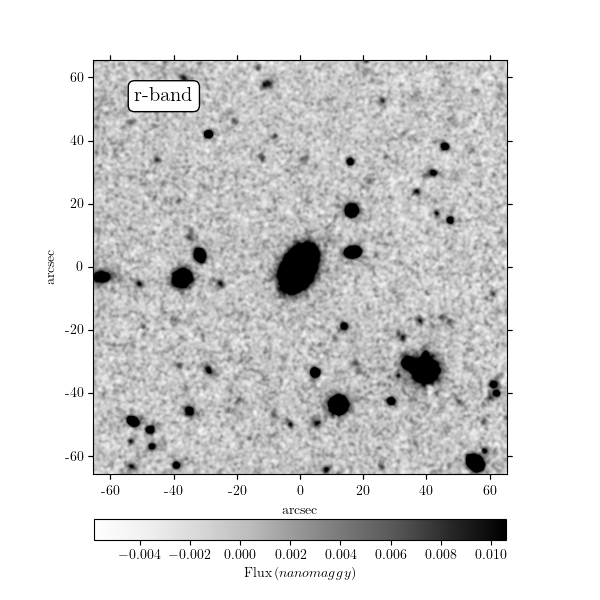}}
{\includegraphics[width=0.3\textwidth]{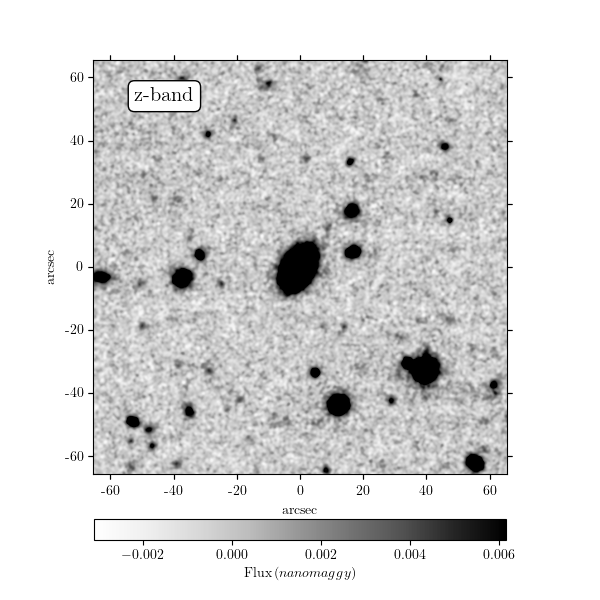}}
\caption{Large-field grayscale color-inverted $g,r,z$ images centered on two galaxies (ID 1 and ID 54). The image is smoothed with a Gaussian of $\sigma$=1.5 to highlight any possible large-scale faint merger features. We show only two galaxies here, but performed this exercise for all galaxies to detect any faint features that are otherwise not seen near to the galaxy.}\label{figure:fig3}
\end{figure*}

\begin{figure}
\centering
{\includegraphics[width=0.5\textwidth]{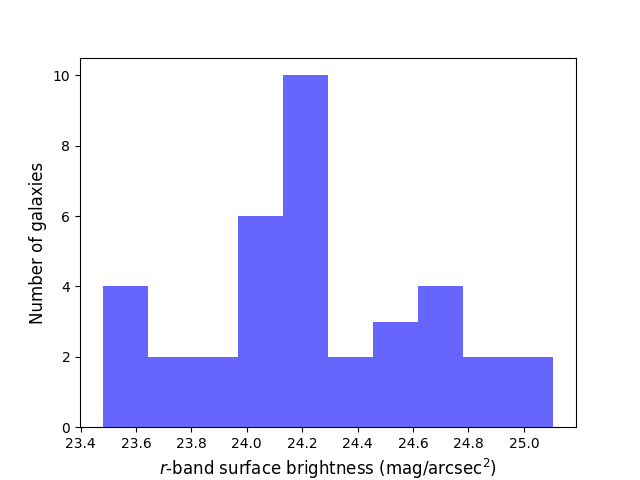}}
\caption{Distribution of the surface brightness of faint features seen around 37 galaxies.}\label{figure:fig4}
\end{figure}

The  color composite and deeper $r$-band imaging analysis reveal 37 galaxies that show signatures of recent merger activity. \citet{George_2017} showed residual images of surface brightness profile fits of 32 galaxies with features arising due to recent interactions. We now attempt to bring out the very faint features by smoothing the pixel noise through running a gaussian of $\sigma$=1.5 pixel. We then checked visually for features in a wider region around the galaxy. We could detect faint features around 15 more galaxies when analysed over a region as wide as 300 $\arcsec$ $\times$ 300 $\arcsec$ centered on the galaxy smoothed image. We demonstrate the technique in a large field grey scale inverted $g,r,z$ images (in relatively strong and weak case) centered on galaxy with ID 1 and ID 54 in Figure~\ref{figure:fig3}. Note the tidal features seen in $g$ which fades from $r$ to $z$ band imaging data. This could be related to different limiting magnitude for $g,r,z$ band images or a stellar population effect inherent to the faint feature. The strongest feature seen are tidal in nature for galaxy with ID 1 and the faintest seen as a one sided diffuse feature in galaxy with ID 54. Table~2 give the details of the galaxies with features. The appearance of galaxy with ID 1 in Figure~\ref{figure:fig3} is very similar to NGC 7252, the post merger galaxy with strong tidal features \citep{George_2023}. We selected the faint features outside the galaxies by visually tracing the faintest contour level at the galaxy outskirts. We then measured the area and the flux within the selected region using SAO DS9 tool \citep{Joye_2003}. The coadd images are flux calibrated with pixels units of nanomaggies, which we converted to AB magnitude system. Surface brightness is computed for the faint features outside 37 galaxies and the distribution of which is shown in Figure~\ref{figure:fig4}. The 15 galaxies for which tidal features are identified from wide field analysis are very faint, for which we didn't attempt to measure the surface brightness.\\

\subsection{Galaxy Color maps}
The SDSS imaging data of 55 star forming blue early-type galaxies had been used in previous works to do the structural analysis and to identify tidal features in residual analysis \citep{George_2017}. The SDSS data is shallow to construct the spatial color maps of the galaxies and hence we used the  legacy survey imaging data which are significantly deeper than SDSS and also had a better point spread function.  The 55 star forming galaxies in the S09 blue early-type galaxy catalogue are observed in  $g$, $r,$ and $z$ bands (except two galaxies that lack $z$-band imaging) as part of the legacy survey. We used $g$ - and $r$ - band imaging data to construct the color maps. These bands are more sensitive to ongoing star formation and prominently show the effect of star formation on the spectral energy distribution in color maps. \\

The optical $g-r$ color maps of 55 galaxies are created using the coadd imaging data. The $g$ - and $r$ -band images were checked for any possible astrometry offsets, which were found to be negligible in all the 55 galaxies. We further checked for any possible differences in resolution between the $g$ - and $r$ -band coadded imaging data that might affect the color maps. We selected stars within the $g$ - and $r$ -band imaging data and fit a Gaussian profile, computing the full width at half maximum (FWHM). The stars in the $g$ - and $r$ -band images are found to have profiles with similar FWHM values.  

The following equation was used to construct the $g-r$ color map of the galaxies:\\

($g-r$)$_{i,j}$ = $-2.5$ $\times$  $log_{10} $($flux_{g}/flux_{r}$)$_{i,j,}$ \\

where $flux_{g}$ $_{i,j}$ and $flux_{r}$ $_{i,j}$ are the values of the i$_{th}$ and j$_{th}$ pixel in the $g$ - and $r$ -band imaging data of a galaxy. We used a Gaussian kernel of $\sigma$=1.5 pixel to suppress the noise level and make the faint features in the galaxy outskirts detectable in the color map. The right panel in Figure~\ref{figure:fig2} shows the $g-r$ color maps of 55 star-forming blue early-type galaxies. Table~2 gives the details we obtained from the visual inspection of the star-forming blue early-type galaxy $g-r$ color maps. The features are separated into cores and extended structures.\\

\begin{figure}
\centering
{\includegraphics[width=0.5\textwidth]{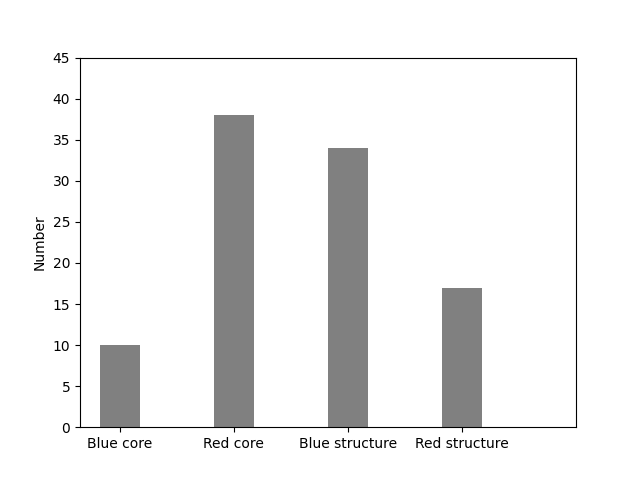}}
\caption{Distribution of the features seen in the $g-r$ color map of galaxies.}\label{figure:fig5}
\end{figure}

The distribution of the features in the color map of the main body of the galaxy are shown in Figure~\ref{figure:fig5}.  The optical color map of 17 galaxies shows red structures in the main body of the galaxies and 34 galaxies with blue structures.
 
\begin{figure}
\centering
{\includegraphics[width=0.5\textwidth]{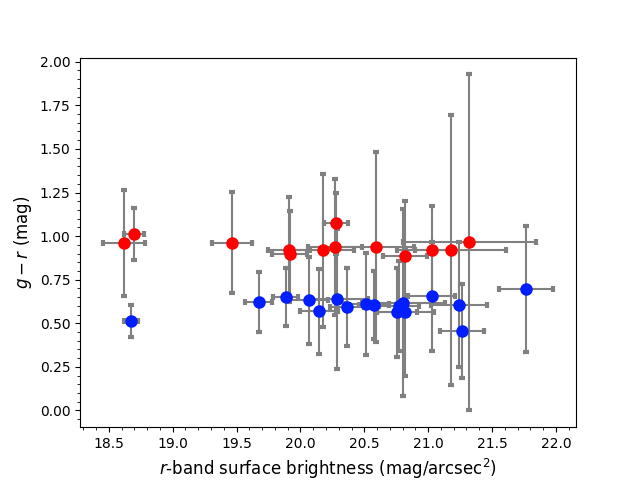}}
\caption{ $g-r$ color of the features seen in the main body of 30 galaxies plotted against the $r$-band surface brightness. The features are color-coded in blue ($g-r$ < 0.75) and red ($g-r$ > 0.75). Seventeen galaxies have a blue color, and 13 galaxies have red patches in the main body.}\label{figure:fig6}
\end{figure}

Ten galaxies have a blue core, and 38 galaxies have a red core region at the center. We note that the central color might be affected by a variation in the point spread function between the g and r images. We therefore did not attempt to interpret the red and blue cores in these galaxies. We estimated the $g-r$ color and $r$-band surface brightness of the distinct color structure in the main body of 30 galaxies that were clearly identified and measured visually by selecting the region of interest. A polygon was drawn, within which the color and surface brightness was computed. The $r$-band $g-r$ color of the patches detected for these 30 galaxies is shown in Figure~\ref{figure:fig6}. We distinguished them into red and blue based on a $g-r$=0.75 color criterion. Thirteen galaxies have a blue structure, and 17 galaxies have a red structure. The blue structure could be due to young stars as a result of recent merger activity, and the red structures could be the attenuation due to dust lanes. Dust lanes in early-type galaxies are common and are understood to be associated with recent interactions \citep{Dokkum_1995, Smith_2012}.

\begin{table}
\centering
\label{galaxymorphdetails}

\begin{tabular}{llcc} %23
\hline 
\hline
Morphology           & E  & S0  & Spiral\\
\hline
Strong merger feature  & 20   & 17   & - \\
Weak merger feature    & 8    &  7   &  -  \\
Total           & 28   &  24 & 1  \\
\hline
\end{tabular}
\caption{\label{t7} Details of the morphology and low surface brightness feature of 55 star-forming blue early-type galaxies. Two galaxies lack detectable features.}
\end{table}

\section{Discussion}

Elliptical galaxies are known to host merger features and  grow through gas-free (dry) mergers involving little star formation \citep{Tal_2009}. \citet{Duc_2015} presented a catalog of 92 ETGs with tidal tails, stellar streams, and shells from the ATLAS3D sample. These ETGs are from low- and medium-density environments with features revealed from deeper observations and data reduction optimized for a low surface brightness detection. Based on a morphological analysis of very deep g-band images that reach low surface  brightness levels of 29 mag/arcsec$^2$ , \citet{Bilek_2020} found features indicative of recent interaction events in 177 nearby massive ETGs.  These are mainly tidal features in the form of stellar streams, shells, and tails, and disturbed outer isophotes. \\

  A detailed structural study based on a surface brightness profile fit and the residual analysis revealed that a fraction of blue ETGs show signatures of recent interactions that might have brought fuel for subsequent star formation \citep{George_2017}. The origin of this star-forming population of blue ETGs might therefore be linked to major and minor merger events in galaxies. Deeper optical imaging data in our study revealed merging signatures in 37 (out of 55) blue early-type galaxies. If the original classification of galaxies were based on deeper imaging data, most of the star-forming blue ETGs would have been classified as post-merger systems and not as early-type galaxies. Faint features are detected in 15 more galaxies from the analysis of a larger area centered on the galaxy. This clearly shows that 94.5 $\%$ (52 out of 55) of the star-forming blue ETGs studied here have signatures of recent interactions. \\

The formation of massive elliptical galaxy is thought to arise from the merger of two equal-mass spiral galaxies \citep{Toomre_1972}, such as the famous post-merger system NGC 7252. This galaxy system has strong tidal features in the form of shells, tidal tails, and halos when analyzed in optical imaging. The current consensus is that the system is a post-merger remnant formed from the recent merger of two similarly massive disk galaxies, and that it hosts significant ongoing star formation from a recent merger event \citep{Schweizer_1982}. The central regions of the galaxy show compact spiral features in high-resolution HST imaging \citep{Whitmore_1993}. When this galaxy system is placed at higher redshift by slightly degrading the image quality, the strong merger features can fade to faint halos as are detected in the blue ETGs presented here \citep{George_2023}. \\

Gas-rich minor mergers can trigger star formation in ETGs \citep{Ge_2020}. Mergers can also enhance the dust content and leave behind dust ripples and dust lanes in ETGs. The dust present in the galaxy can also affect the radial trends in color gradients, which otherwise mostly depend on metallicity variations throughout the galaxy stellar population \citep{Schweizer_1992,Goudfrooij_1995,Smith_2012,Kaviraj_2012,Duc_2015,Kokusho_2019}. The red color variations seen in the case of star-forming blue ETGs can be explained with the presence of dust lanes that attenuate optical light. The blue ETGs studied here show merger features indicative of recent accretion, which could be gas-rich minor mergers that bring significant gas content, which in turn triggers new star formation. The galaxies with an elliptical morphology could be undergoing an equal-mass spiral galaxy merger, forming post-merger systems with significant star formation, which will eventually fade to become a normal elliptical galaxy. Almost all (except for one) galaxies are on the main sequence and lie in the starburst region of the plot.  This implies that the star-forming blue ETGs studied here are not recently quenched spiral galaxies with a disk that fades to reveal only the bulge.\\

Recently, \citet{Keel_2022} studied a subsample of six blue early-type galaxies using high-resolution imaging from the Hubble Space Telescope. We note that two of the six galaxies are star forming and are included in our analysis here (ID 30 and ID 46). These galaxies are designated Mkn 888 and CGCG-315-014 in Fig. 5 of \citet{Keel_2022}. The galaxies show a merger signature throughout our optical color-composite and deep r-band imaging data. The color map also shows a blue central region and a surrounding stripe with red color. The outer disk shows faint features with blue color that extends to the faint features seen in the  r-band imaging. The blue patch seen in the center could be the spiral structure that is resolved in HST imaging (as in the case of NGC 7252). We note that galaxy CGCG-315-014 (ID 46) is undergoing an ongoing tidal interaction with a neighboring galaxy (see also \citep{George_2015}). The spiral structure at the center is very clear in the HST imaging for this galaxy. The blue color along the faint features outside the galaxy could be the star formation associated with a very recent gas-rich wet merger. We therefore speculate that in this case, spiral structures at the centers of the galaxies with a blue patch in the optical color maps are remnants of post-merger activity. Ten of the galaxies in the sample of 55 we studied here have a central blue patch (see Fig. 4) and might host a spiral structure at the center that would be unresolved in ground-based optical imaging data. All these 10 galaxies show merger features seen in r-band imaging data. This supports this hypothesis. \\

Post-merger galaxies have been studied in detail in the local Universe, and many of them have an elliptical morphology that depends on the initial conditions that set the merger. The orientation, mass, morphology, and gas content of the progenitor galaxies decide the fate of the post-merger system. A wet merger involving significant gas content of two equal-mass spiral galaxies can produce a star-forming elliptical galaxy such as those we presented here. In this sense, blue elliptical galaxies are a transient phase in the lifecycle of an elliptical galaxy. This is in accordance with the hierarchical merging scenario of galaxy formation and evolution. We can expect to see more such systems when we look back in time when the Universe was younger, galaxy mergers were more frequent, and significant gas content was available within the galaxies. We cannot make general conclusions on the star-forming blue early-type galaxies based on the 55 galaxies. However, it is quite tempting to use this fraction of galaxies to understand the formation of these systems. We found that one galaxy shows a spiral morphology in deeper-imaging data. Of the remaining 54 galaxies, 37 show signatures of recent merger events in deeper r-band imaging, and another 15 show very diffuse emission in a detailed image analysis. The optical $g-r$ color map of the galaxies shows blue and red structures. The blue structure might be due to ongoing star formation, and the red structure might be due to dust lanes, both of which are associated with recent interactions.  This indicates that a recent accretion event that involves gas-rich galaxies with ongoing star formation might have caused the ongoing star formation in blue ETGs.

\onecolumn 

\begin{table*}
\centering
\label{galaxydetails}
\tabcolsep=0.05cm
\caption{\label{t5}Details of the legacy survey imaging analysis of 55 star-forming blue early-type galaxies}
%\small
%\tiny  % Switch from 12pt to 11pt; otherwise, table won't fit
\fontsize{7}{11}\selectfont
\setlength\LTleft{0pt}            % default: \parindent
\setlength\LTright{0pt}           % default: \fill
%\begin{longtable}{cccccllllllllllllllllll} %23
\begin{longtable}{cccccccccccll} %9
%\begin{tabular}{cccccccccll} %9
%\hline\hline
%\hline
ID & S09 & RA & DEC  & $z$  &   M$_r$ &  $u-r$ &  SFR$_{H\alpha}$ & log$_{10}$M$\star$ & SFR$_{SED}$ & Morph &  RGB and r-band & Color map\\
   &  ID & h:m:s  & d:m:s &   &  (mag) &  (mag) &  (M$\odot$/yr) & M$\odot$ & (M$\odot$/yr) &   & feature remarks  & feature remarks\\
 \hline
1 &   2     & 11:23:27.0 & $-$00:42:48.8 & 0.04084 & $-$20.81 & 1.606 & 4.5 & 10.39 & 2.79 & E & tidal tail$\star$ & blue core  \\
2 & 4     & 15:23:47.1 & $-$00:38:23.0 & 0.03747 & $-$21.0  & 1.998 & 2.5  & 10.58 & 2.44 & E & $-$ & red ring around center\\
3 & 5     & 12:08:23.5 & $+$00:06:37.0 & 0.04081 & $-$21.5  & 2.038 & 3.5   &   10.76 & 2.73 & E & diffuse at north & blue core   \\
4 & 7    & 01:41:43.2 & $+$13:40:32.8 & 0.04539 & $-$21.81 & 1.432 & 12.0   &  10.73 & 10.21& E & diffuse at south& blue core and blue structure $\dagger$ \\
5 & 8     & 01:03:58.7 & $+$15:14:50.1 & 0.04176 & $-$21.4  & 2.261 & 7.0  & $-$  &   $-$ & S0 & diffuse at west& red core and red structure $\dagger$\\
6 & 14   & 12:35:02.6 & $+$66:22:33.4 & 0.04684 & $-$21.67 & 1.959 & 6.1 & 10.83 & 6.19 & E& faint diffuse at east & red core and blue disky structure $\dagger$\\
7 & 20    & 08:29:09.1 & $+$52:49:06.9 & 0.04842 & $-$21.0  & 1.835 & 2.1 & 10.47 & 2.61  & E& faint at south east&  blue core and blue structure $\dagger$\\
8 & 23    & 12:06:47.2 & $+$01:17:09.8 & 0.04124 & $-$21.18 & 2.207 & 1.2 & 10.69 & 1.83 & S0 & faint at north& red core, blue structure $\dagger$\\
9 & 30    & 15:17:19.7 & $+$03:19:18.9 & 0.03749 & $-$20.74 & 2.131 & 0.73 & 10.46 & 1.02 & S0& $-$ & flat blue disk\\
10 & 41   & 10:16:28.4 & $+$03:35:02.7 & 0.04848 & $-$21.72 & 2.38  & 6.1 & 10.90 & 2.24 & S0&  all around &  red core, red structure $\dagger$ \\
11 & 50    & 13:57:07.5 & $+$05:15:06.8 & 0.03967 & $-$21.86 & 2.158 & 6.6 & 10.88 &  3.72 & E&  all around & red core\\ 
12 & 52    & 14:51:15.7 & $+$62:00:14.6 & 0.04306 & $-$21.45 & 2.173 & 3.9 & 10.88 & 5.33 & S0& one sided &  red core, red structure $\dagger$\\
13 & 56    & 17:23:24.9 & $+$27:48:46.3 & 0.04845 & $-$21.46 & 2.017 & 3.2 & 10.73 & 2.95 & E& one sided diffuse halo$\star$ &  blue core, blue ring structure $\dagger$\\
14 & 58   & 14:06:56.4 & $-$01:35:41.0 & 0.02916 & $-$21.37 & 2.127 & 5.7 & 10.71 & 3.79 & S0& one sided diffuse halo$\star$ &   red core\\
15 & 61    & 22:15:16.2 & $-$09:15:47.6 & 0.03843 & $-$21.61 & 1.707 & 21.0 & $-$  &$-$ &E&  all around & blue core, red ring structure $\dagger$\\
16 & 66    & 10:54:37.9 & $+$55:39:46.0 & 0.04787 & $-$20.89 & 1.976 & 6.5 & $-$ &$-$ & E& all around & red core, blue structure $\dagger$\\
17   &  68    & 03:01:26.2 & $-$00:04:25.5 & 0.04285 & $-$21.1  & 2.156 & 3.2 & 10.71 & 2.11& S0&  all around & red core, red elongated structure $\dagger$\\
18   &  72    & 09:13:23.7 & $+$43:58:34.2 & 0.04292 & $-$21.45 & 2.176 & 14.0 & 10.86 &  6.11 & E&  all around &red core\\
19   &  77    & 12:20:37.4 & $+$56:28:46.2 & 0.04381 & $-$20.84 & 2.238 & 0.53 & 10.57 & 0.86 & E& one sided diffuse halo$\star$ &  red core, blue structure \\
20   &  84    & 07:54:20.6 & $+$25:51:33.2 & 0.04167 & $-$21.06 & 2.011 & 1.7 & 10.51 & 1.57 & E& one sided &  red core,blue ring  \\
21   &  86    & 08:53:11.4 & $+$37:08:06.5 & 0.0498  & $-$21.59 & 2.063 & 10.0 & 10.99 & 3.93 & E& one sided diffuse halo$\star$ & red core \\
22   &  92    & 12:20:23.1 & $+$08:51:37.1 & 0.04894 & $-$20.88 & 2.086 & 3.6 & 10.62 & 2.81 & S0& one sided diffuse halo$\star$ & red core, red structure $\dagger$\\
23   &  101   & 14:02:48.8 & $+$52:30:00.8 & 0.04361 & $-$20.77 & 1.893 & 1.4 & $-$&$-$ & S0 & spiral feature outside galaxy & red core, blue structure $\dagger$\\
24   &  102   & 14:07:47.2 & $+$52:38:09.7 & 0.04381 & $-$20.78 & 1.812 & 4.2 & 10.44 & 3.08 & S0 &  all around & red core, blue structure  $\dagger$\\
25   &  103   & 15:50:00.5 & $+$41:58:11.2 & 0.03391 & $-$20.8  & 1.879 & 3.6 & 10.63 & 1.69 & E& one sided diffuse halo$\star$ & red core, blue structure $\dagger$\\
26   &  105   & 16:51:16.7 & $+$28:06:52.5 & 0.04724 & $-$21.59 & 1.921 & 4.0 & 10.81 & 4.14 & S0 & faint  all around & blue color structure from center to south $\dagger$\\
27   &  108   & 07:47:23.1 & $+$22:20:41.3 & 0.04549 & $-$20.9  & 2.256 & 0.45 & 10.66 & 0.80 & E& one sided diffuse halo$\star$ & red core \\
28   &  119   & 15:53:35.6 & $+$32:18:20.6 & 0.04985 & $-$21.07 & 1.789 & 3.2 & 10.41 &  2.69 & S0&  all around & blue core, blue structure $\dagger$\\
29   &  121   & 13:47:47.7 & $+$11:16:27.0 & 0.03942 & $-$21.2  & 1.911 & 5.5 & 10.60 & 3.63 & S0& one sided & red core, blue structure $\dagger$ \\
30   &  124   & 16:44:30.8 & $+$19:56:26.7 & 0.023   & $-$20.71 & 1.873 & 5.2 & 10.36 & 1.96 & S0&  all around & blue core, red structure $\dagger$\\
31   &  129   & 07:59:12.4 & $+$53:33:26.0 & 0.03479 & $-$20.92 & 1.657 & 13.0 & 10.41 & 5.7 & E&  all around & red core, blue structure $\dagger$\\ 
32   &  130   & 08:10:20.1 & $+$56:12:26.3 & 0.04623 & $-$20.78 & 2.13  & 1.0 & 10.60 & 1.34 & E&  all around & red core, blue structure\\
33   &  137   & 07:56:36.3 & $+$18:44:17.7 & 0.03988 & $-$21.53 & 2.088 & 3.9 & 10.80 & 2.13  & S0& one sided diffuse halo$\star$ & red core, blue structure\\
34   &  139   & 07:56:08.7 & $+$17:22:50.5 & 0.02899 & $-$20.73 & 2.211 & 1.6 & 10.60 & 1.06 & S0& one sided diffuse halo$\star$ & red core, red structure $\dagger$\\
35   &  145   & 15:18:09.6 & $+$25:42:11.5 & 0.0326  & $-$20.85 & 1.96  & 6.6 & 10.44 & 3.77 & Spiral & $-$ & red core, blue structure $\dagger$\\
36   &  146   & 16:07:54.0 & $+$20:03:03.8 & 0.03165 & $-$20.73 & 1.487 & 4.8 & 10.32 & 2.25 & E& faint  all around & blue core \\
37   &  147   & 13:26:20.8 & $+$31:41:59.9 & 0.04999 & $-$21.89 & 1.93  & 6.7 & 11.01 &  6.78& E& one sided diffuse halo$\star$ & red core, blue structure $\dagger$\\
38   &  148   & 10:20:34.9 & $+$29:14:10.8 & 0.04846 & $-$20.96 & 1.998 & 1.0 & 10.50 & 1.07 & S0& one sided diffuse halo$\star$ & blue structure\\
39   &  149   & 11:31:22.0 & $+$32:42:22.9 & 0.03368 & $-$21.61 & 2.015 & 6.3 & 10.76 & 4.33 & S0& faint all around & red core, blue structure\\
40   &  151   & 15:44:51.5 & $+$17:51:22.5 & 0.04521 & $-$21.32 & 1.994 & 3.7 & 10.78 & 3.85 & E& faint  all around & red core, blue structure\\
41   &  157   & 16:04:39.4 & $+$16:44:43.6 & 0.04599 & $-$20.74 & 2.308 & 0.54 & 10.48 & 0.29 & S0& one sided diffuse halo$\star$ &  red core, red structure \\
42   &  160   & 10:25:24.7 & $+$27:25:06.3 & 0.04973 & $-$20.98 & 2.256 & 3.1 & 10.68 & 1.41  & S0& faint  all around & red core, red structure $\dagger$\\
43   &  172   & 08:17:56.3 & $+$47:07:19.5 & 0.03901 & $-$21.06 & 2.309 & 8.1 & 10.85 & 1.21 & S0&  north and south & blue core, red structure $\dagger$\\
44   &  175   & 09:24:29.6 & $+$53:41:37.8 & 0.0459  & $-$20.8  & 1.997 & 0.63 & 10.49 & 1.47 & E& one sided diffuse halo$\star$ & red core, blue structure \\
45   &  177   & 13:01:41.4 & $+$04:40:49.9 & 0.03836 & $-$21.4  & 2.101 & 1.5 & 10.74 & 2.39 & E& blue diffuse on disk & red core, blue structure $\dagger$\\
46   &  180   & 12:06:17.0 & $+$63:38:19.0 & 0.03974 & $-$21.26 & 1.686 & 18.0 & 10.67 &  18.84 & E& faint on east and south & blue core, red structure $\dagger$ \\
47   &  182   & 14:32:22.7 & $+$56:51:08.4 & 0.04302 & $-$21.74 & 1.864 & 6.1 & 10.92 & 5.00 & E& faint all around & red core, blue structure \\
48   &  190   & 14:14:33.2 & $+$40:45:22.9 & 0.04185 & $-$20.86 & 1.614 & 5.2 &  $-$& $-$& S0& faint all around & red core, blue structure $\dagger$\\
49   &  192   & 11:52:05.0 & $+$45:57:06.6 & 0.04316 & $-$20.72 & 1.842 & 1.7 & 10.48 & 1.61 & E& diffuse one sided & red core, blue structure\\
50   &  195   & 14:53:23.4 & $+$39:04:13.6 & 0.03153 & $-$20.88 & 2.134 & 1.2 & 10.51 & 0.78 & E&  all around & flat blue disk \\
51   &  202   & 14:17:32.6 & $+$36:20:19.1 & 0.04712 & $-$20.93 & 1.916 & 1.6 & 10.51 & 2.50 & E& faint all around & red core, blue structure\\
52   &  206   & 14:37:33.0 & $+$08:04:43.0 & 0.04987 & $-$21.62 & 2.353 & 3.6 & 10.90 & 3.30 & S0& faint all around & red core, red structure $\dagger$\\
53   &  209   & 16:18:18.7 & $+$34:06:40.1 & 0.04733 & $-$21.58 & 2.307 & 2.9 & 10.79 & 2.21 & S0&  all around & red core, red structure $\dagger$\\
54   &  213   & 08:43:46.7 & $+$31:34:52.6 & 0.04756 & $-$20.7  & 2.11  & 7.3 & 10.25 & 3.58 & S0& one sided feature$\star$ &  red core, elongated blue structure \\
55   &  215   & 10:26:54.6 & $-$00:32:29.4 & 0.03463 & $-$21.38 & 2.137 & 7.6 & 10.71 & 4.74 & E&  all around & blue central region, red outskirts 
%with progressively redder and then blue outer regions $\dagger$\\
%\hline
%\end{tabular}
\end{longtable}
\tablefoot{Col. (1): Object ID assigned for the galaxies used in this work; Col. (2): ID for the galaxies as in S09; Col. (3) and (4): galaxy coordinates (epoch J2000); Col. (5): spectroscopic redshift of blue ETGs from SDSS; Col. (6): $r$-band absolute magnitude; Col. (7) : $u-r$ colors; Col. (8): H$\alpha$ SFRs; Col. (9): log stellar mass;  Col. (10):  SFRs from the spectral energy distribution; Col. (11), (12) and (13): early-type morphology (E or S0) assigned by visual check of deeper imaging data, details obtained from the visual inspection of the $rgz$ color composite, $r$-band imaging and $g-r$ color map of 55 blue star-forming early-type galaxies. $\star$ Feature seen in large field smoothed image. $\dagger$ $g-r$ color structure used in Figure~\ref{figure:fig6} .}
\end{table*}
%\end{sidewaystable*}
%\end{adjustbox}

%\twocolumn

\section{Summary}

We presented a detailed imaging and color map analysis from the deeper $g,r,z$ legacy survey imaging data of 55 star-forming blue ETGs. These galaxies contribute a small fraction of the low-redshift ETG population. All the galaxies except for one fall on the starburst or star-forming main sequence in the M$\star$-SFR plot of local Universe galaxies. We found that the deeper $r$-band imaging data of the 37 galaxies show features that might be due to recent merger events. We were able to identify tidal features in 15 other galaxies based on a larger field analysis of the smoothed imaging data. The optical $g-r$ color map reveals diverse features such as a blue and red core and red and blue patches in the main body of galaxy. We estimated the surface brightness of the features seen outside the galaxies and color structures seen in the main body of the galaxy. The blue and red structures indicate a recent wet merger event that brought fuel for star formation in the main body of the galaxy. The star formation in these galaxies will most likely quench when the recently acquired fuel from a wet merger is consumed, and the galaxies will move from the main sequence to a passive population of red and dead galaxies such as normal early-type galaxies in the local Universe. The star-forming blue early-type phase can be a phase in the formation and evolution of normal early-type galaxy as revealed from their low number fraction in the local Universe.

\begin{acknowledgements}
The Legacy Surveys consist of three individual and complementary projects: the Dark Energy Camera Legacy Survey (DECaLS; Proposal ID 2014B-0404; PIs: David Schlegel and Arjun Dey), the Beijing-Arizona Sky Survey (BASS; NOAO Prop. ID 2015A-0801; PIs: Zhou Xu and Xiaohui Fan), and the Mayall z-band Legacy Survey (MzLS; Prop. ID 2016A-0453; PI: Arjun Dey). DECaLS, BASS and MzLS together include data obtained, respectively, at the Blanco telescope, Cerro Tololo Inter-American Observatory, NSF’s NOIRLab; the Bok telescope, Steward Observatory, University of Arizona; and the Mayall telescope, Kitt Peak National Observatory, NOIRLab. Pipeline processing and analyses of the data were supported by NOIRLab and the Lawrence Berkeley National Laboratory (LBNL). The Legacy Surveys project is honored to be permitted to conduct astronomical research on Iolkam Du’ag (Kitt Peak), a mountain with particular significance to the Tohono O’odham Nation.
NOIRLab is operated by the Association of Universities for Research in Astronomy (AURA) under a cooperative agreement with the National Science Foundation. LBNL is managed by the Regents of the University of California under contract to the U.S. Department of Energy.
This project used data obtained with the Dark Energy Camera (DECam), which was constructed by the Dark Energy Survey (DES) collaboration. Funding for the DES Projects has been provided by the U.S. Department of Energy, the U.S. National Science Foundation, the Ministry of Science and Education of Spain, the Science and Technology Facilities Council of the United Kingdom, the Higher Education Funding Council for England, the National Center for Supercomputing Applications at the University of Illinois at Urbana-Champaign, the Kavli Institute of Cosmological Physics at the University of Chicago, Center for Cosmology and Astro-Particle Physics at the Ohio State University, the Mitchell Institute for Fundamental Physics and Astronomy at Texas A\&M University, Financiadora de Estudos e Projetos, Fundacao Carlos Chagas Filho de Amparo, Financiadora de Estudos e Projetos, Fundacao Carlos Chagas Filho de Amparo a Pesquisa do Estado do Rio de Janeiro, Conselho Nacional de Desenvolvimento Cientifico e Tecnologico and the Ministerio da Ciencia, Tecnologia e Inovacao, the Deutsche Forschungsgemeinschaft and the Collaborating Institutions in the Dark Energy Survey. The Collaborating Institutions are Argonne National Laboratory, the University of California at Santa Cruz, the University of Cambridge, Centro de Investigaciones Energeticas, Medioambientales y Tecnologicas-Madrid, the University of Chicago, University College London, the DES-Brazil Consortium, the University of Edinburgh, the Eidgenossische Technische Hochschule (ETH) Zurich, Fermi National Accelerator Laboratory, the University of Illinois at Urbana-Champaign, the Institut de Ciencies de l’Espai (IEEC/CSIC), the Institut de Fisica d’Altes Energies, Lawrence Berkeley National Laboratory, the Ludwig Maximilians Universitat Munchen and the associated Excellence Cluster Universe, the University of Michigan, NSF’s NOIRLab, the University of Nottingham, the Ohio State University, the University of Pennsylvania, the University of Portsmouth, SLAC National Accelerator Laboratory, Stanford University, the University of Sussex, and Texas A\&M University.
BASS is a key project of the Telescope Access Program (TAP), which has been funded by the National Astronomical Observatories of China, the Chinese Academy of Sciences (the Strategic Priority Research Program “The Emergence of Cosmological Structures” Grant  XDB09000000), and the Special Fund for Astronomy from the Ministry of Finance. The BASS is also supported by the External Cooperation Program of Chinese Academy of Sciences (Grant  114A11KYSB20160057), and Chinese National Natural Science Foundation (Grant  12120101003,  11433005).
The Legacy Survey team makes use of data products from the Near-Earth Object Wide-field Infrared Survey Explorer (NEOWISE), which is a project of the Jet Propulsion Laboratory/California Institute of Technology. NEOWISE is funded by the National Aeronautics and Space Administration.
The Legacy Surveys imaging of the DESI footprint is supported by the Director, Office of Science, Office of High Energy Physics of the U.S. Department of Energy under Contract No. DE-AC02-05CH1123, by the National Energy Research Scientific Computing Center, a DOE Office of Science User Facility under the same contract; and by the U.S. National Science Foundation, Division of Astronomical Sciences under Contract No. AST-0950945 to NOAO.
\end{acknowledgements}

%__________________________________________________________________

%
   
%
%______________________________________________________________

%-------------------------------------------------------------------

\appendix
\renewcommand\thefigure{\thesection.\arabic{figure}}  
\section{}
\setcounter{figure}{0}

\begin{figure*}
\centering
\begin{multicols}{3}
    \includegraphics[width=6.0cm]{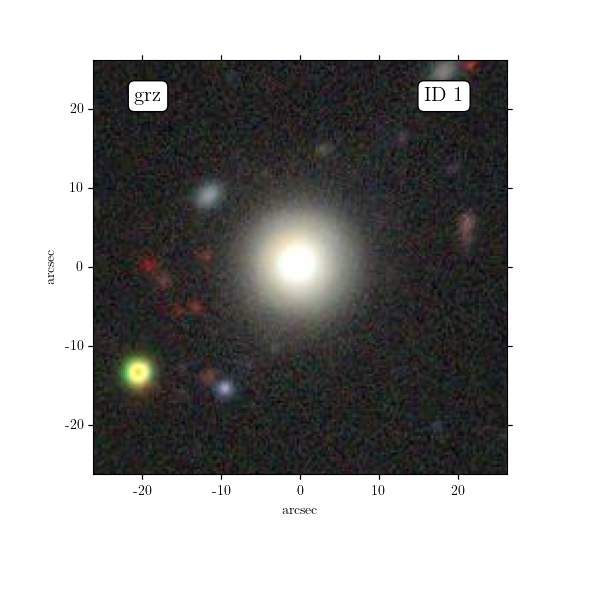}\par 
    \includegraphics[width=6.0cm]{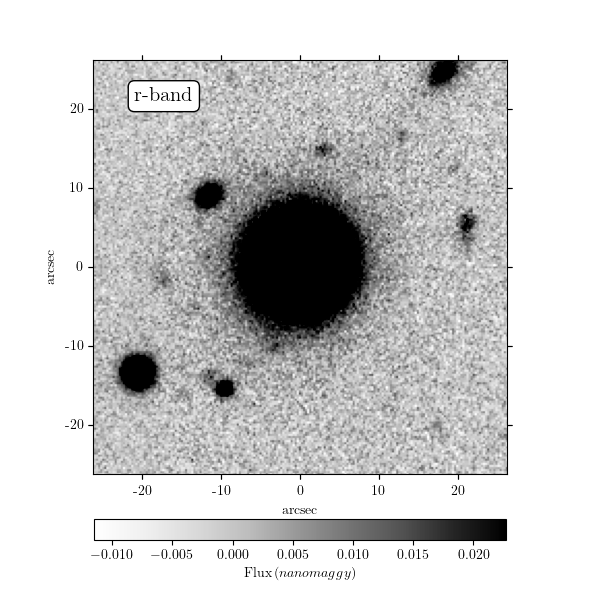}\par 
    \includegraphics[width=6.0cm]{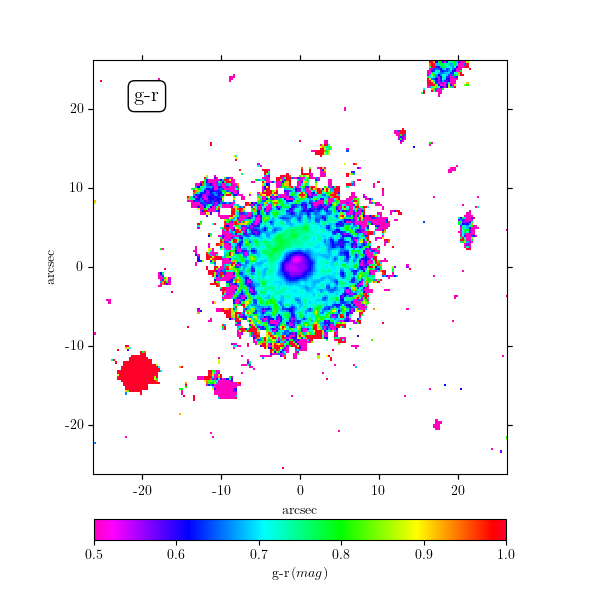}\par
 \end{multicols}
\begin{multicols}{3}
    \includegraphics[width=6.0cm]{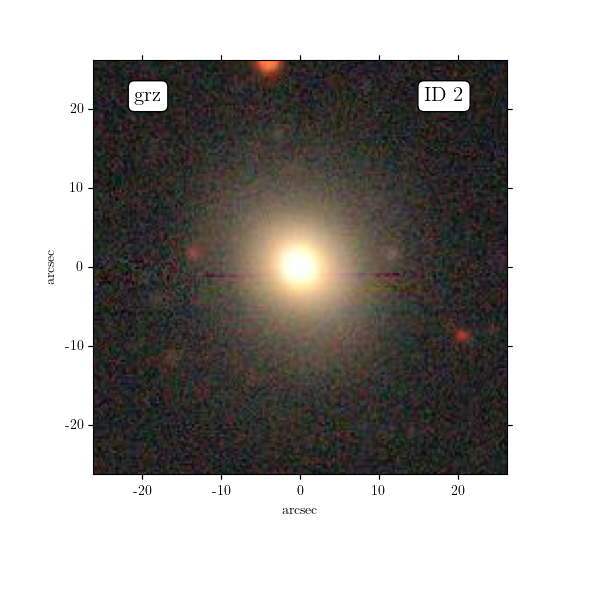}\par 
    \includegraphics[width=6.0cm]{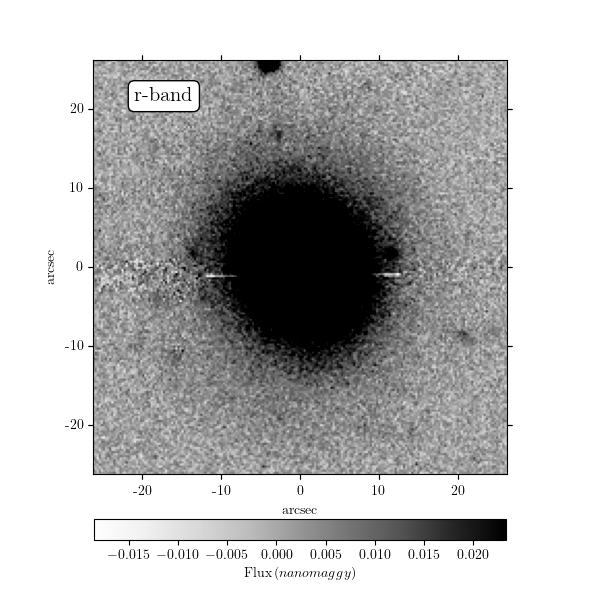}\par 
    \includegraphics[width=6.0cm]{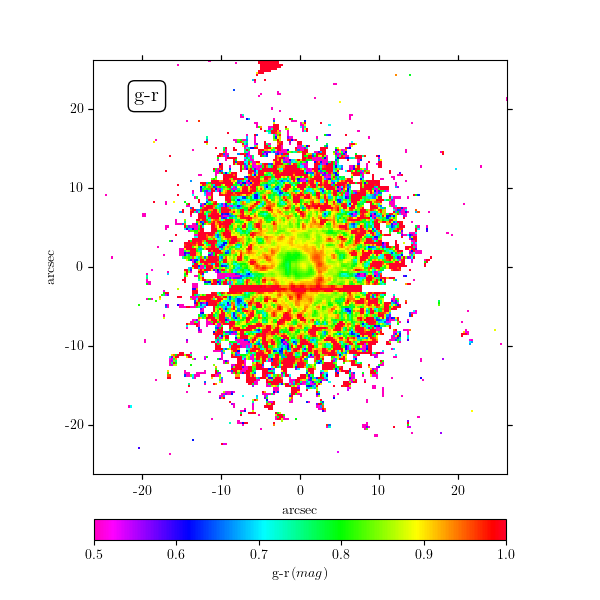}\par
 \end{multicols}
 \begin{multicols}{3}
    \includegraphics[width=6.0cm]{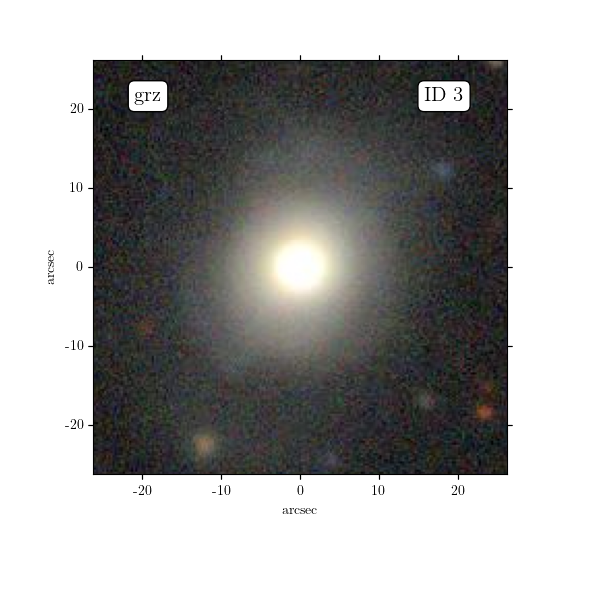}\par 
    \includegraphics[width=6.0cm]{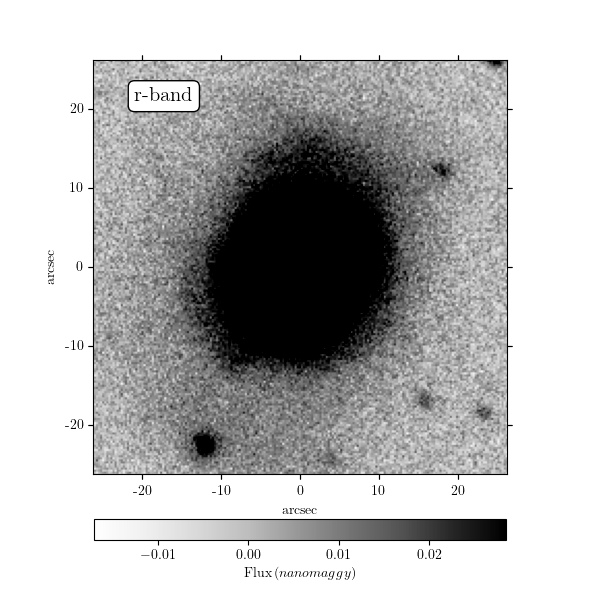}\par 
    \includegraphics[width=6.0cm]{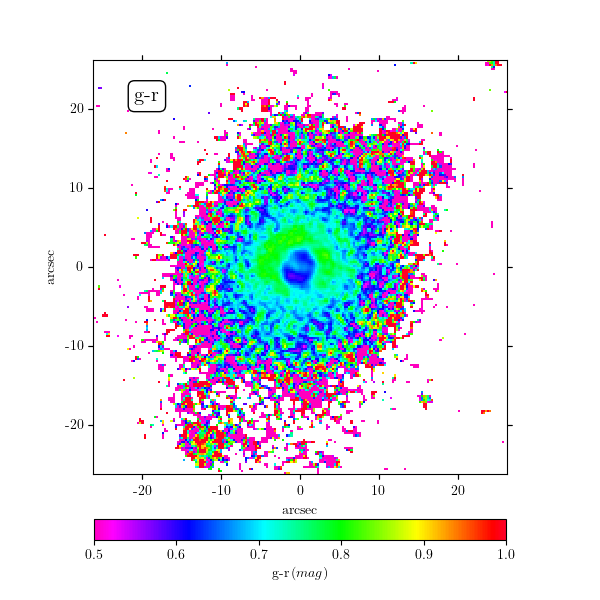}\par
 \end{multicols}
  \begin{multicols}{3}
    \includegraphics[width=6.0cm]{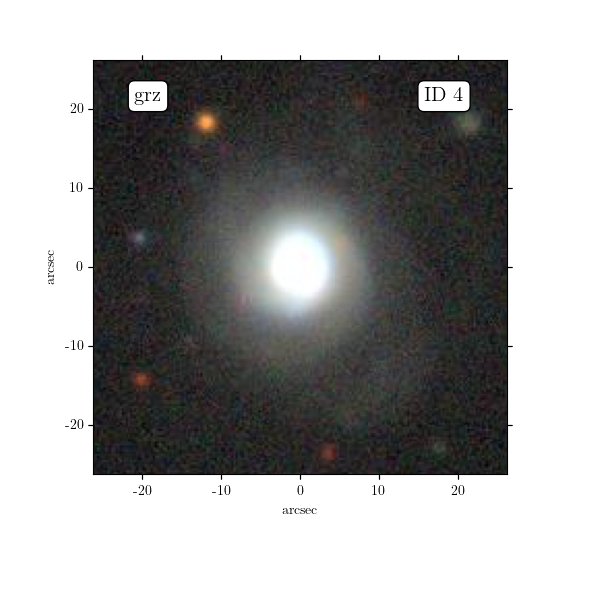}\par 
    \includegraphics[width=6.0cm]{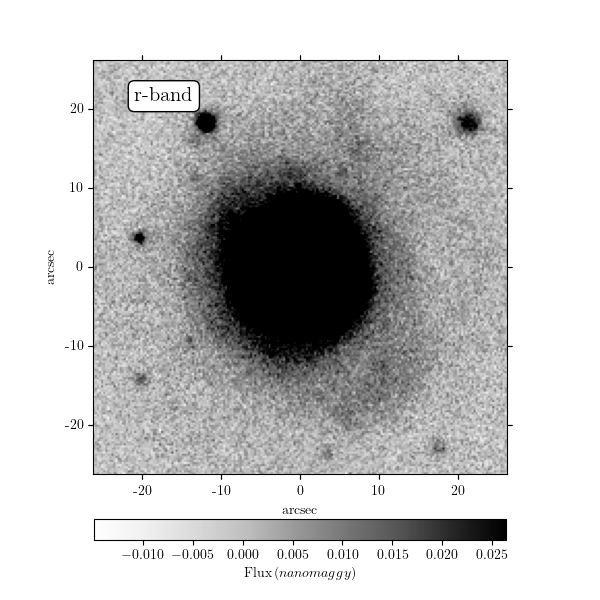}\par 
    \includegraphics[width=6.0cm]{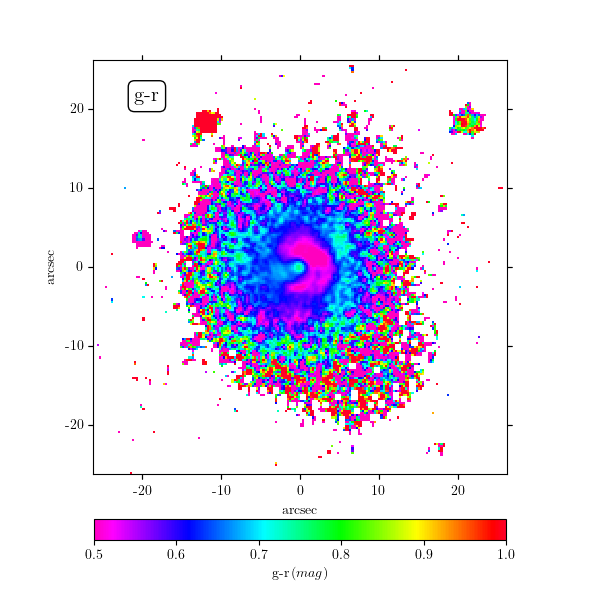}\par
 \end{multicols}

 \end{figure*}

 \begin{figure*}
%\ContinuedFloat 
 \begin{multicols}{3}
    \includegraphics[width=6.0cm]{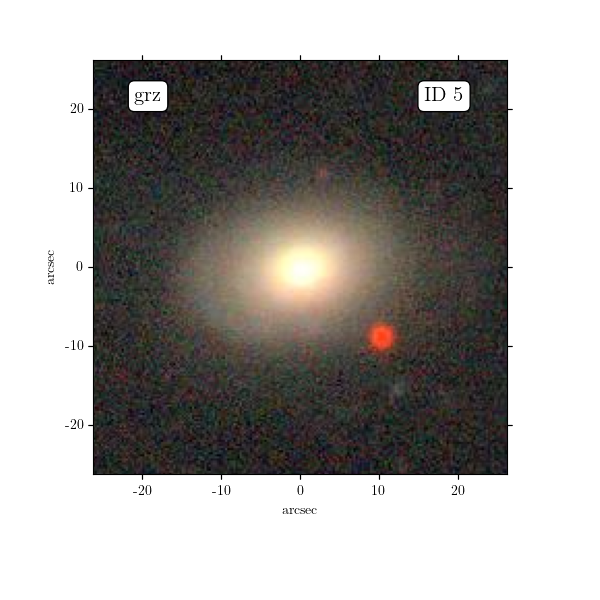}\par 
    \includegraphics[width=6.0cm]{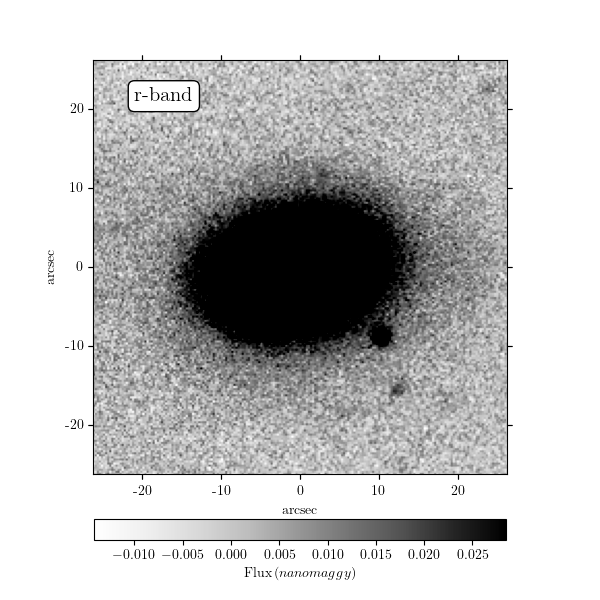}\par 
    \includegraphics[width=6.0cm]{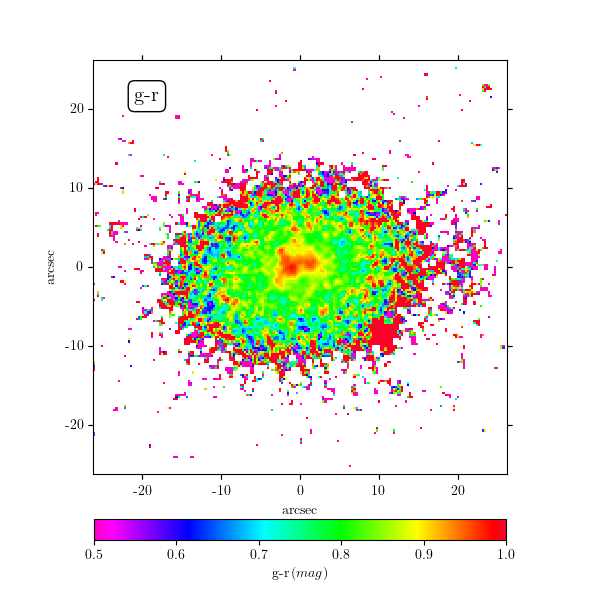}\par
 \end{multicols}
 \begin{multicols}{3}
    \includegraphics[width=6.0cm]{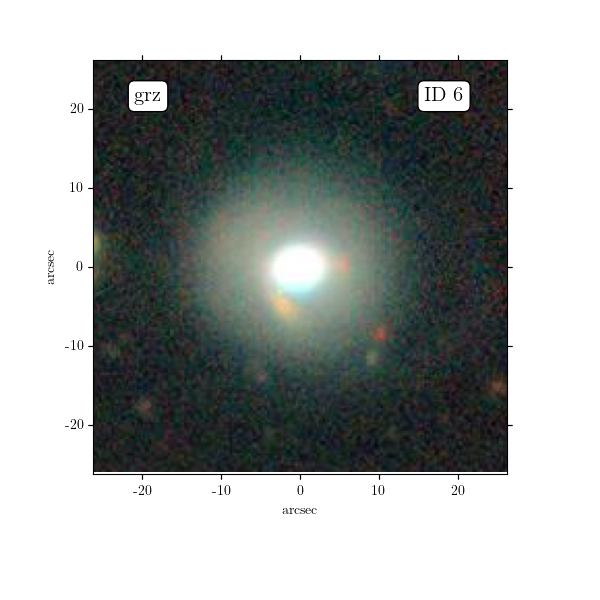}\par 
    \includegraphics[width=6.0cm]{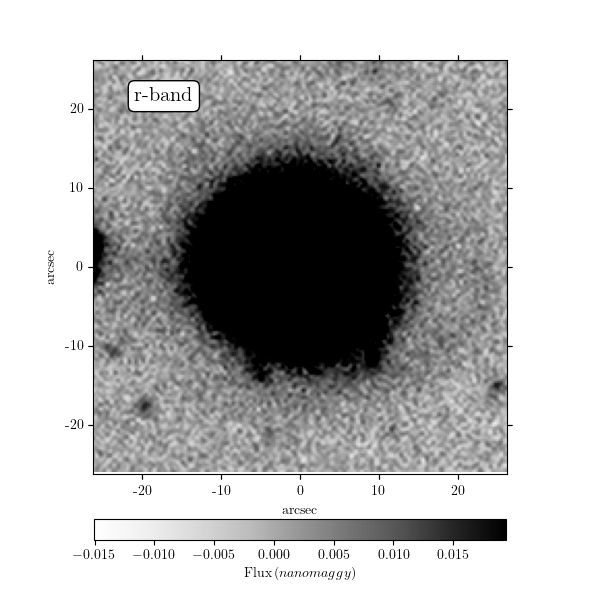}\par 
    \includegraphics[width=6.0cm]{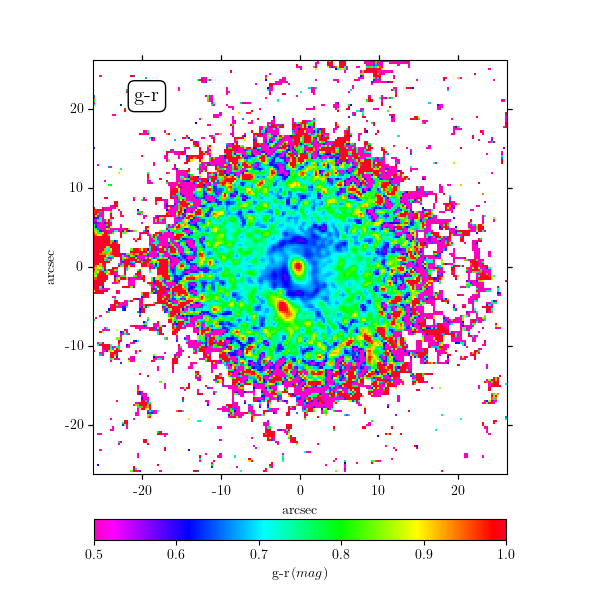}\par
 \end{multicols}
 \begin{multicols}{3}
    \includegraphics[width=6.0cm]{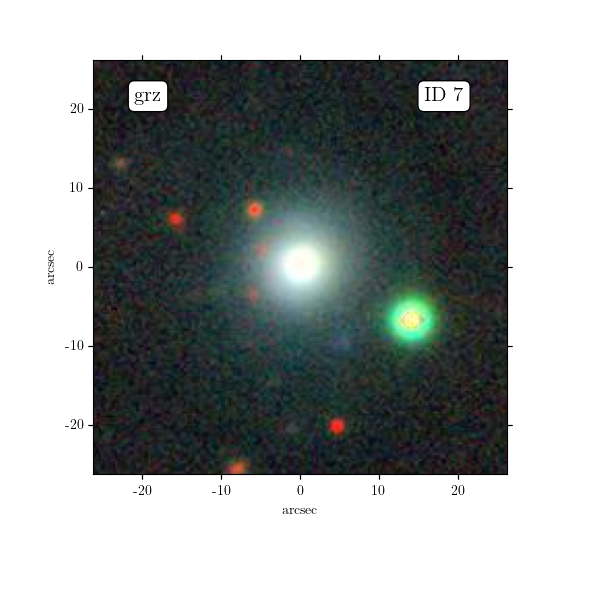}\par 
    \includegraphics[width=6.0cm]{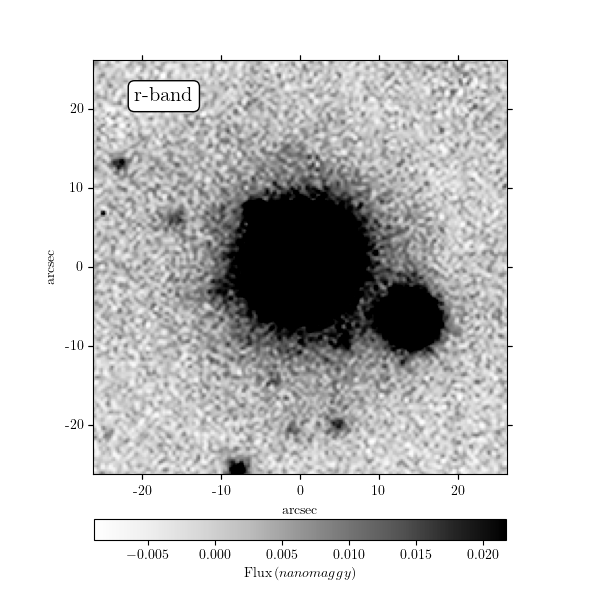}\par 
    \includegraphics[width=6.0cm]{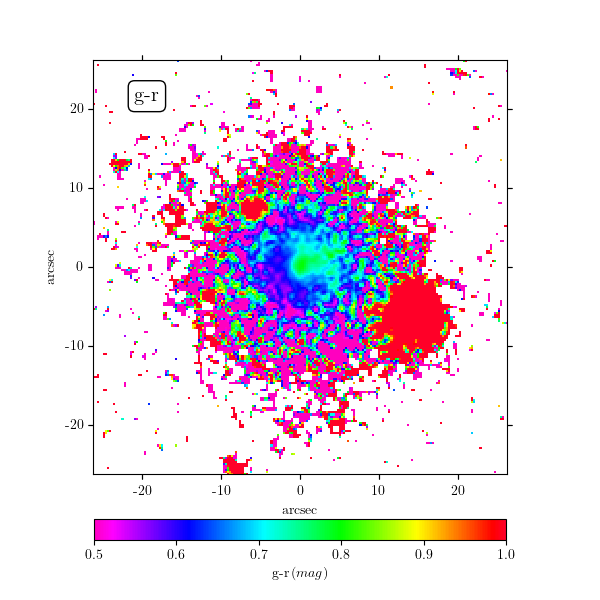}\par
 \end{multicols}
  \begin{multicols}{3}
     \includegraphics[width=6.0cm]{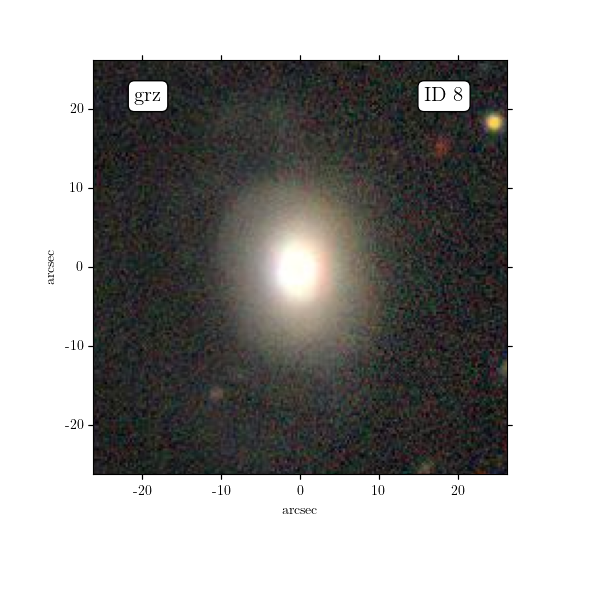}\par 
    \includegraphics[width=6.0cm]{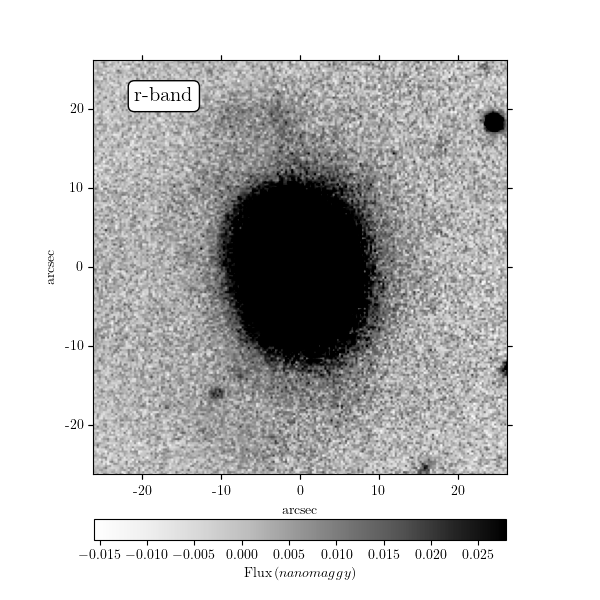}\par 
    \includegraphics[width=6.0cm]{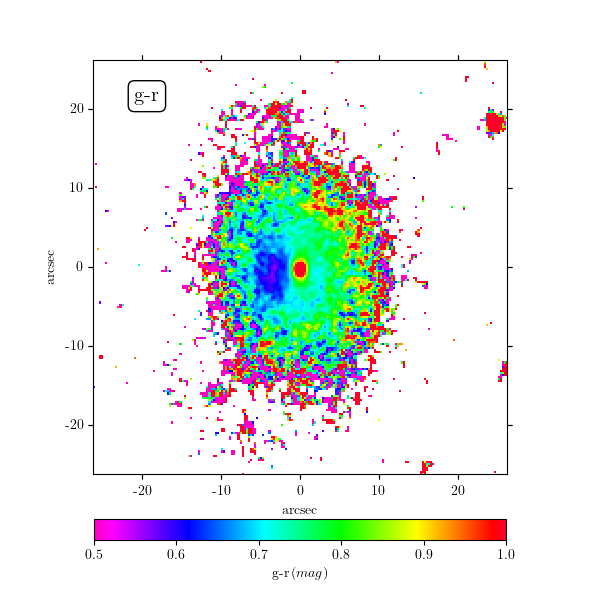}\par
 \end{multicols}
  \end{figure*}

\begin{figure*}
%\ContinuedFloat 
 \begin{multicols}{3}
    \includegraphics[width=6.0cm]{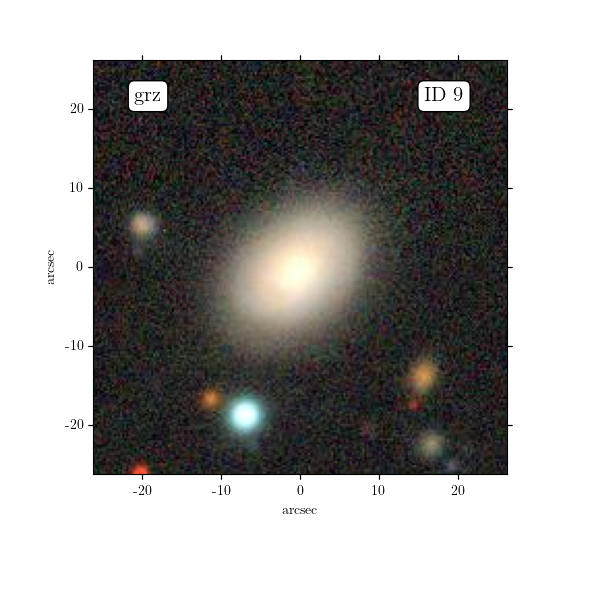}\par 
    \includegraphics[width=6.0cm]{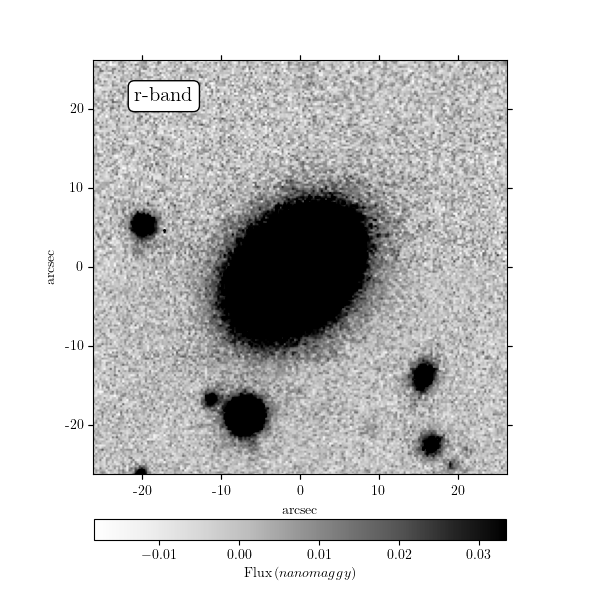}\par 
    \includegraphics[width=6.0cm]{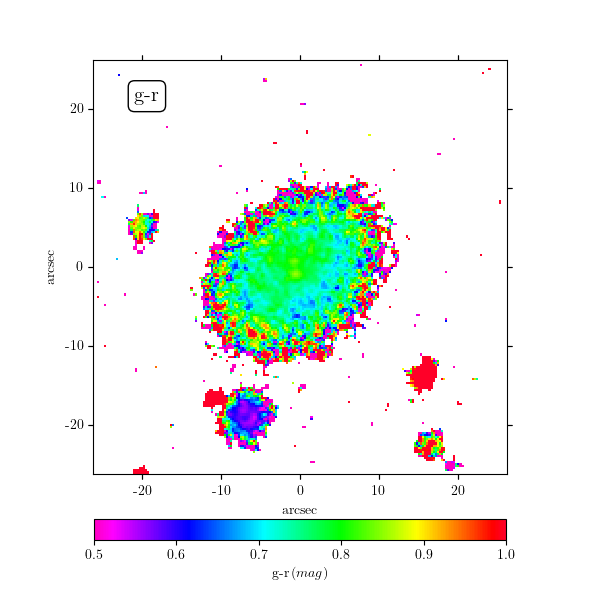}\par
 \end{multicols}
 \begin{multicols}{3}
    \includegraphics[width=6.0cm]{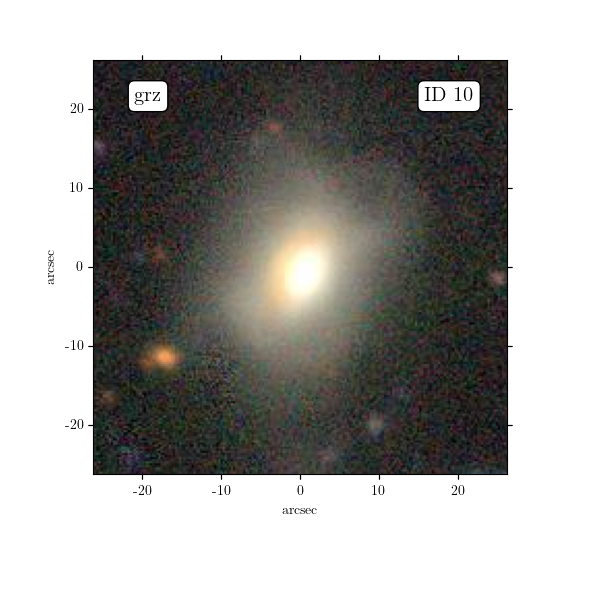}\par 
    \includegraphics[width=6.0cm]{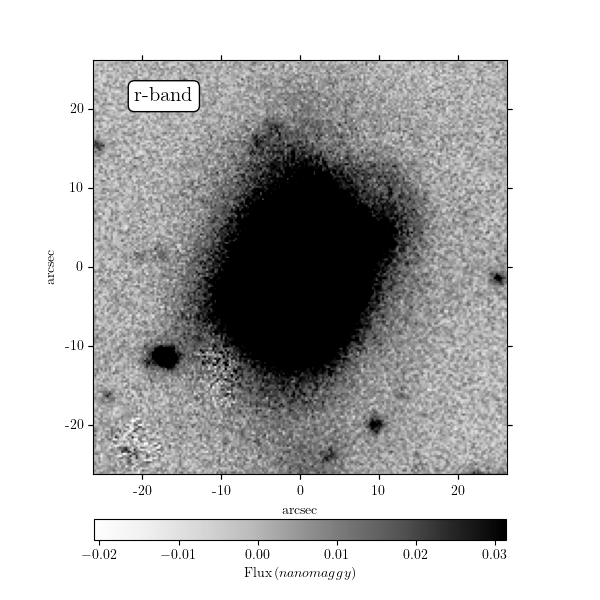}\par 
    \includegraphics[width=6.0cm]{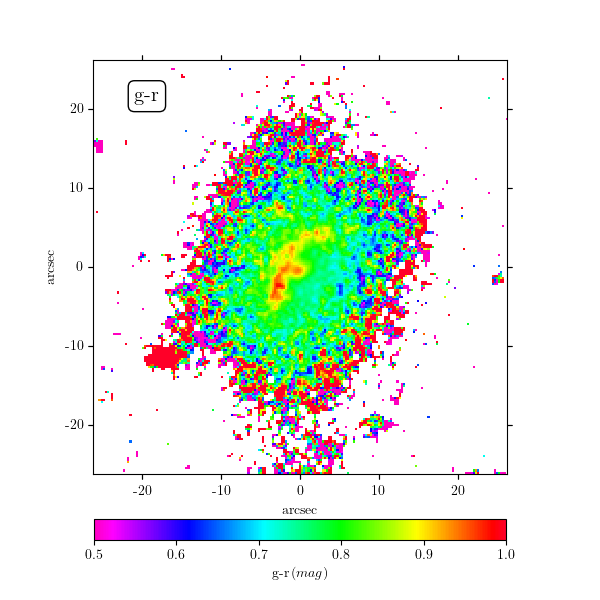}\par
 \end{multicols}
 \begin{multicols}{3}
    \includegraphics[width=6.0cm]{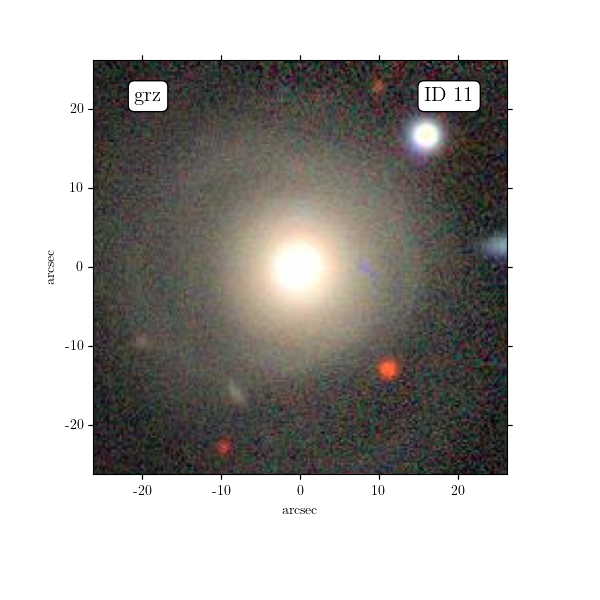}\par 
    \includegraphics[width=6.0cm]{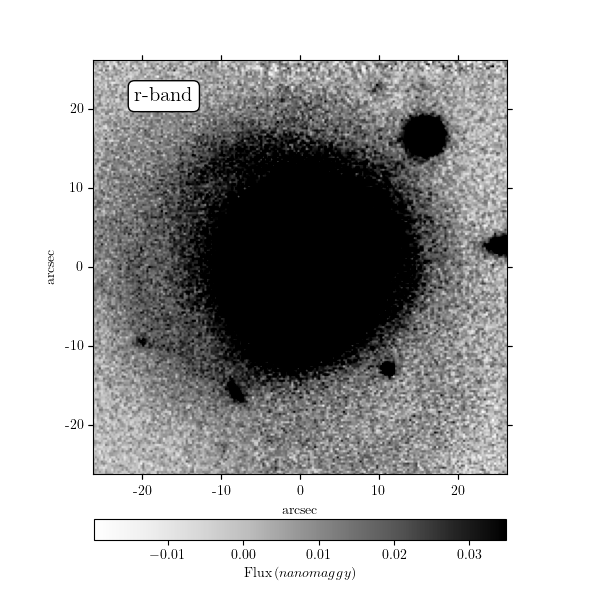}\par 
    \includegraphics[width=6.0cm]{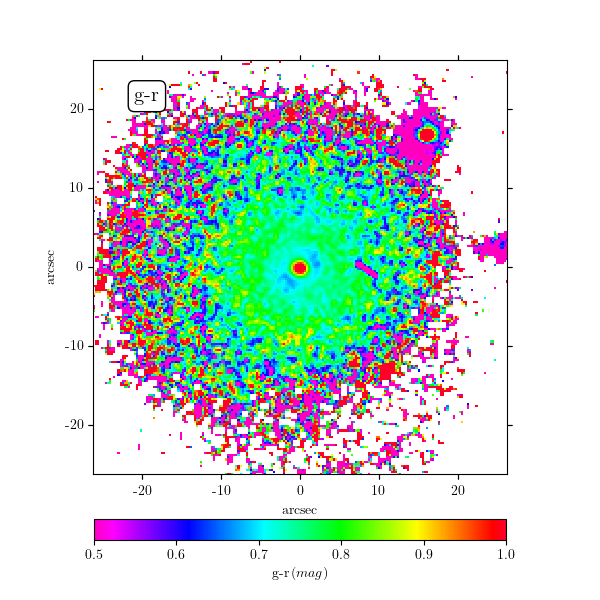}\par
 \end{multicols}
  \begin{multicols}{3}
    \includegraphics[width=6.0cm]{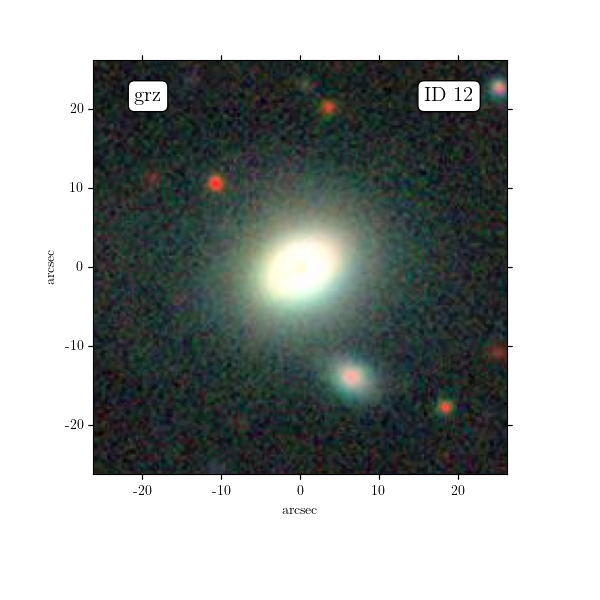}\par 
    \includegraphics[width=6.0cm]{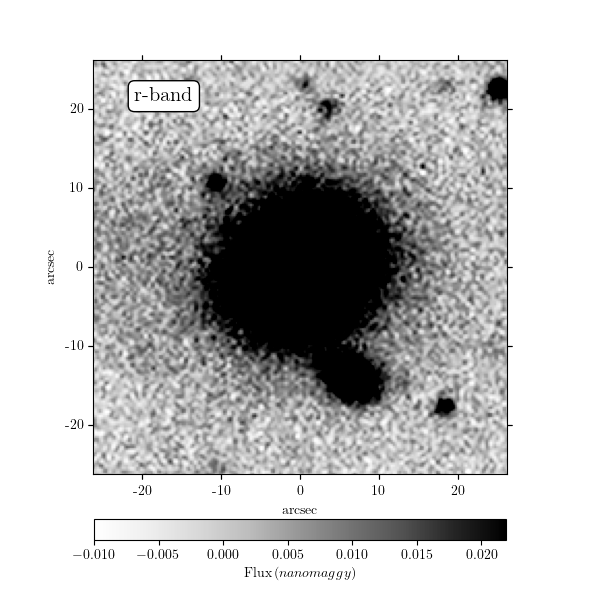}\par 
    \includegraphics[width=6.0cm]{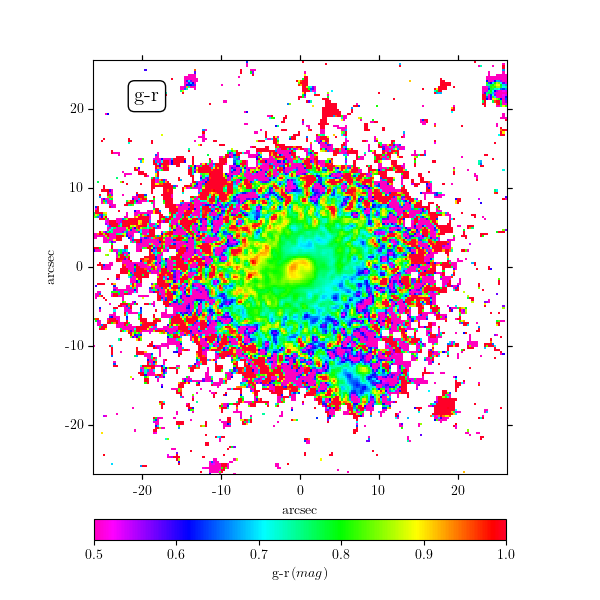}\par
 \end{multicols}
  \end{figure*}

\begin{figure*}
%\ContinuedFloat 
 \begin{multicols}{3}
    \includegraphics[width=6.0cm]{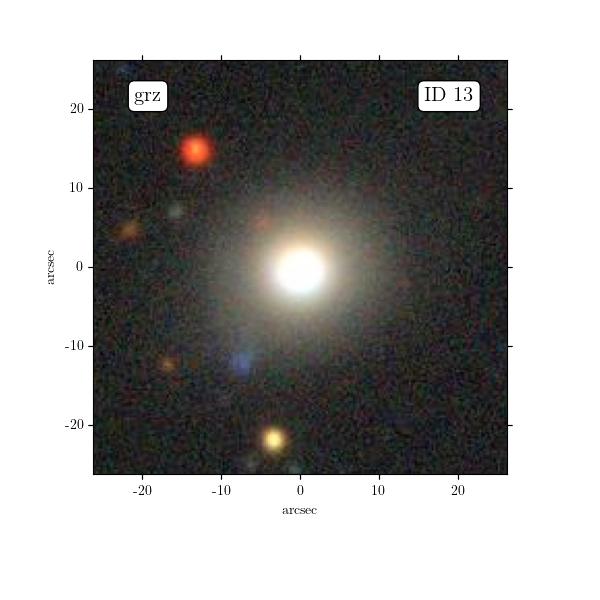}\par 
    \includegraphics[width=6.0cm]{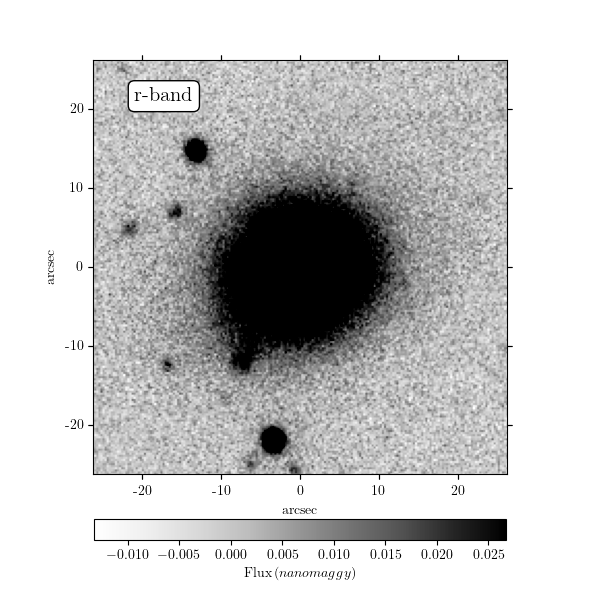}\par 
    \includegraphics[width=6.0cm]{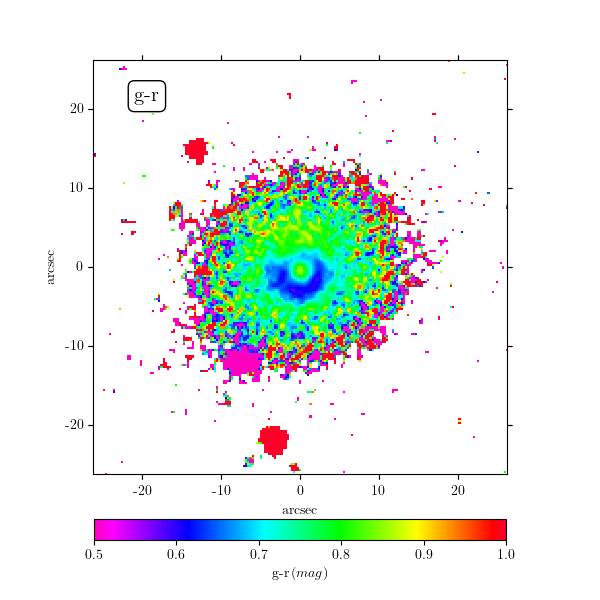}\par
 \end{multicols}
 \begin{multicols}{3}
    \includegraphics[width=6.0cm]{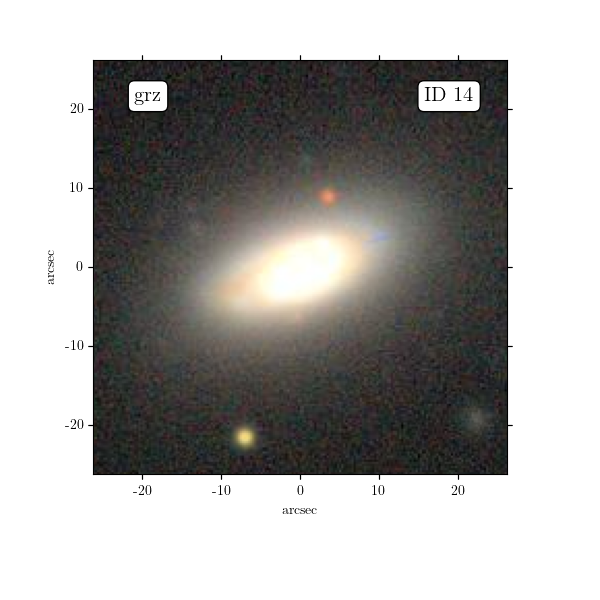}\par 
    \includegraphics[width=6.0cm]{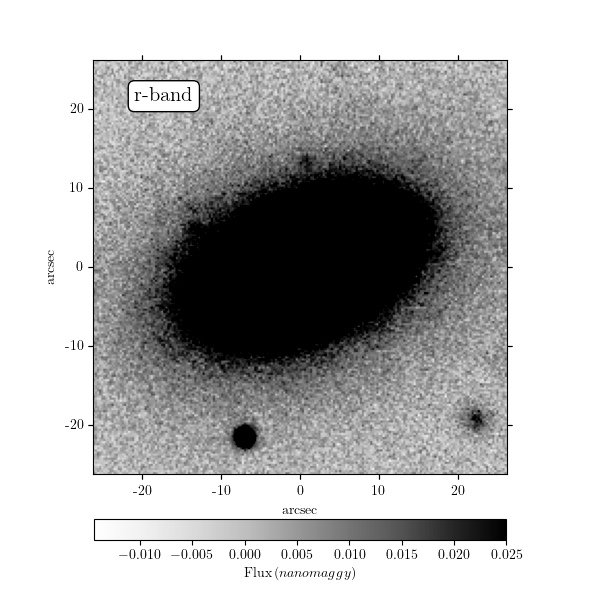}\par 
    \includegraphics[width=6.0cm]{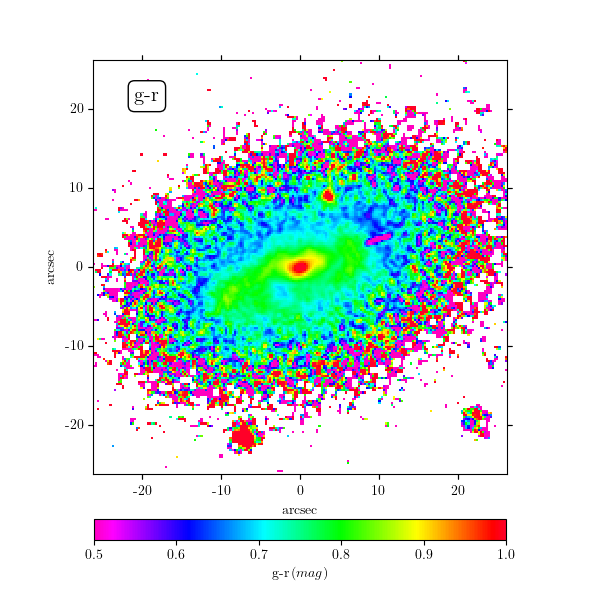}\par
 \end{multicols}
 \begin{multicols}{3}
    \includegraphics[width=6.0cm]{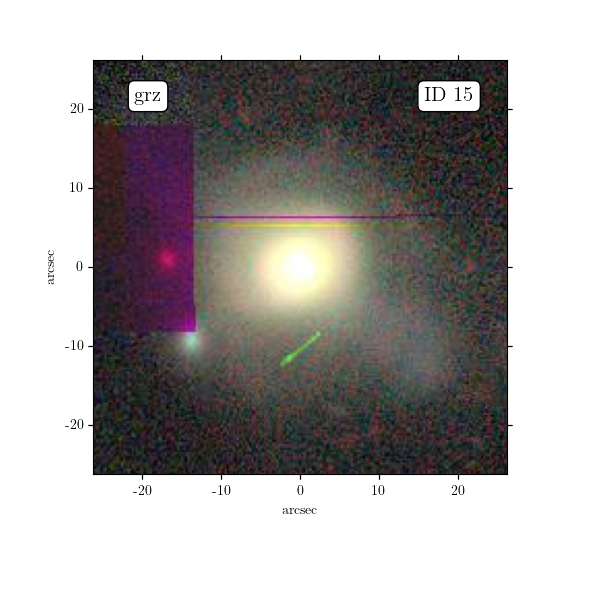}\par 
    \includegraphics[width=6.0cm]{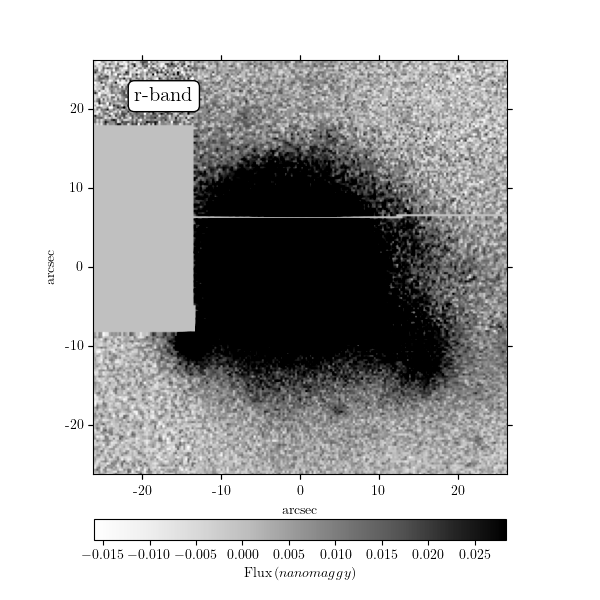}\par 
    \includegraphics[width=6.0cm]{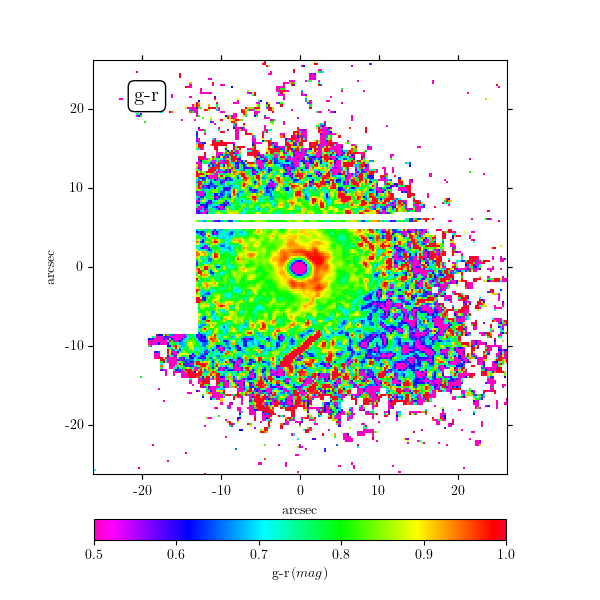}\par
 \end{multicols}
  \begin{multicols}{3}
    \includegraphics[width=6.0cm]{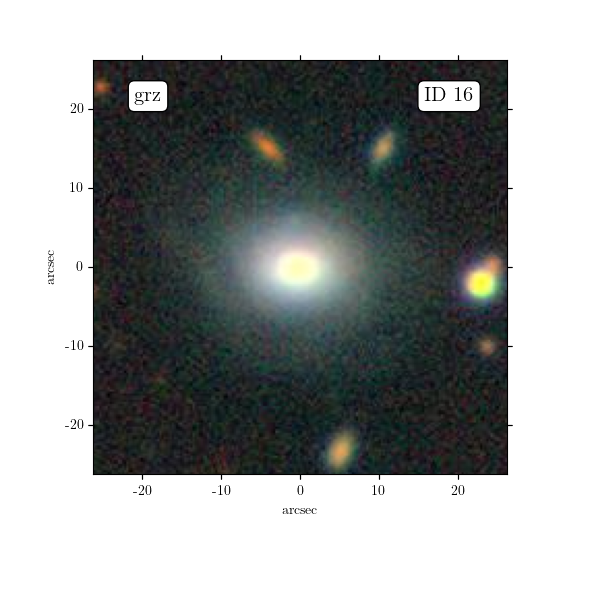}\par 
    \includegraphics[width=6.0cm]{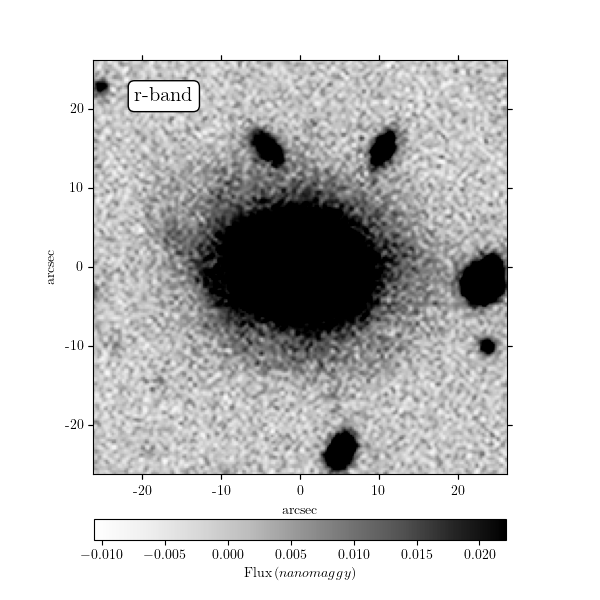}\par 
    \includegraphics[width=6.0cm]{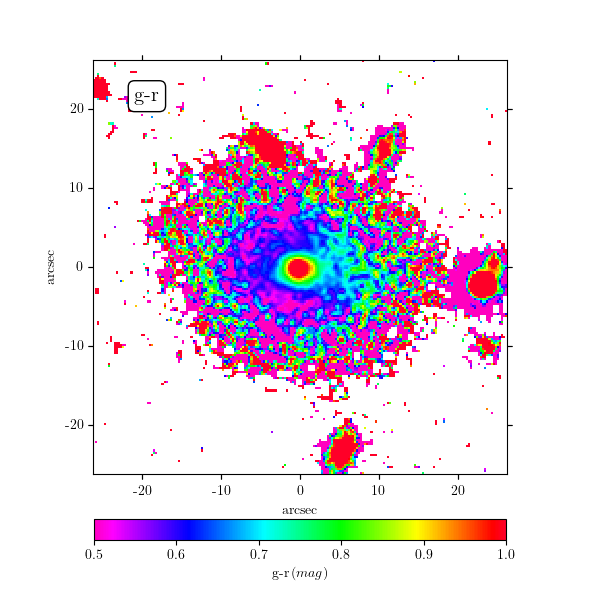}\par
 \end{multicols}
  \end{figure*}

  \begin{figure*}
%\ContinuedFloat 
 \begin{multicols}{3}
    \includegraphics[width=6.0cm]{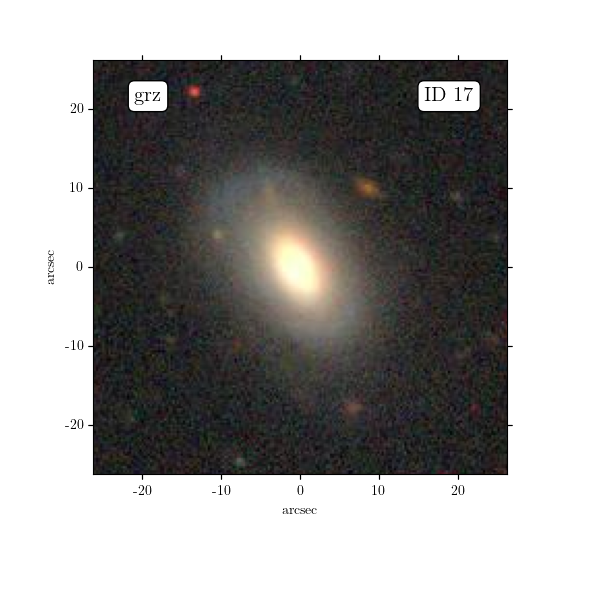}\par 
    \includegraphics[width=6.0cm]{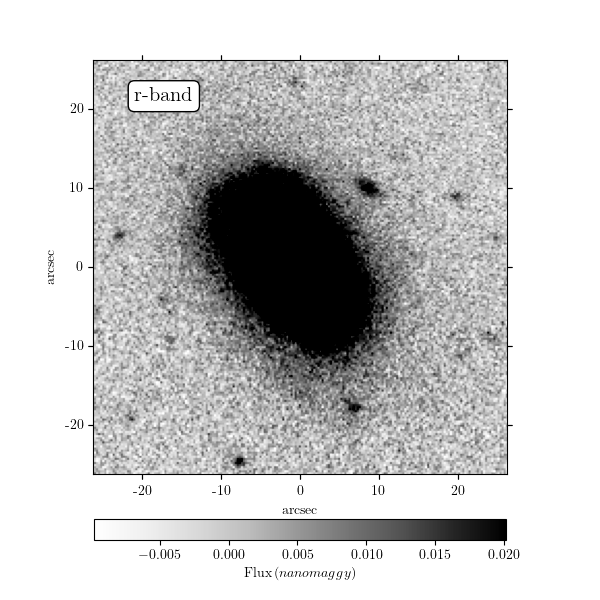}\par 
    \includegraphics[width=6.0cm]{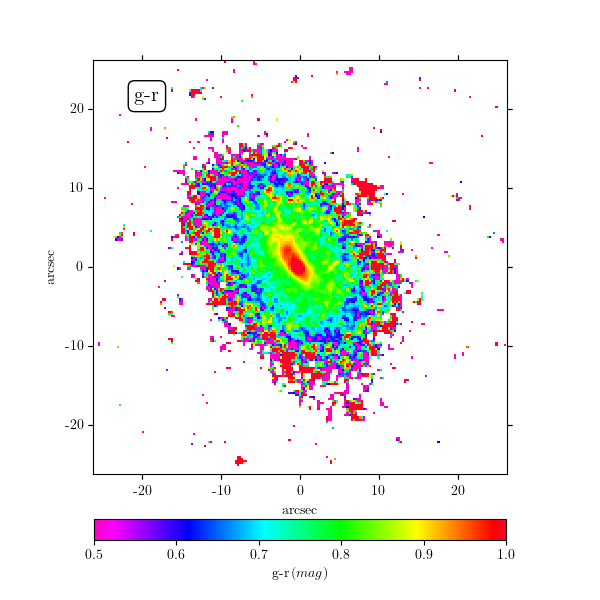}\par
 \end{multicols}
 \begin{multicols}{3}
    \includegraphics[width=6.0cm]{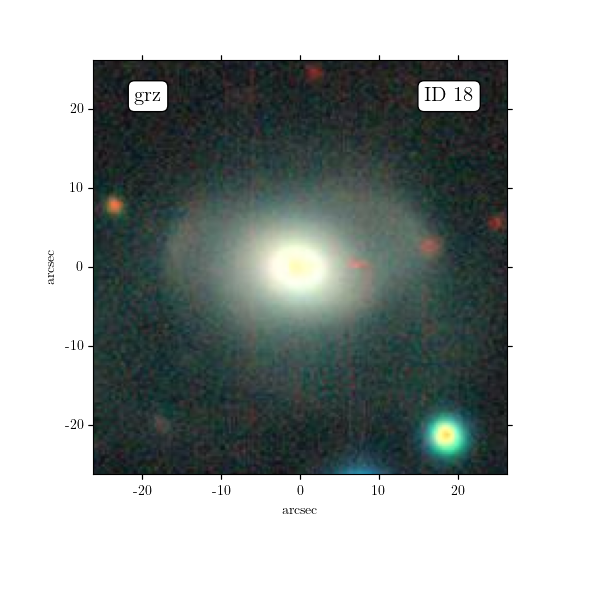}\par 
    \includegraphics[width=6.0cm]{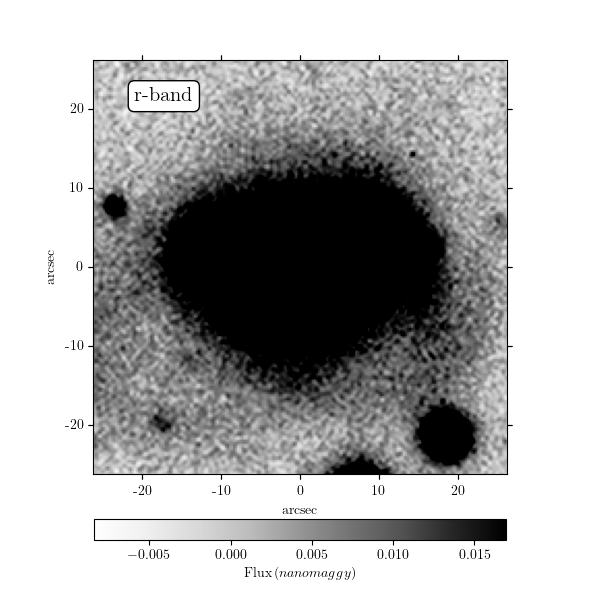}\par 
    \includegraphics[width=6.0cm]{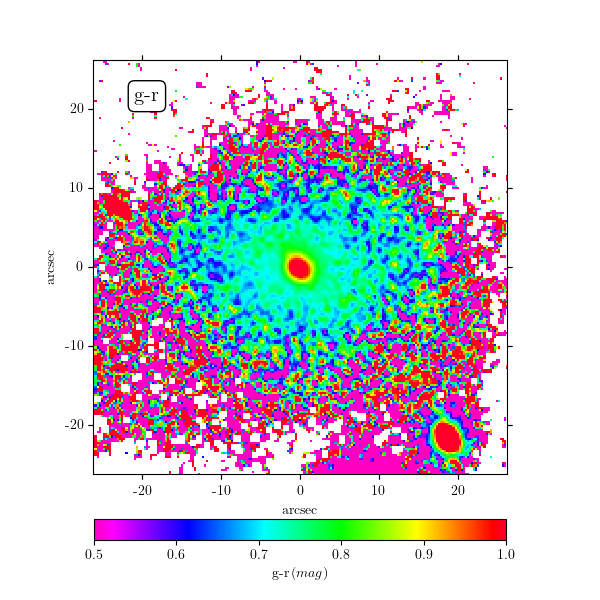}\par
 \end{multicols}
 \begin{multicols}{3}
    \includegraphics[width=6.0cm]{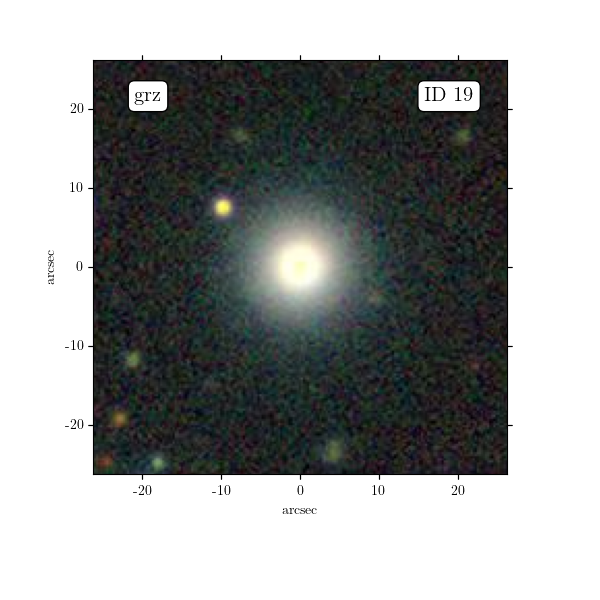}\par 
    \includegraphics[width=6.0cm]{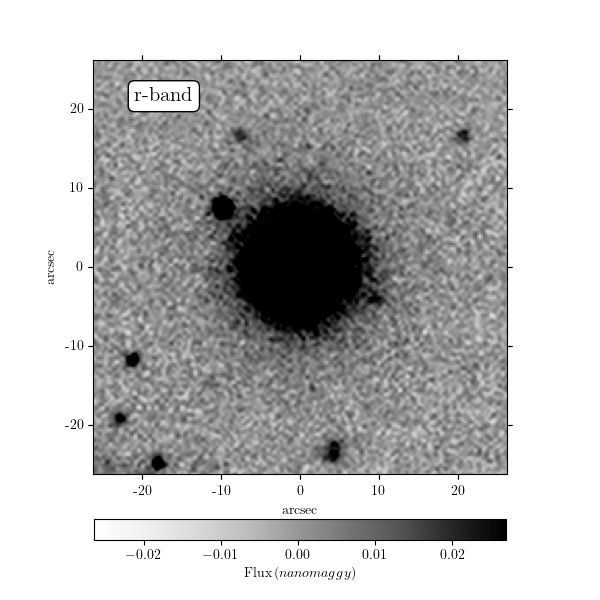}\par 
    \includegraphics[width=6.0cm]{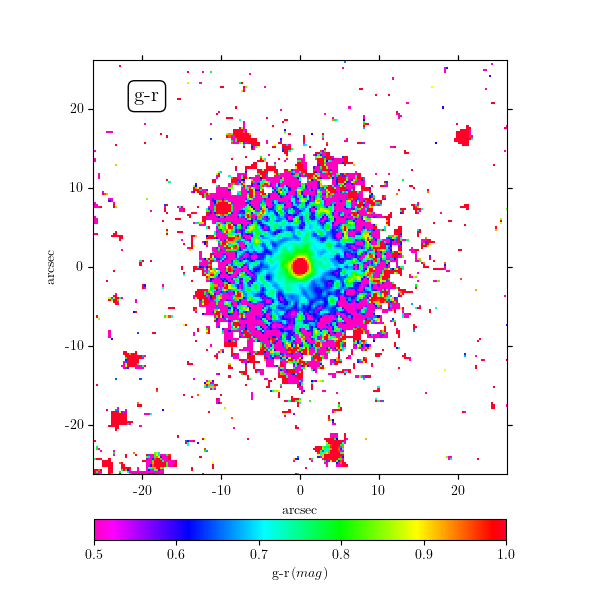}\par
 \end{multicols}
  \begin{multicols}{3}
    \includegraphics[width=6.0cm]{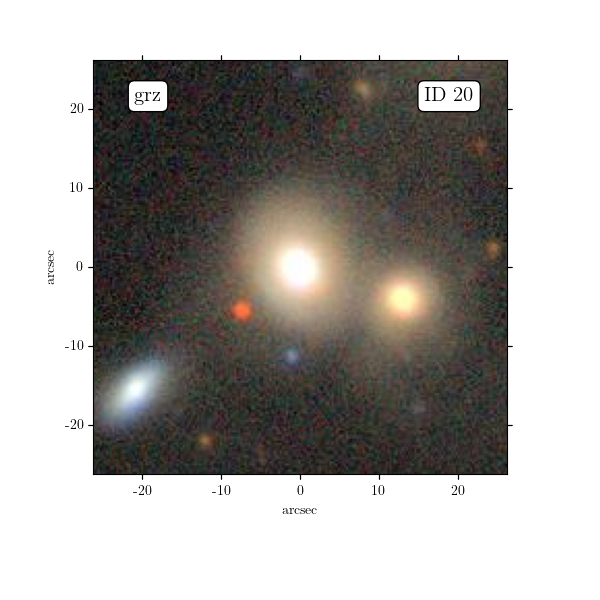}\par 
    \includegraphics[width=6.0cm]{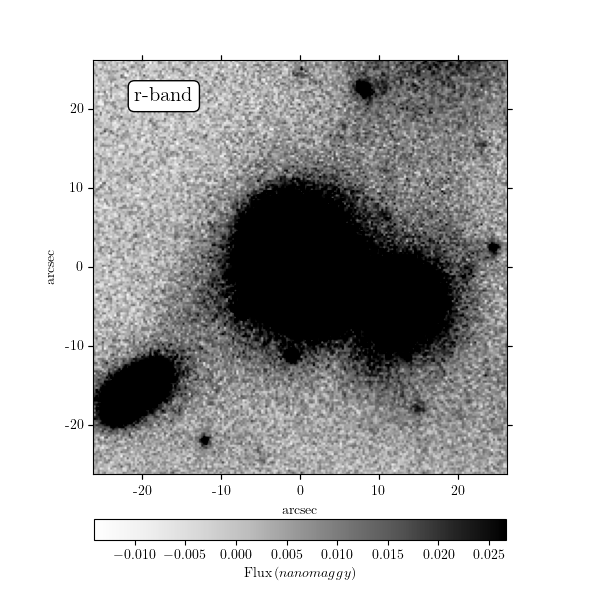}\par 
    \includegraphics[width=6.0cm]{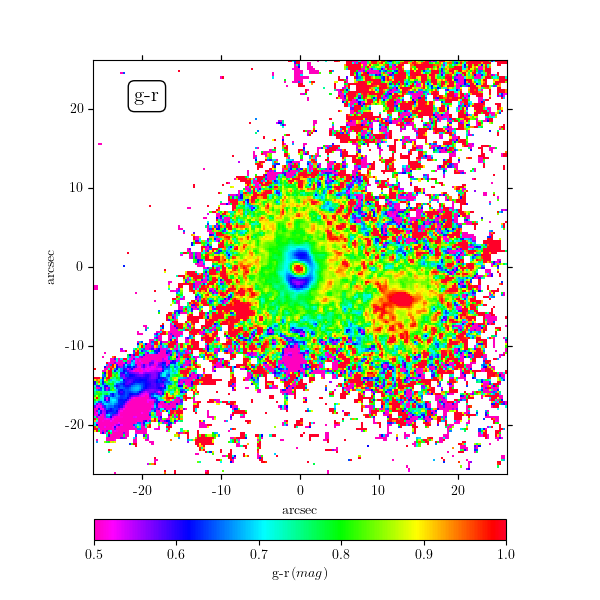}\par
 \end{multicols}
  \end{figure*}

  \begin{figure*}
%\ContinuedFloat 
 \begin{multicols}{3}
    \includegraphics[width=6.0cm]{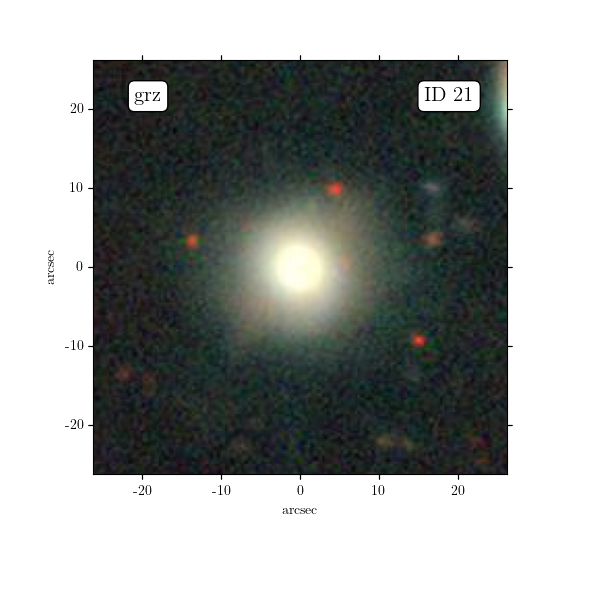}\par 
    \includegraphics[width=6.0cm]{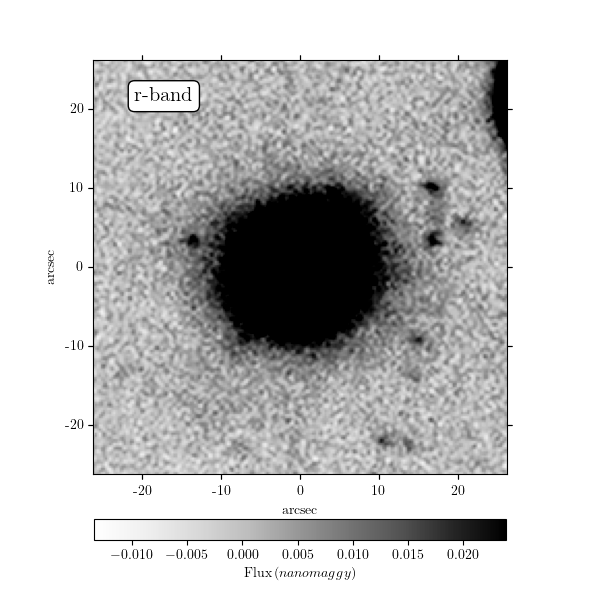}\par 
    \includegraphics[width=6.0cm]{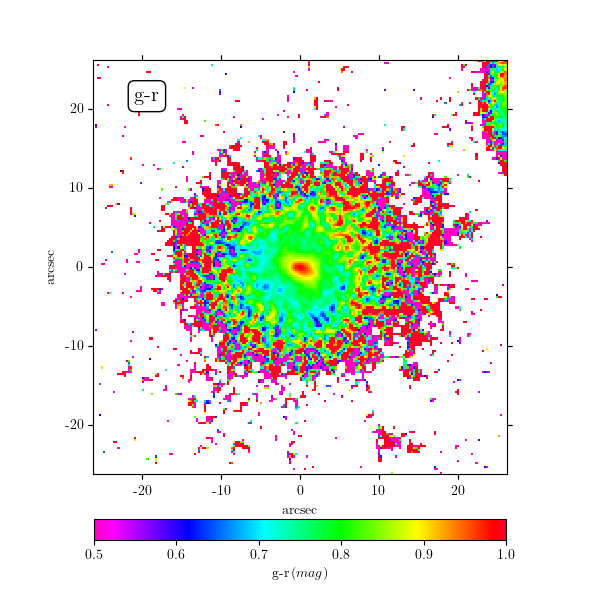}\par
 \end{multicols}
 \begin{multicols}{3}
    \includegraphics[width=6.0cm]{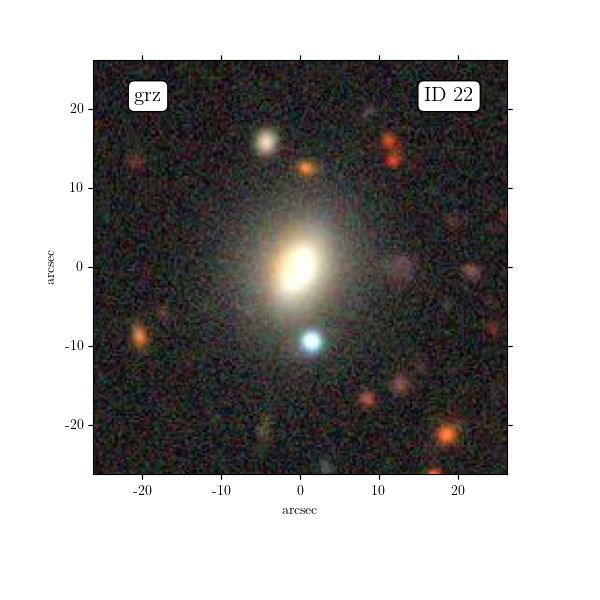}\par 
    \includegraphics[width=6.0cm]{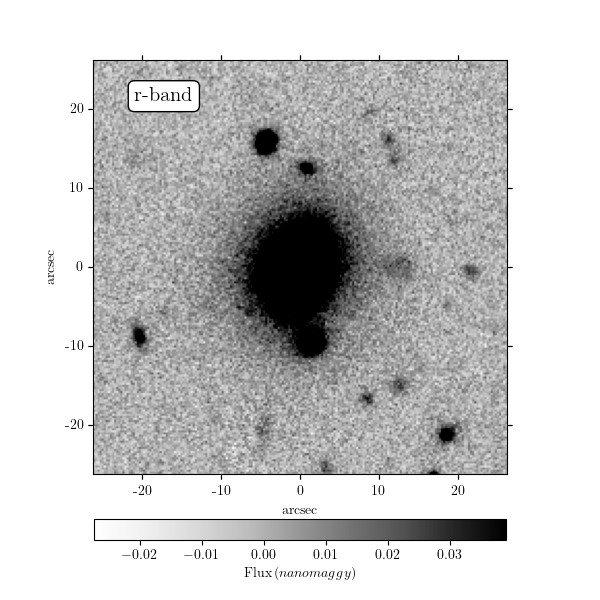}\par 
    \includegraphics[width=6.0cm]{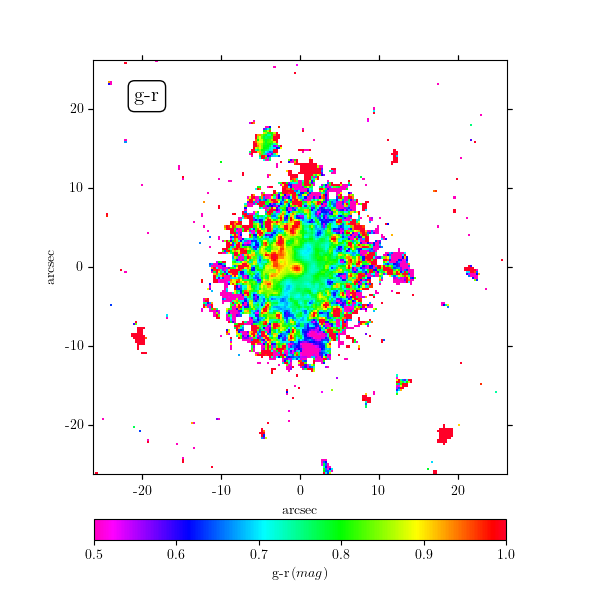}\par
 \end{multicols}
 \begin{multicols}{3}
    \includegraphics[width=6.0cm]{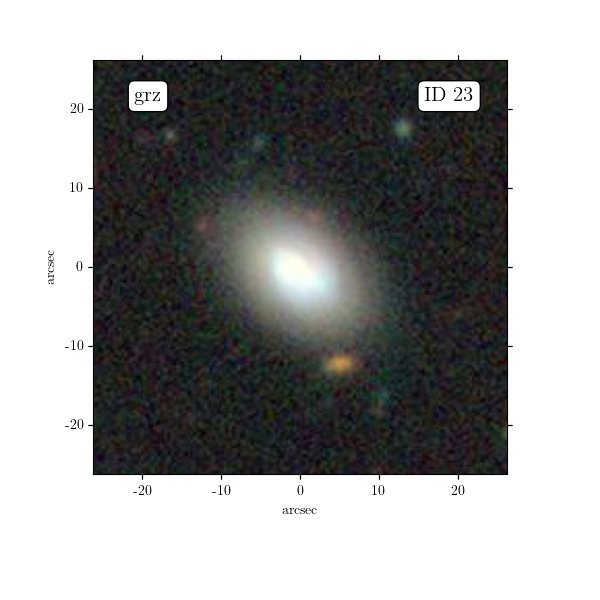}\par 
    \includegraphics[width=6.0cm]{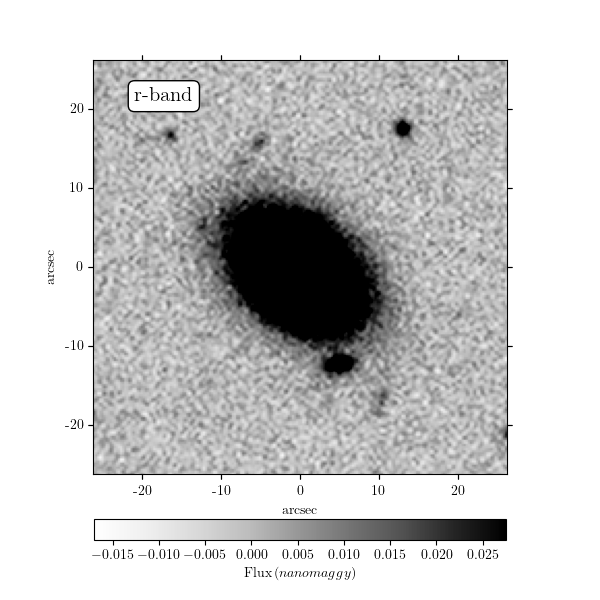}\par 
    \includegraphics[width=6.0cm]{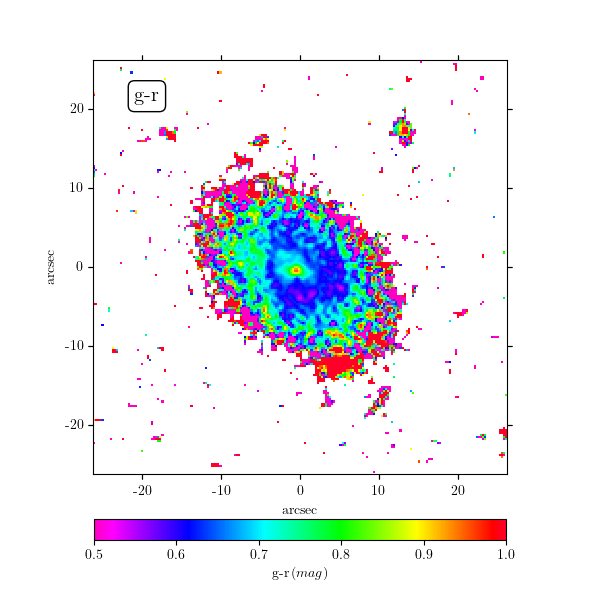}\par
 \end{multicols}
  \begin{multicols}{3}
    \includegraphics[width=6.0cm]{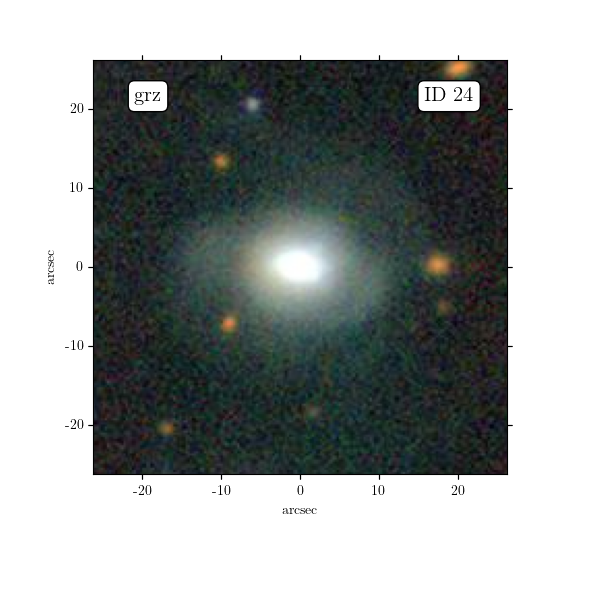}\par 
    \includegraphics[width=6.0cm]{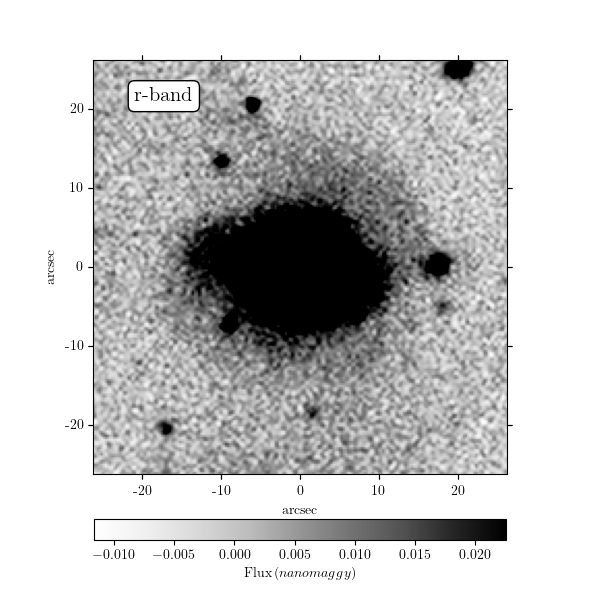}\par 
    \includegraphics[width=6.0cm]{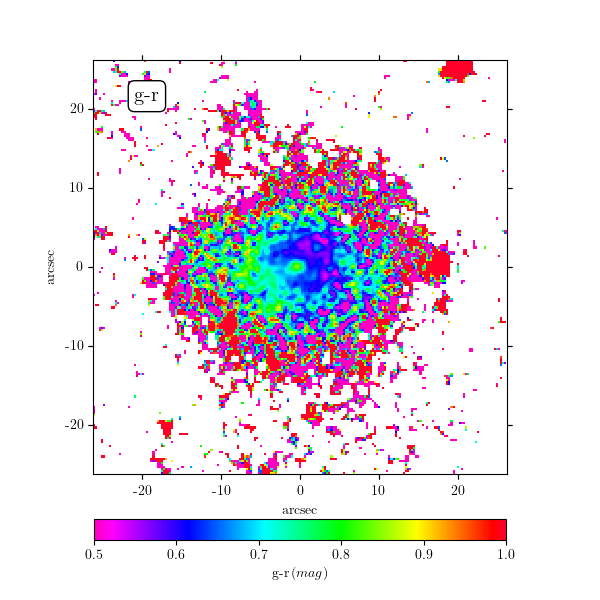}\par
 \end{multicols}
  \end{figure*}

  \begin{figure*}
%\ContinuedFloat 
 \begin{multicols}{3}
    \includegraphics[width=6.0cm]{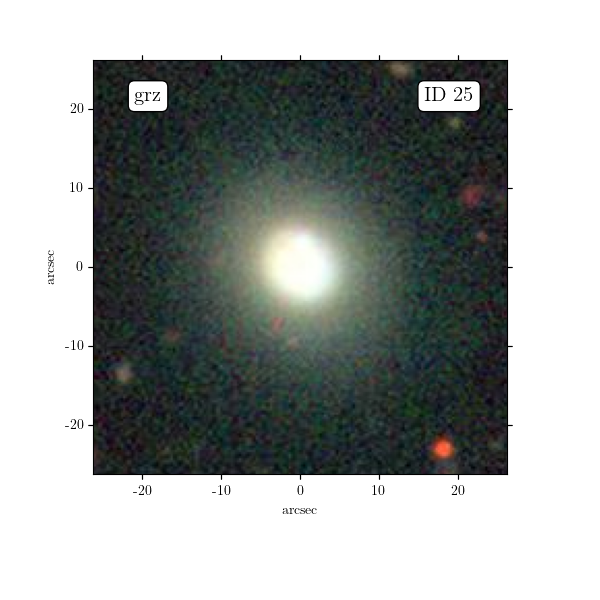}\par 
    \includegraphics[width=6.0cm]{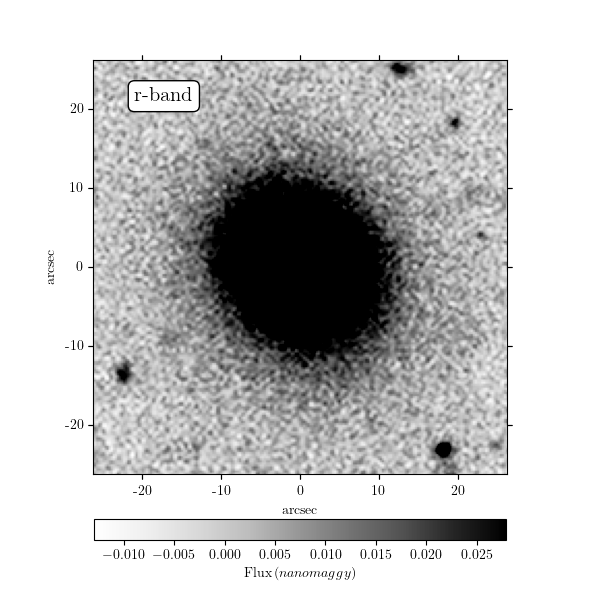}\par 
    \includegraphics[width=6.0cm]{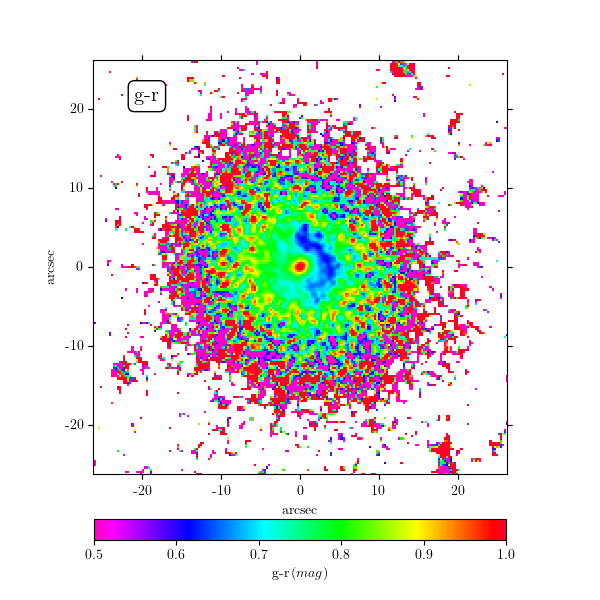}\par
 \end{multicols}
 \begin{multicols}{3}
    \includegraphics[width=6.0cm]{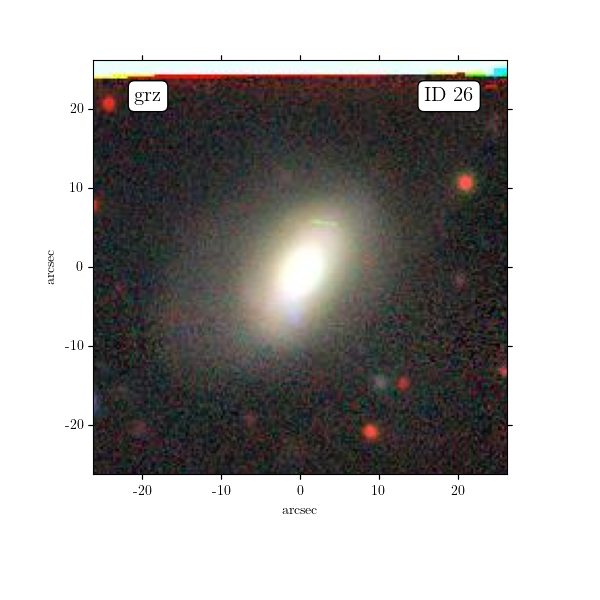}\par 
    \includegraphics[width=6.0cm]{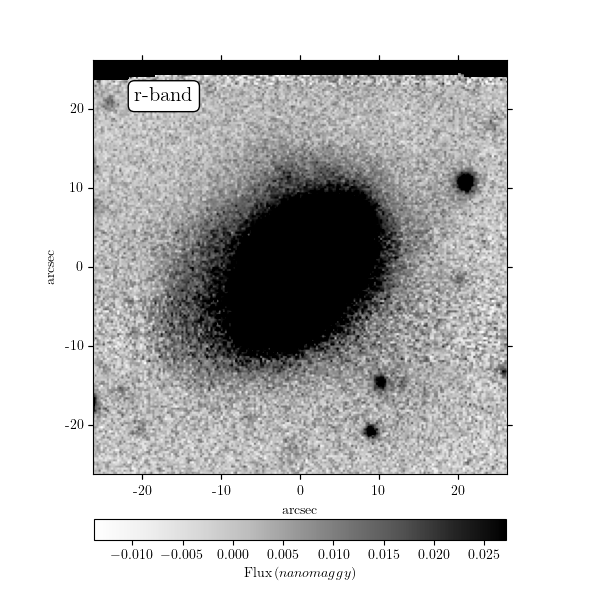}\par 
    \includegraphics[width=6.0cm]{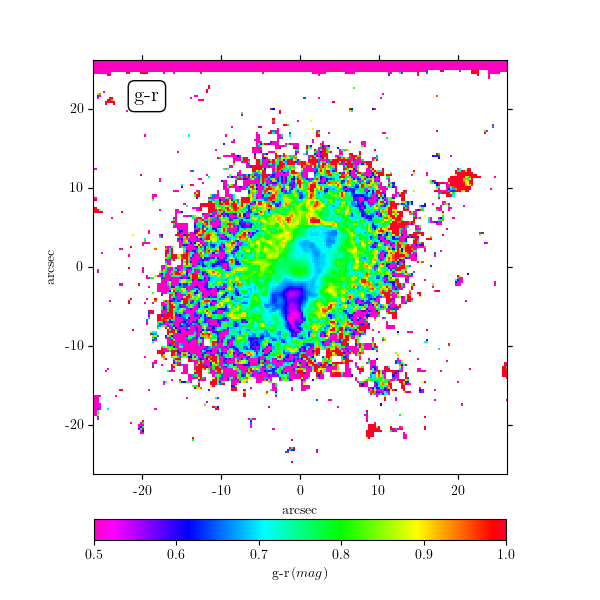}\par
 \end{multicols}
 \begin{multicols}{3}
    \includegraphics[width=6.0cm]{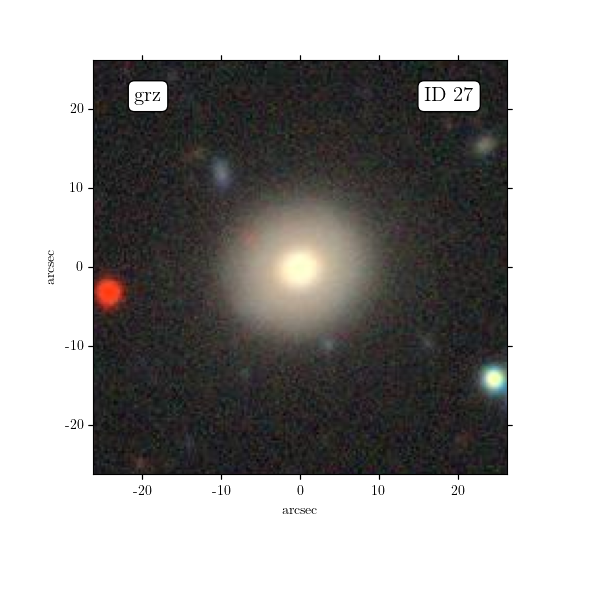}\par 
    \includegraphics[width=6.0cm]{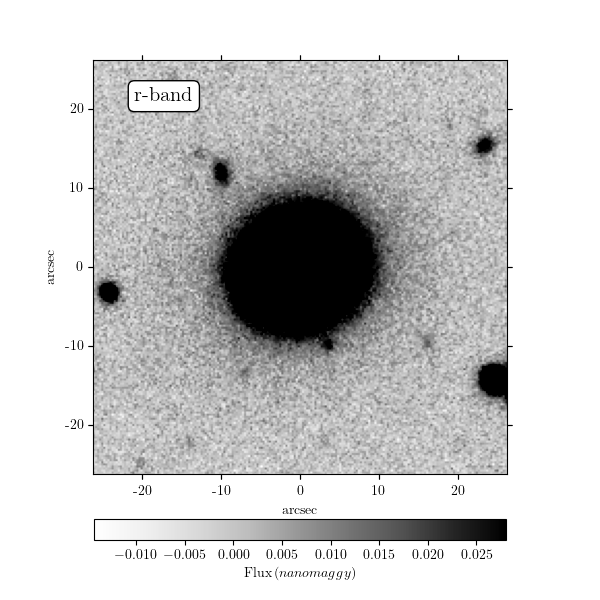}\par 
    \includegraphics[width=6.0cm]{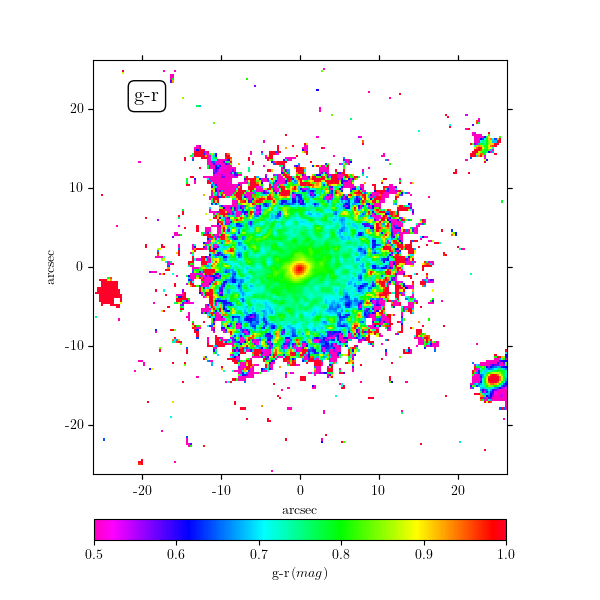}\par
 \end{multicols}
  \begin{multicols}{3}
    \includegraphics[width=6.0cm]{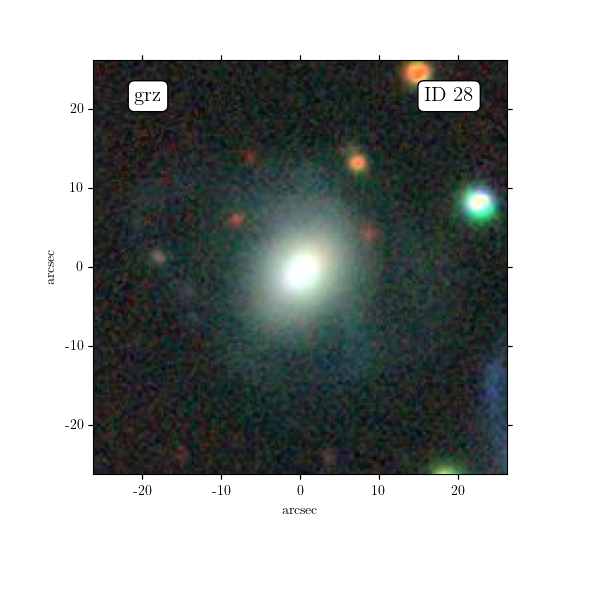}\par 
    \includegraphics[width=6.0cm]{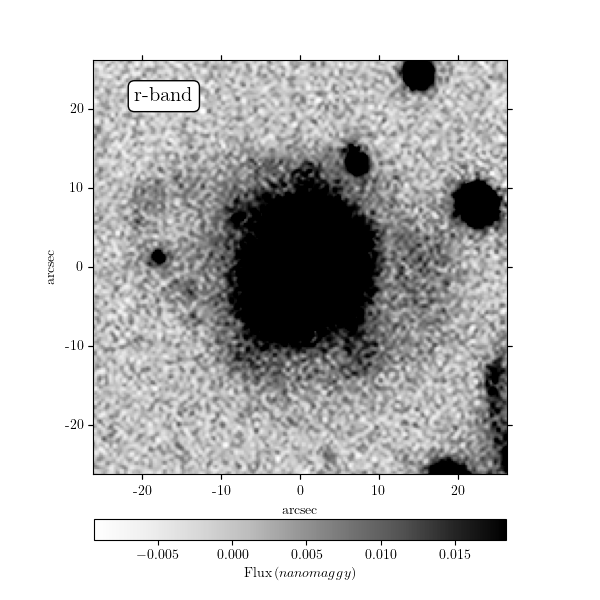}\par 
    \includegraphics[width=6.0cm]{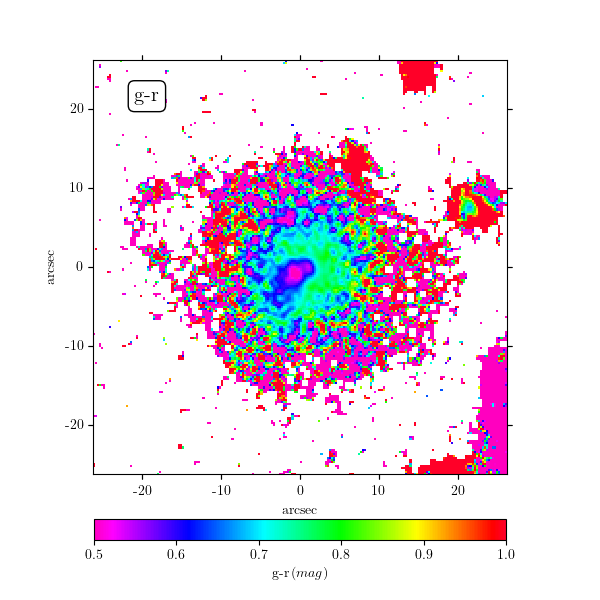}\par
 \end{multicols}
  \end{figure*}

  \begin{figure*}
%\ContinuedFloat 
 \begin{multicols}{3}
    \includegraphics[width=6.0cm]{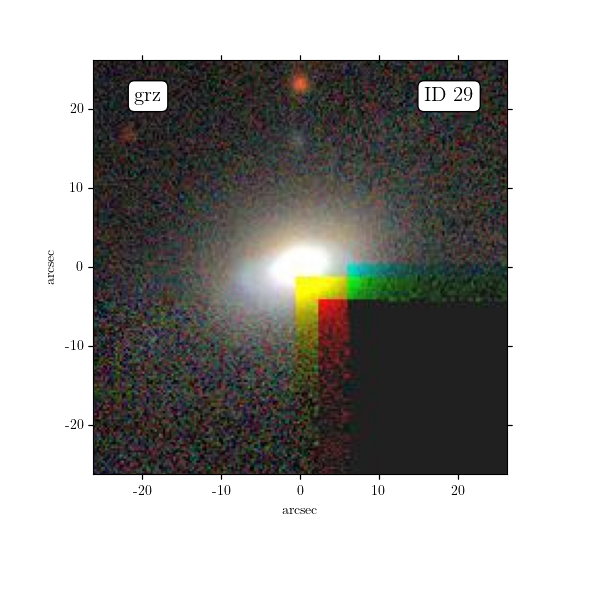}\par 
    \includegraphics[width=6.0cm]{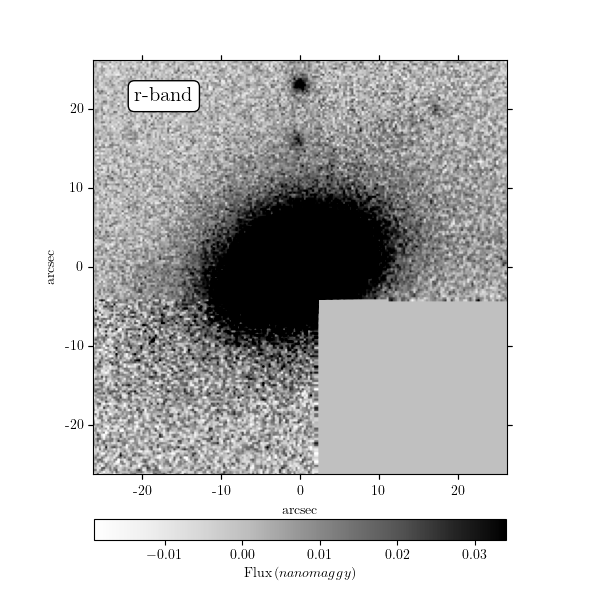}\par 
    \includegraphics[width=6.0cm]{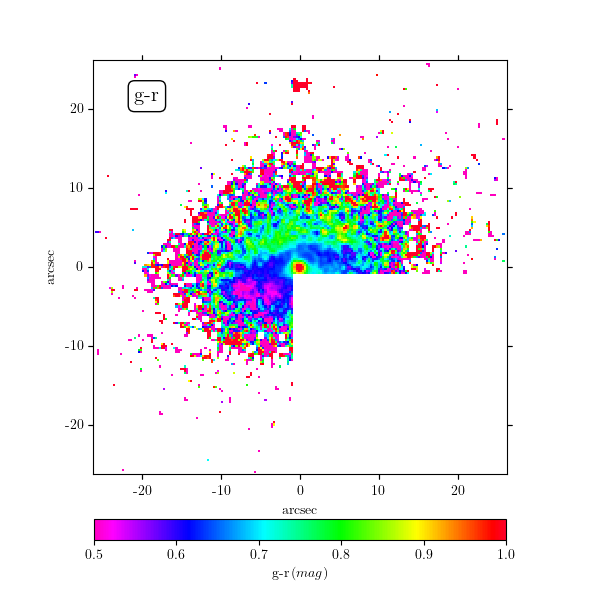}\par
 \end{multicols}
 \begin{multicols}{3}
    \includegraphics[width=6.0cm]{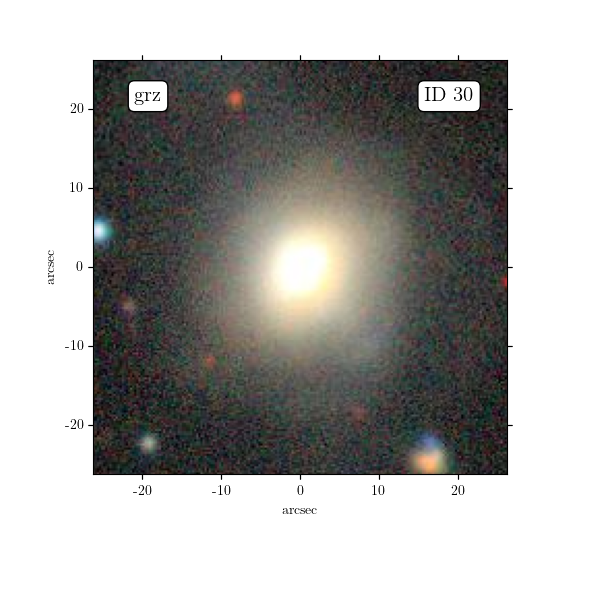}\par 
    \includegraphics[width=6.0cm]{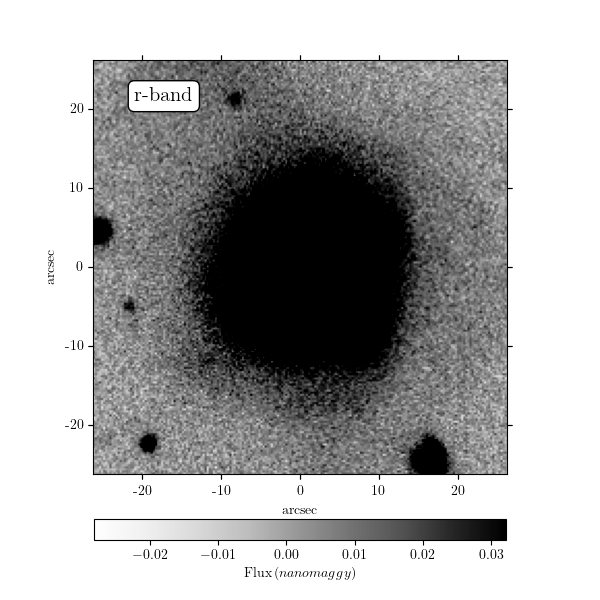}\par 
    \includegraphics[width=6.0cm]{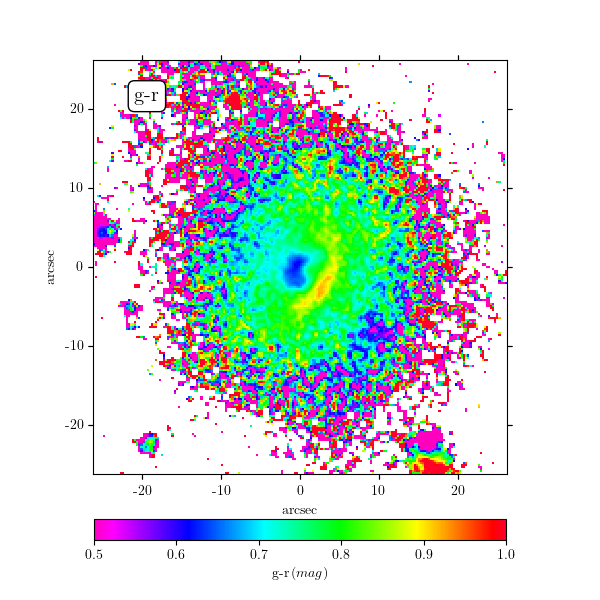}\par
 \end{multicols}
 \begin{multicols}{3}
    \includegraphics[width=6.0cm]{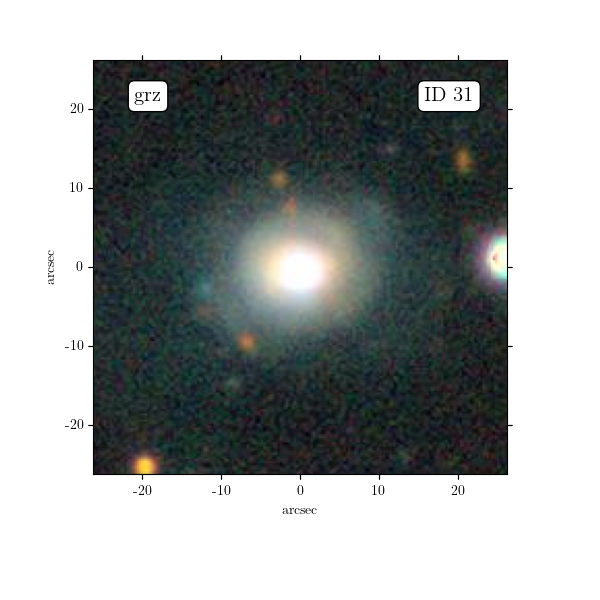}\par 
    \includegraphics[width=6.0cm]{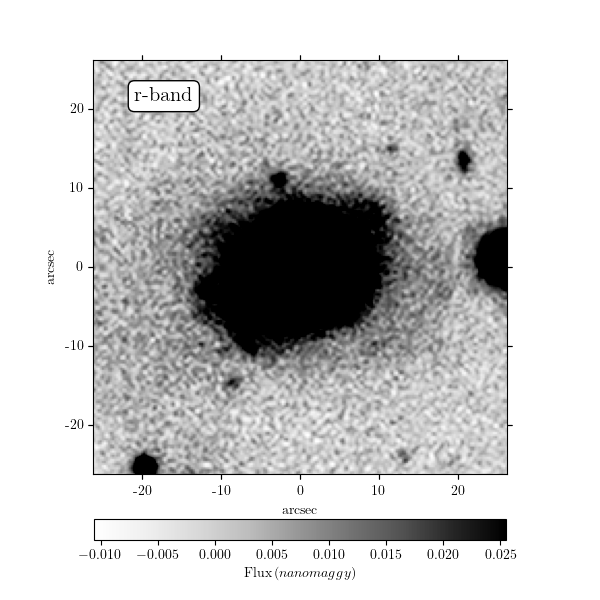}\par 
    \includegraphics[width=6.0cm]{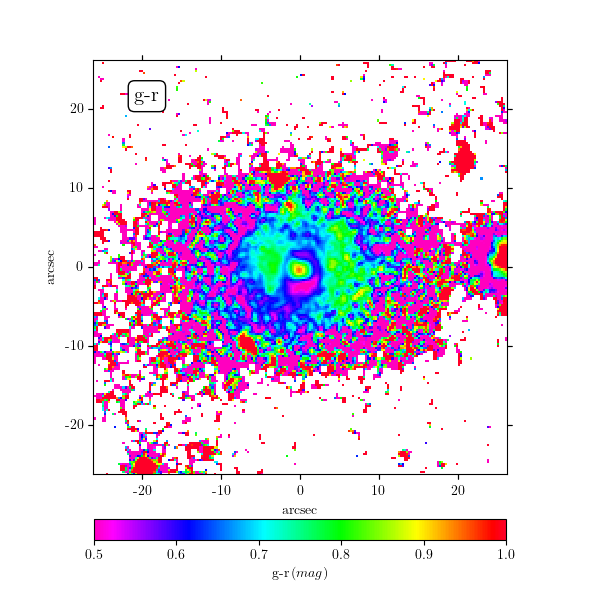}\par
 \end{multicols}
  \begin{multicols}{3}
    \includegraphics[width=6.0cm]{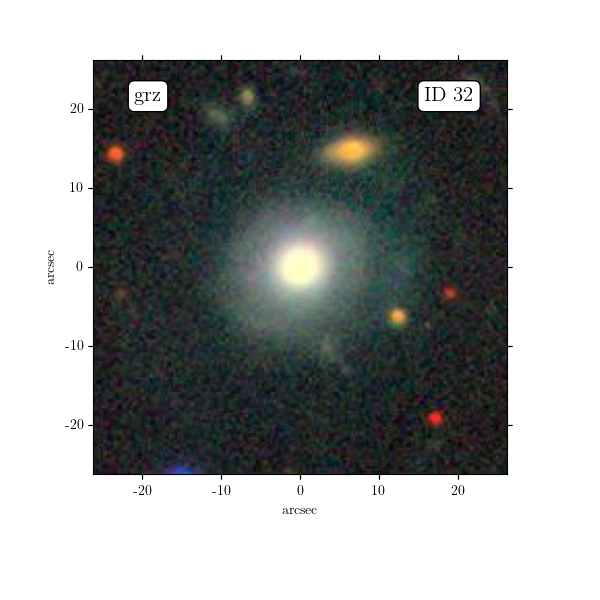}\par 
    \includegraphics[width=6.0cm]{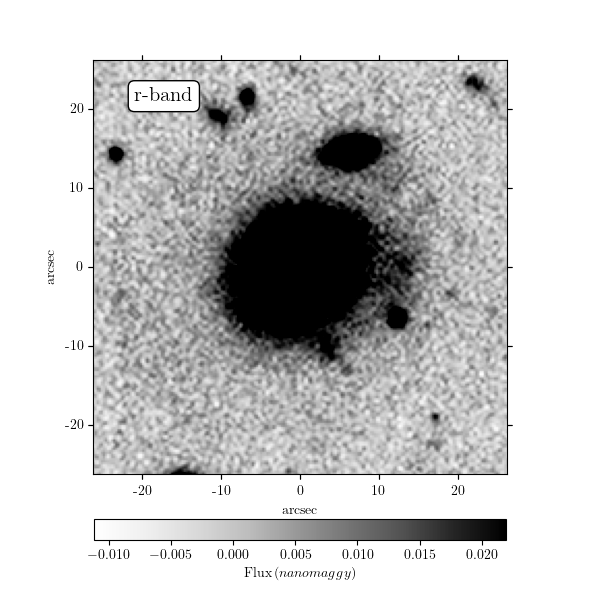}\par 
    \includegraphics[width=6.0cm]{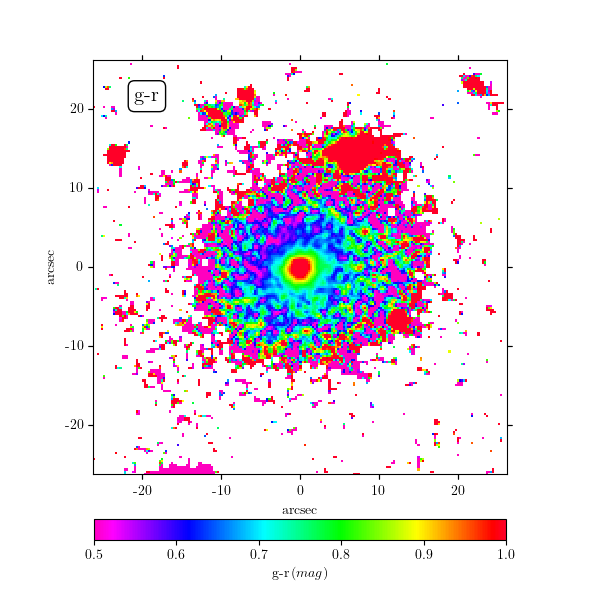}\par
 \end{multicols}
  \end{figure*}

  \begin{figure*}
%\ContinuedFloat 
 \begin{multicols}{3}
    \includegraphics[width=6.0cm]{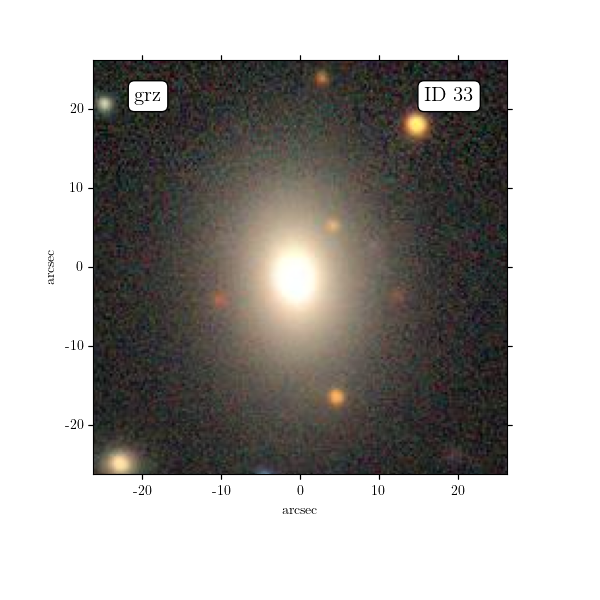}\par 
    \includegraphics[width=6.0cm]{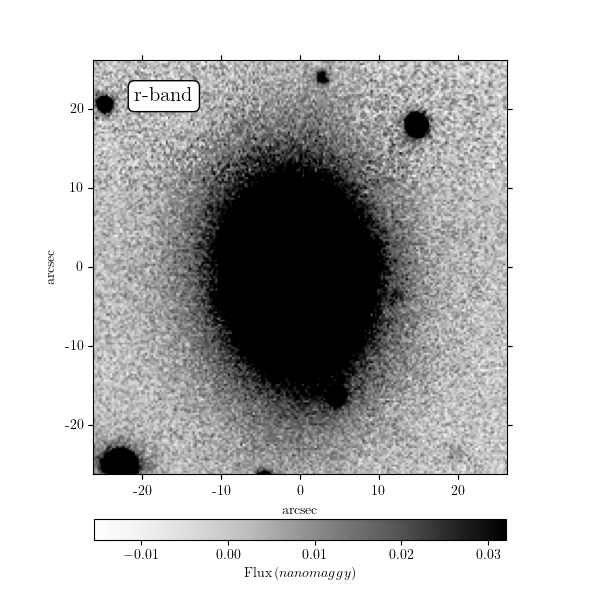}\par 
    \includegraphics[width=6.0cm]{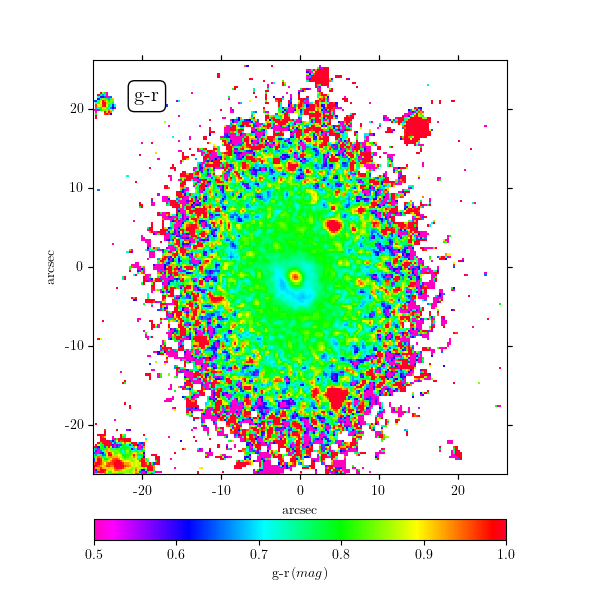}\par
 \end{multicols}
 \begin{multicols}{3}
    \includegraphics[width=6.0cm]{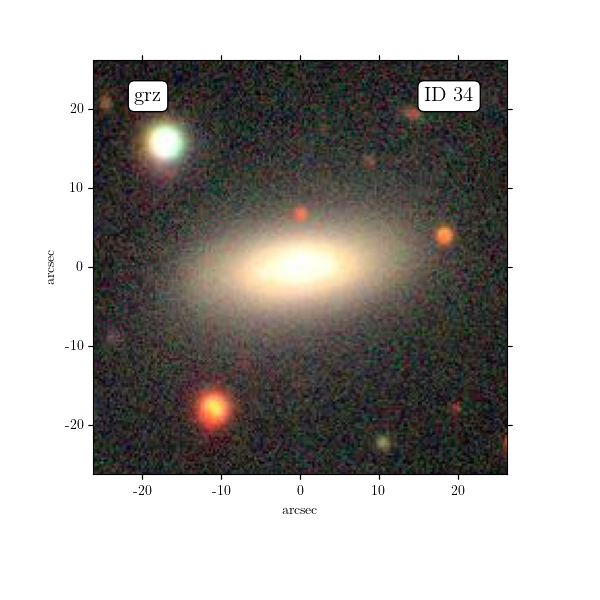}\par 
    \includegraphics[width=6.0cm]{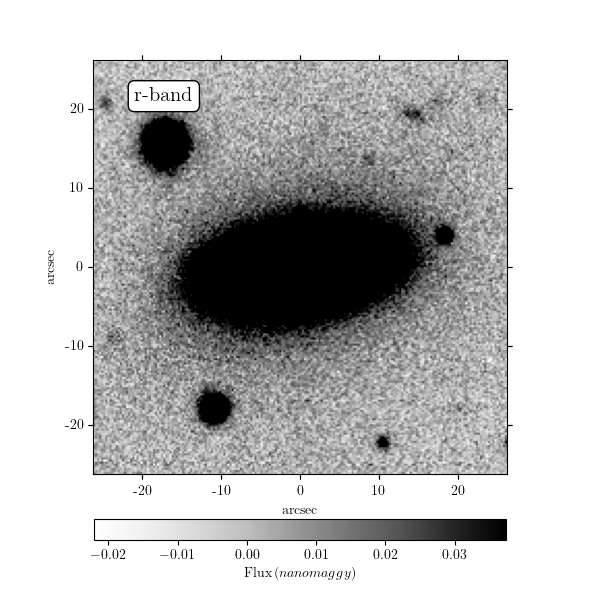}\par 
    \includegraphics[width=6.0cm]{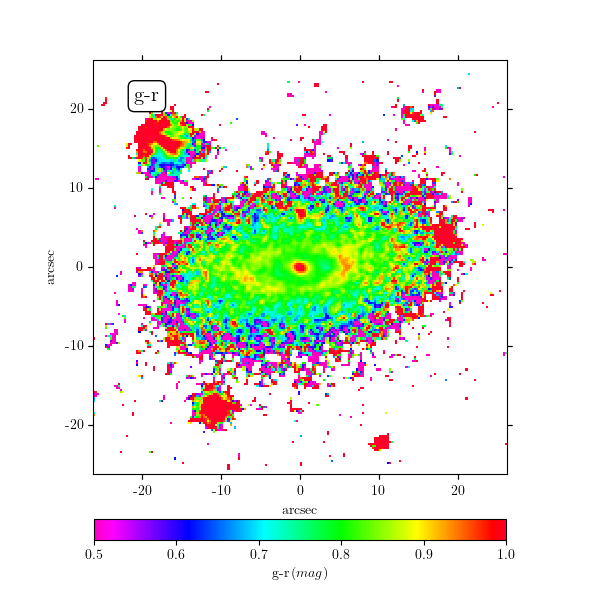}\par
 \end{multicols}
 \begin{multicols}{3}
    \includegraphics[width=6.0cm]{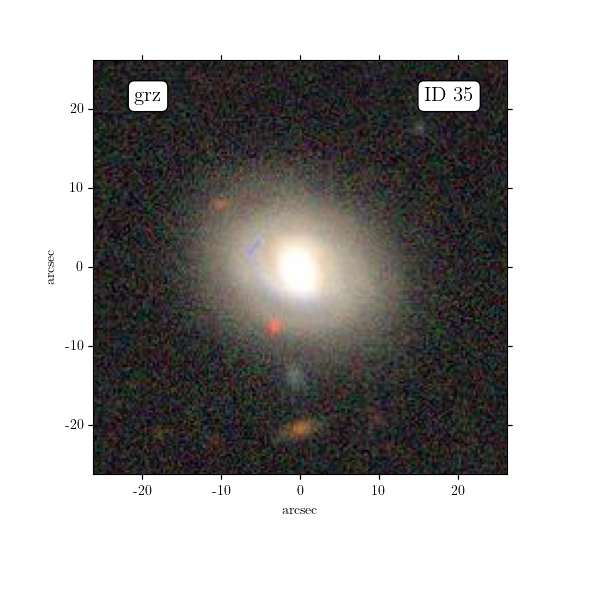}\par 
    \includegraphics[width=6.0cm]{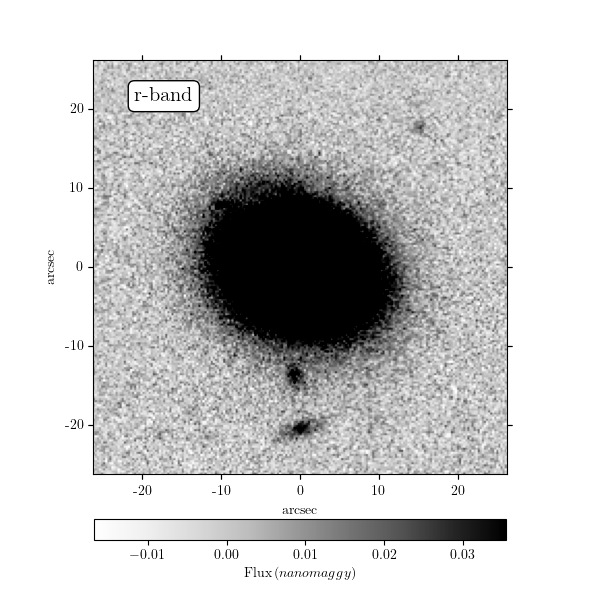}\par 
    \includegraphics[width=6.0cm]{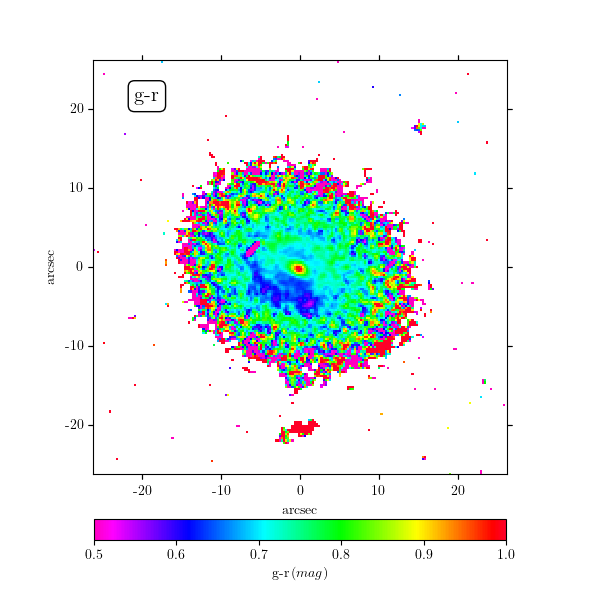}\par
 \end{multicols}
  \begin{multicols}{3}
    \includegraphics[width=6.0cm]{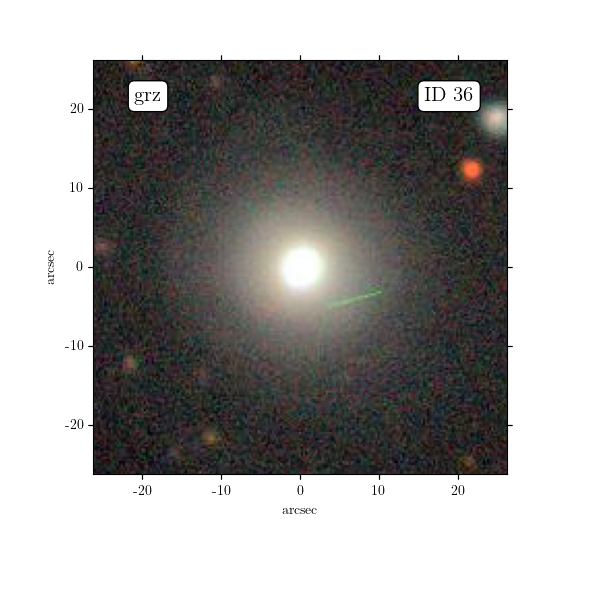}\par 
    \includegraphics[width=6.0cm]{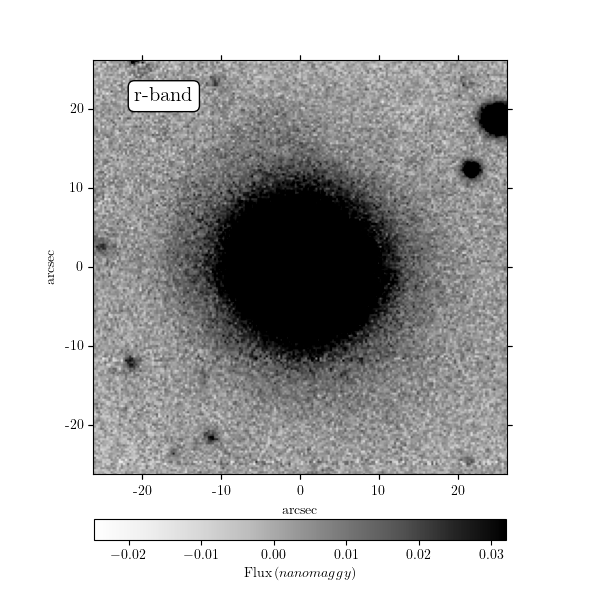}\par 
    \includegraphics[width=6.0cm]{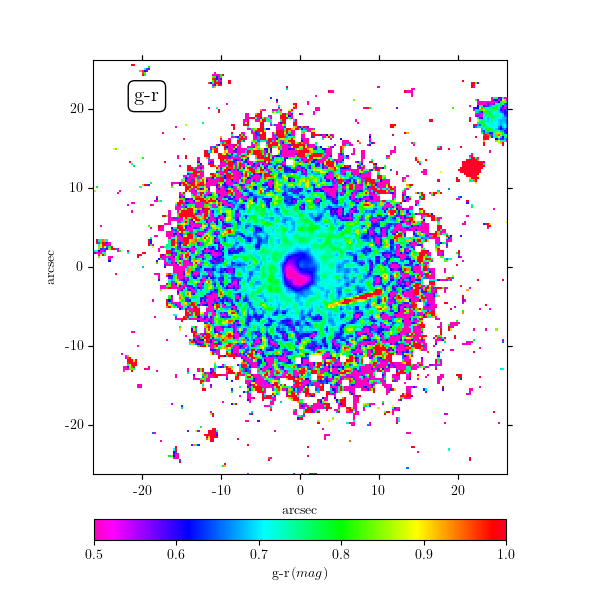}\par
 \end{multicols}
  \end{figure*}

  \begin{figure*}
%\ContinuedFloat 
 \begin{multicols}{3}
    \includegraphics[width=6.0cm]{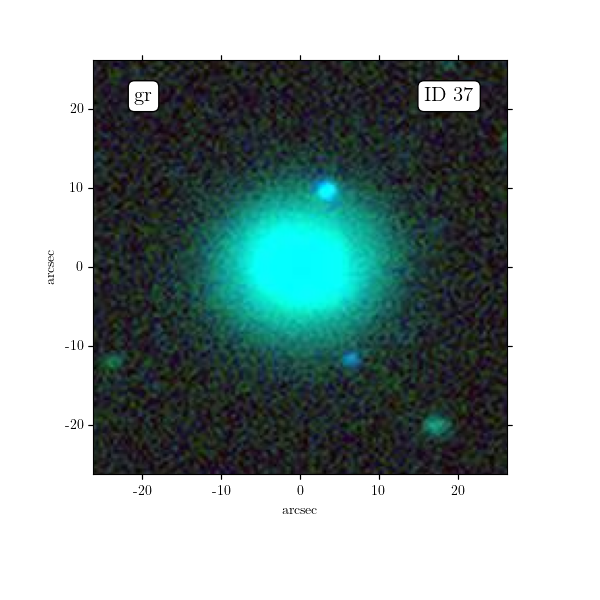}\par 
    \includegraphics[width=6.0cm]{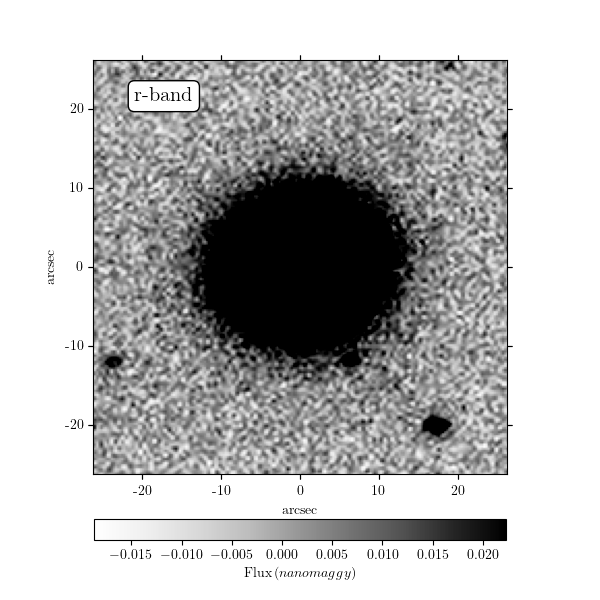}\par 
    \includegraphics[width=6.0cm]{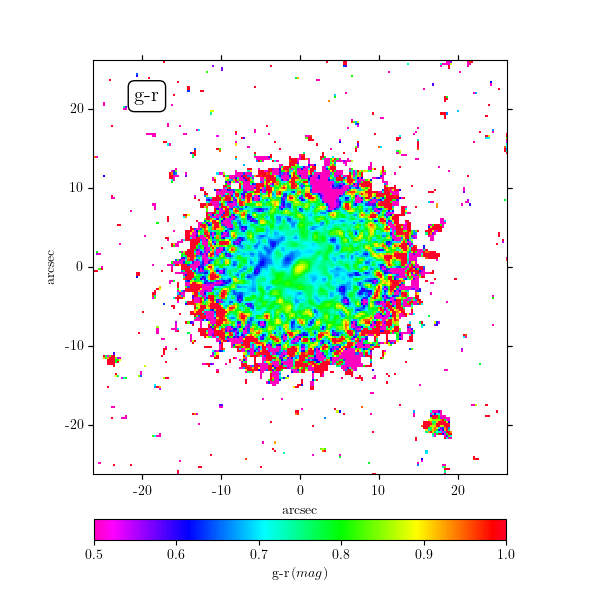}\par
 \end{multicols}
 \begin{multicols}{3}
    \includegraphics[width=6.0cm]{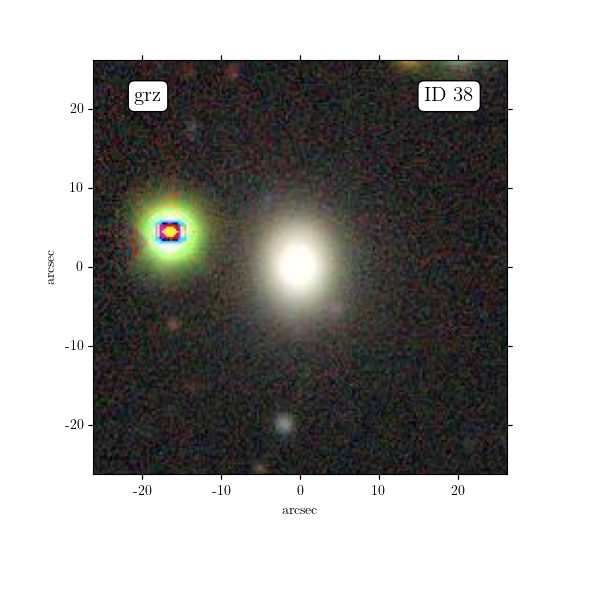}\par 
    \includegraphics[width=6.0cm]{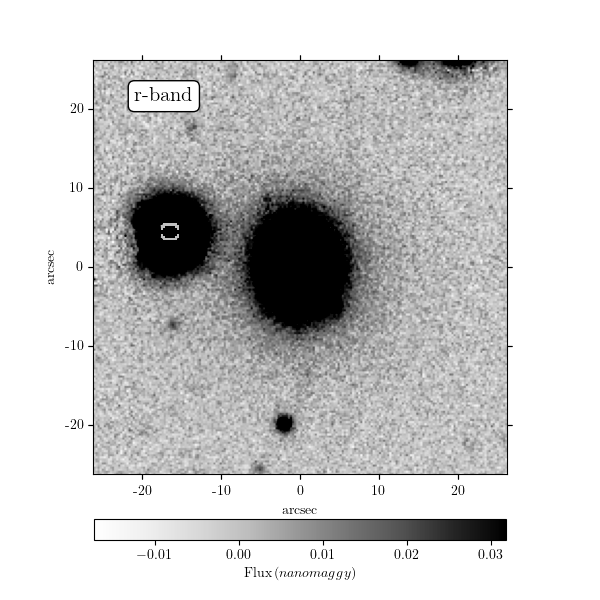}\par 
    \includegraphics[width=6.0cm]{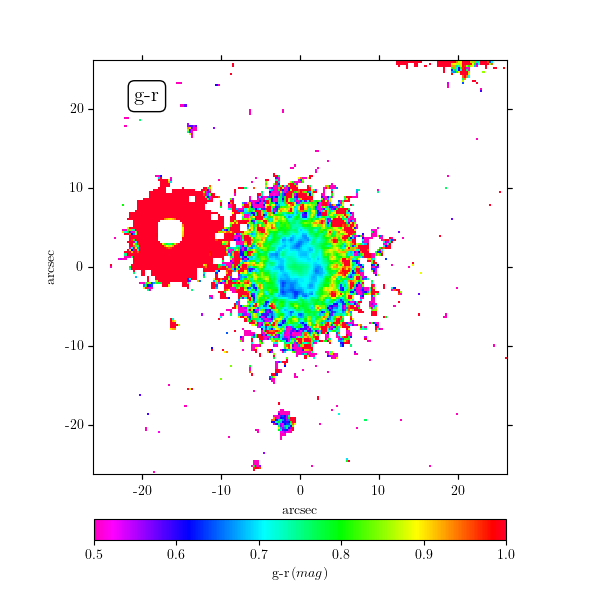}\par
 \end{multicols}
 \begin{multicols}{3}
    \includegraphics[width=6.0cm]{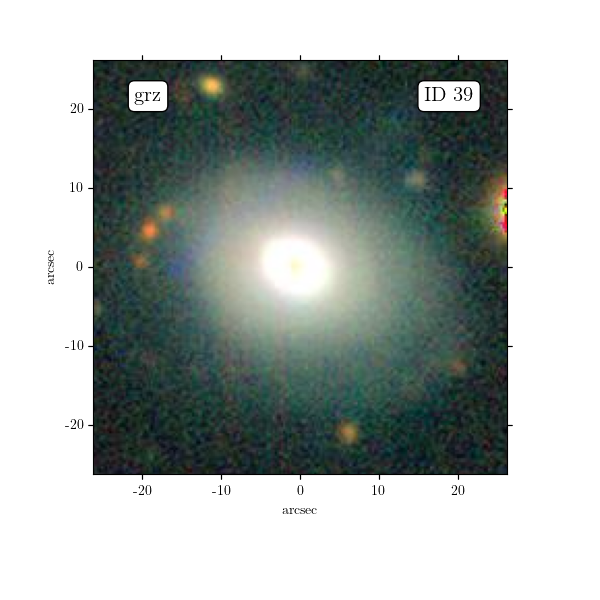}\par 
    \includegraphics[width=6.0cm]{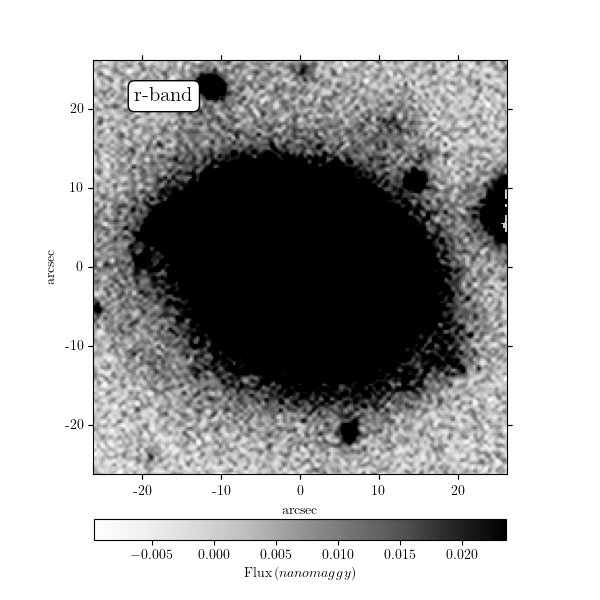}\par 
    \includegraphics[width=6.0cm]{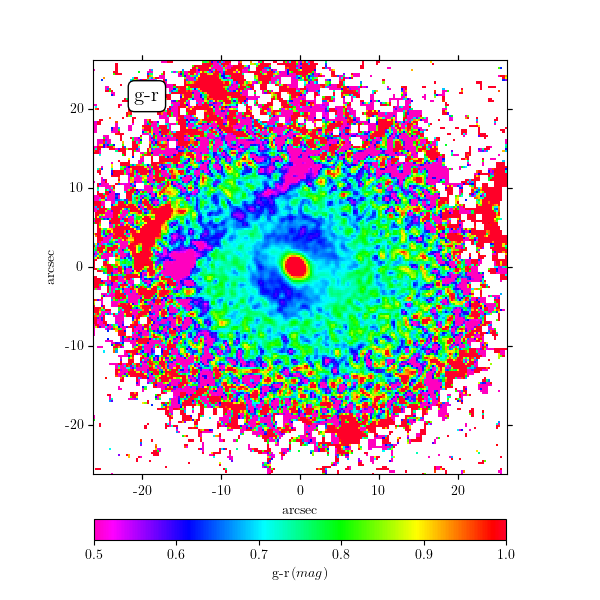}\par
 \end{multicols}
  \begin{multicols}{3}
    \includegraphics[width=6.0cm]{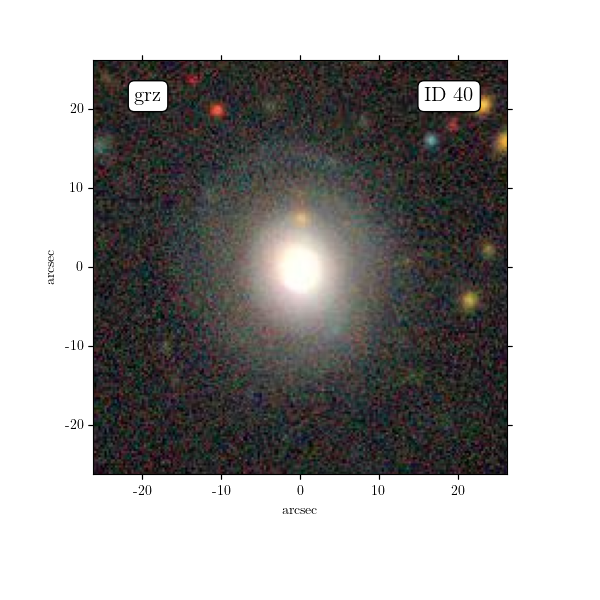}\par 
    \includegraphics[width=6.0cm]{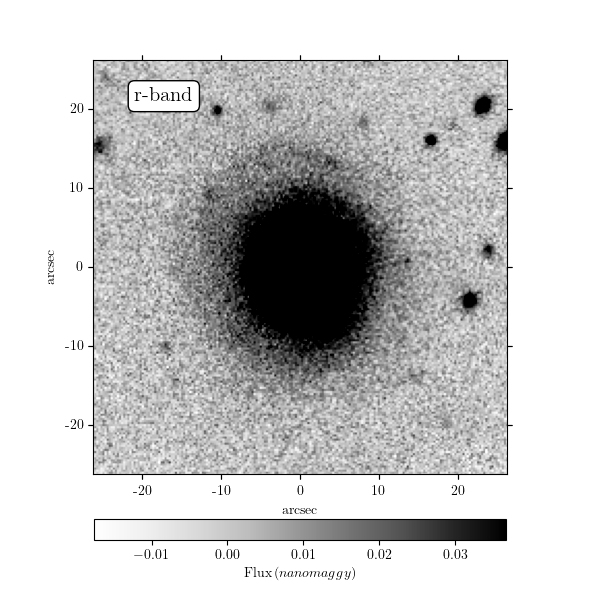}\par 
    \includegraphics[width=6.0cm]{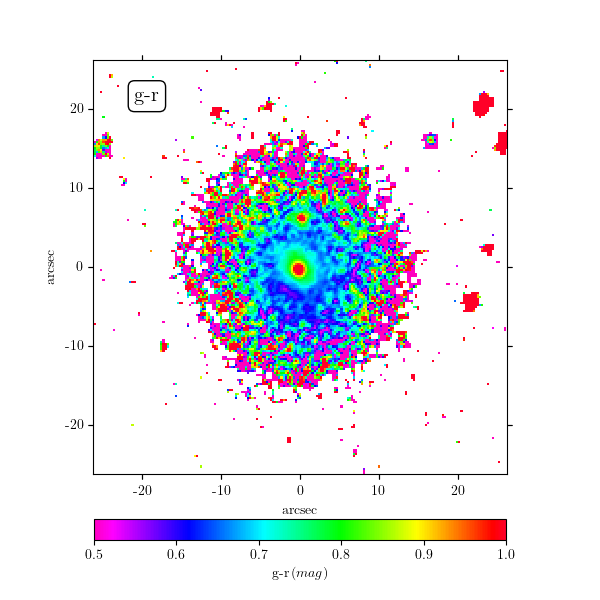}\par
 \end{multicols}
  \end{figure*}

  \begin{figure*}
%\ContinuedFloat 
 \begin{multicols}{3}
    \includegraphics[width=6.0cm]{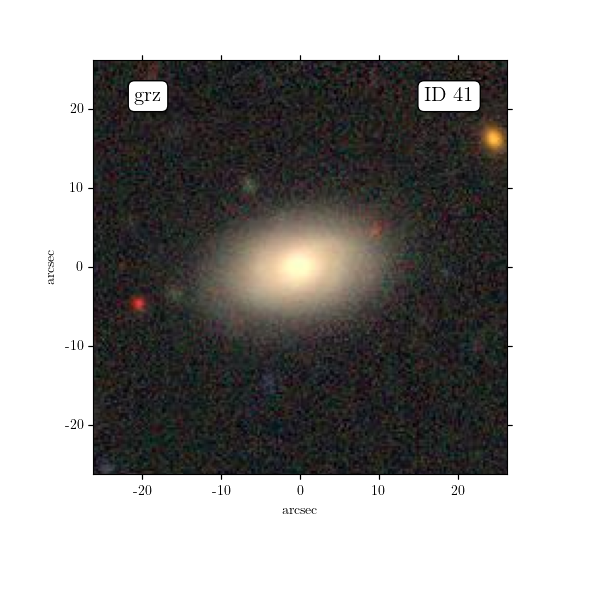}\par 
    \includegraphics[width=6.0cm]{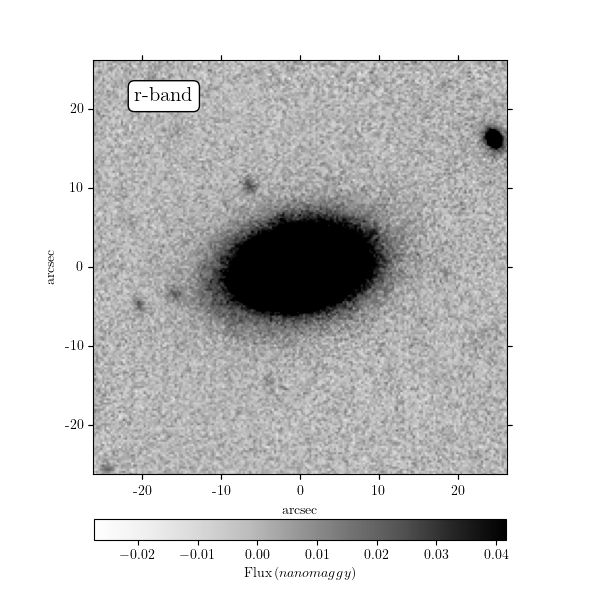}\par 
    \includegraphics[width=6.0cm]{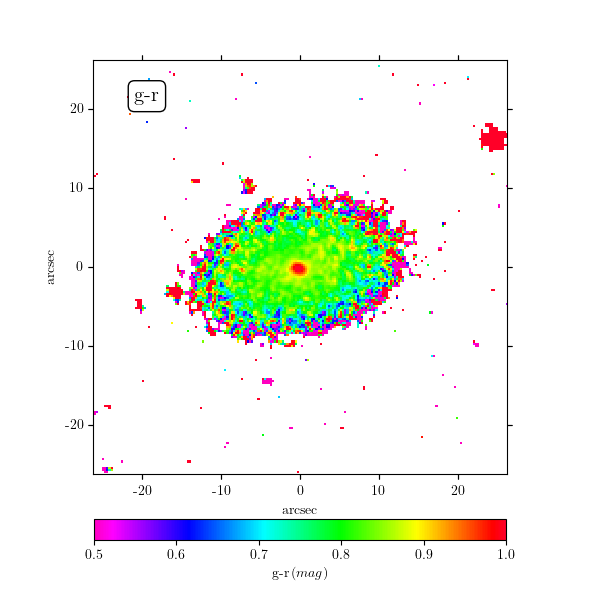}\par
 \end{multicols}
 \begin{multicols}{3}
    \includegraphics[width=6.0cm]{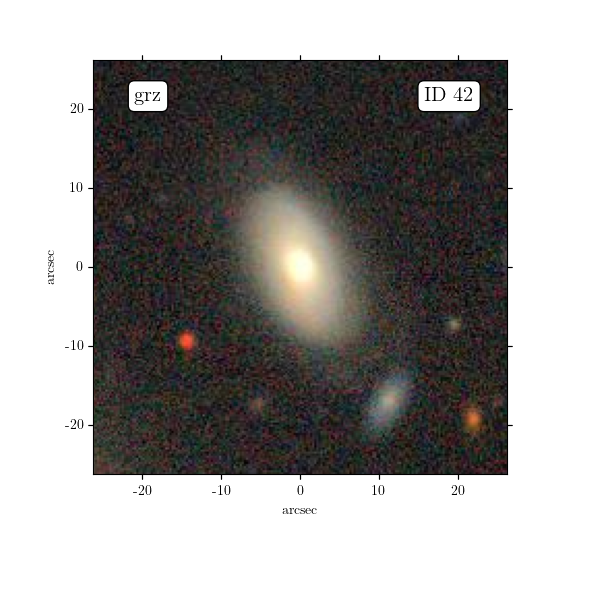}\par 
    \includegraphics[width=6.0cm]{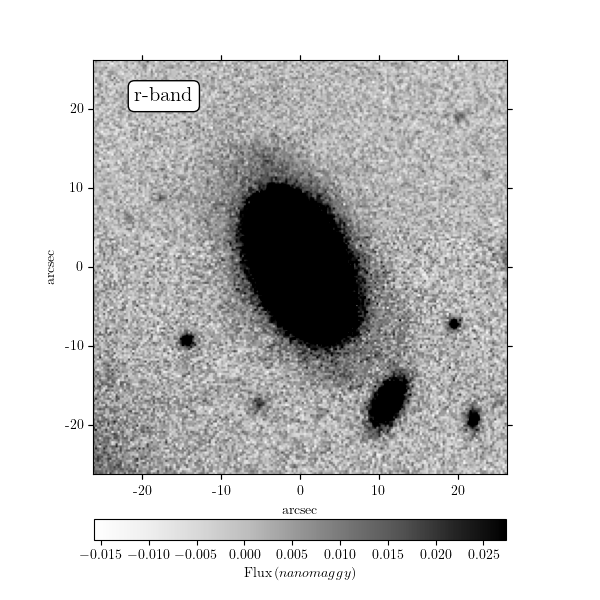}\par 
    \includegraphics[width=6.0cm]{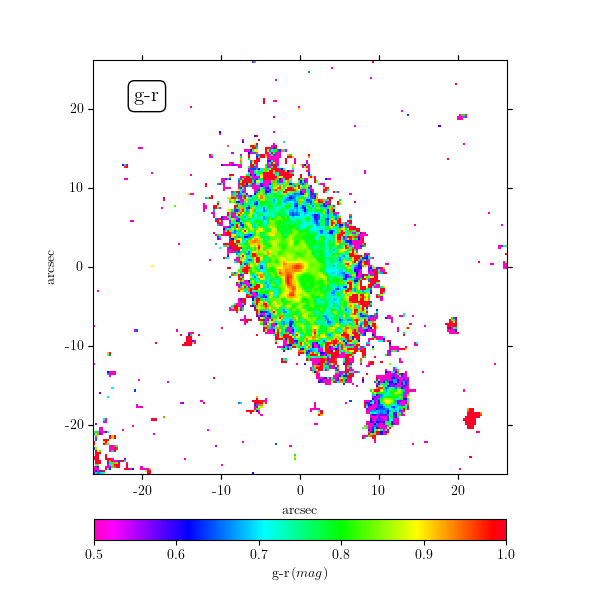}\par
 \end{multicols}
 \begin{multicols}{3}
    \includegraphics[width=6.0cm]{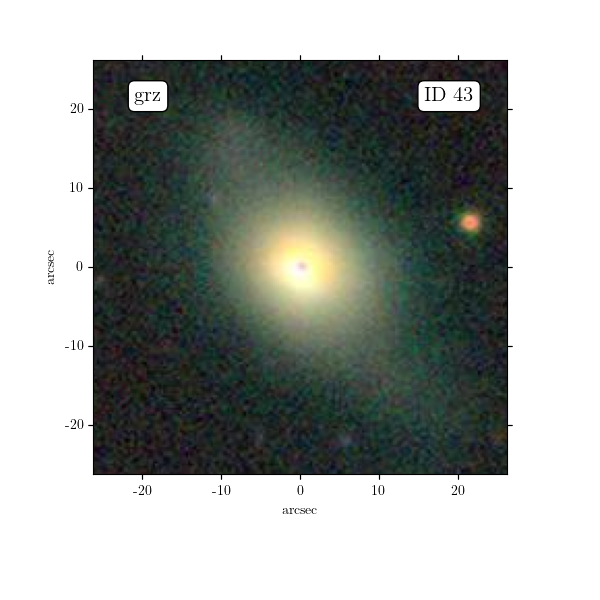}\par 
    \includegraphics[width=6.0cm]{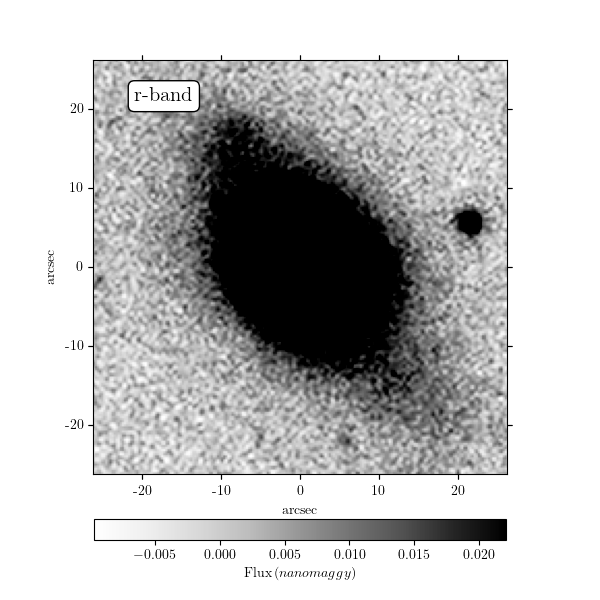}\par 
    \includegraphics[width=6.0cm]{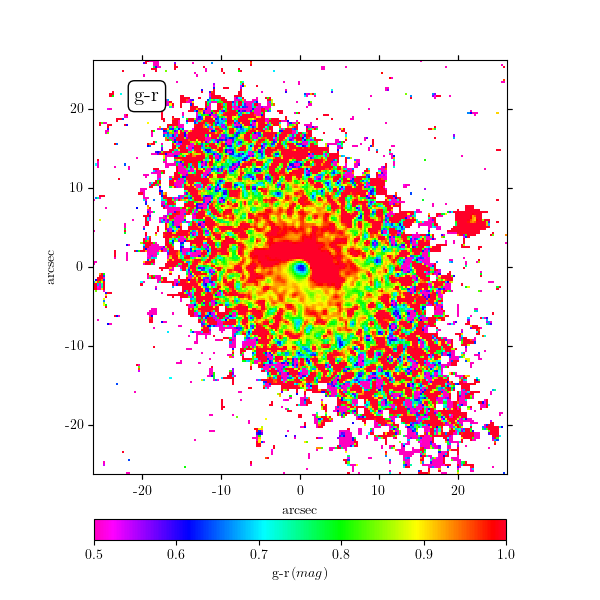}\par
 \end{multicols}
  \begin{multicols}{3}
    \includegraphics[width=6.0cm]{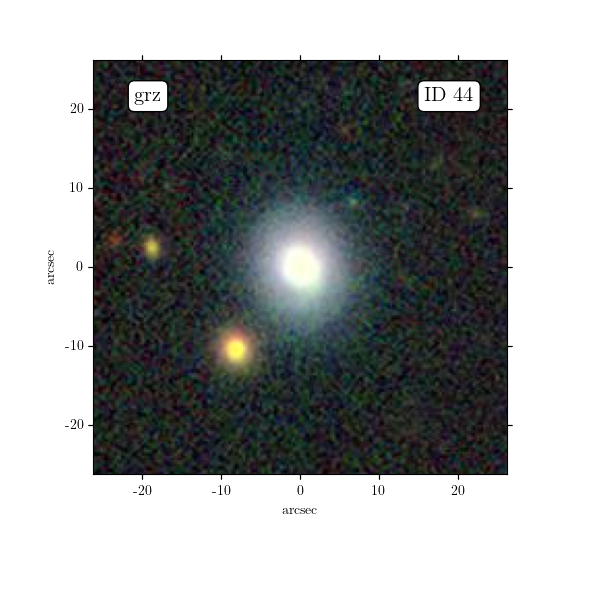}\par 
    \includegraphics[width=6.0cm]{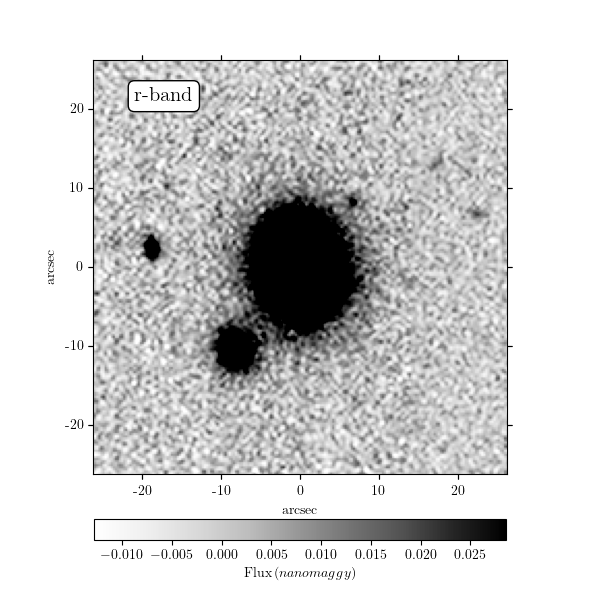}\par 
    \includegraphics[width=6.0cm]{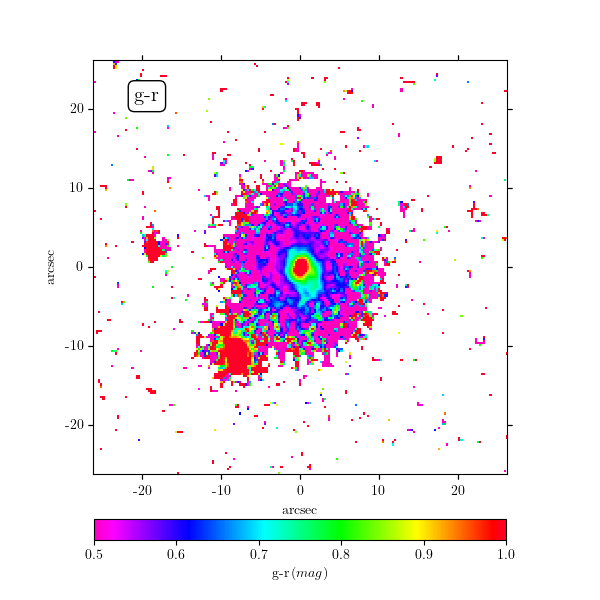}\par
 \end{multicols}
  \end{figure*}

  \begin{figure*}
%\ContinuedFloat 
 \begin{multicols}{3}
    \includegraphics[width=6.0cm]{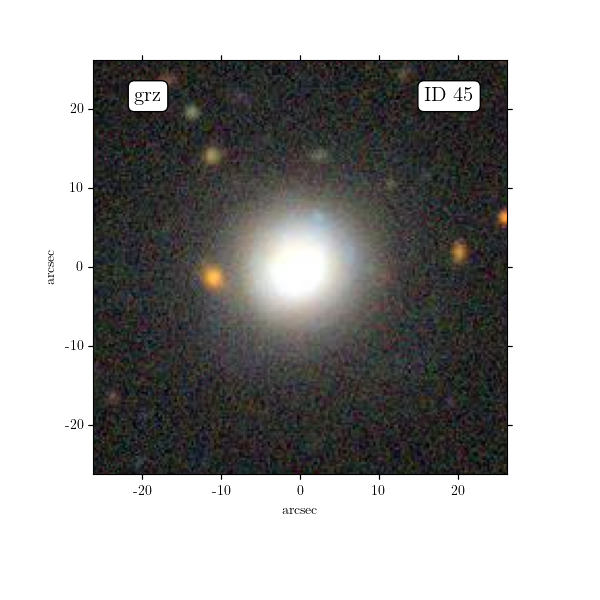}\par 
    \includegraphics[width=6.0cm]{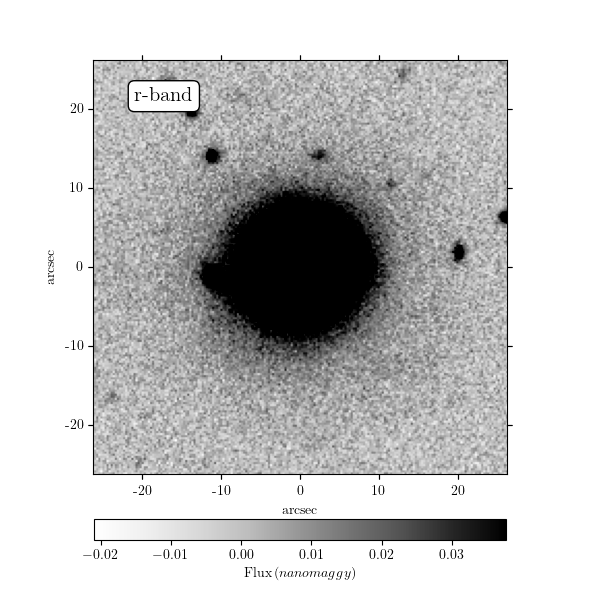}\par 
    \includegraphics[width=6.0cm]{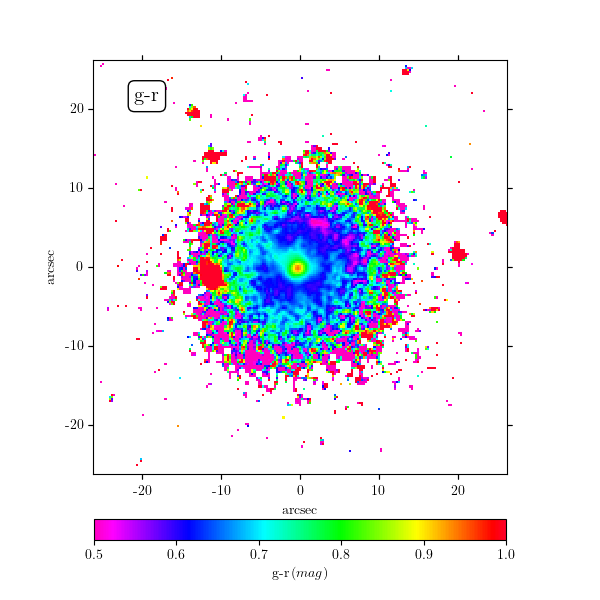}\par
 \end{multicols}
 \begin{multicols}{3}
    \includegraphics[width=6.0cm]{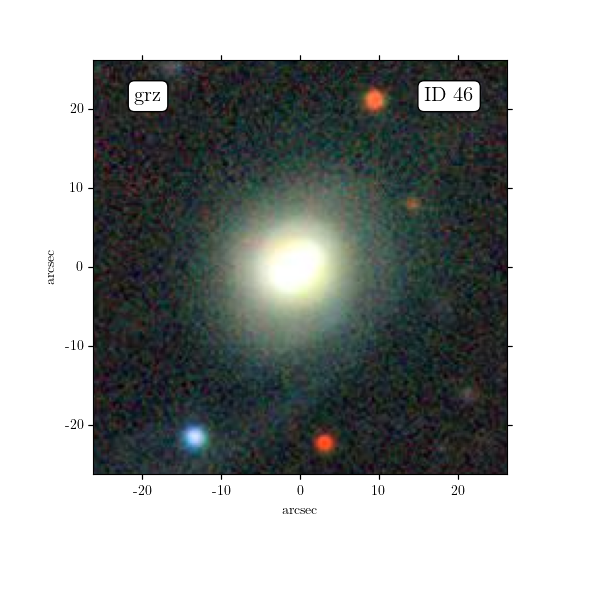}\par 
    \includegraphics[width=6.0cm]{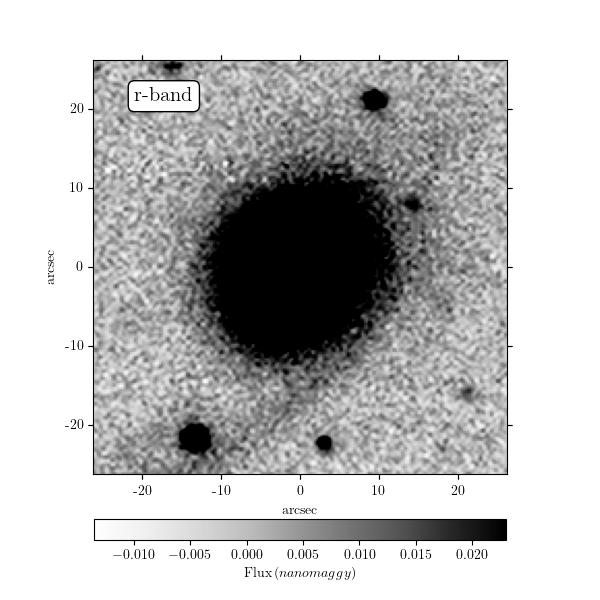}\par 
    \includegraphics[width=6.0cm]{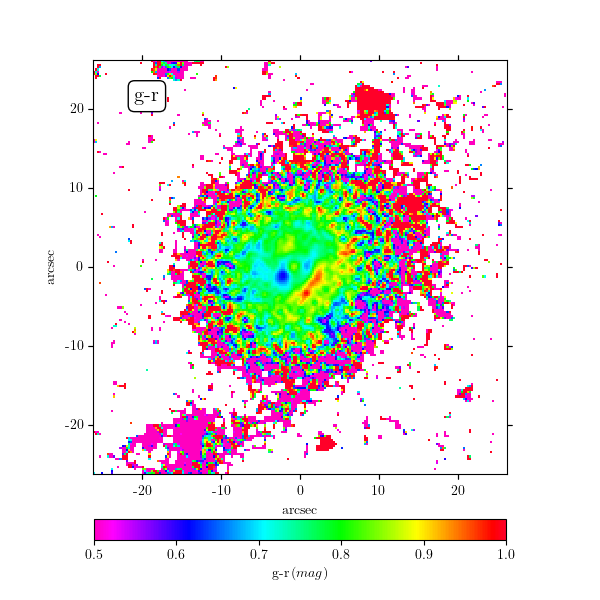}\par
 \end{multicols}
 \begin{multicols}{3}
    \includegraphics[width=6.0cm]{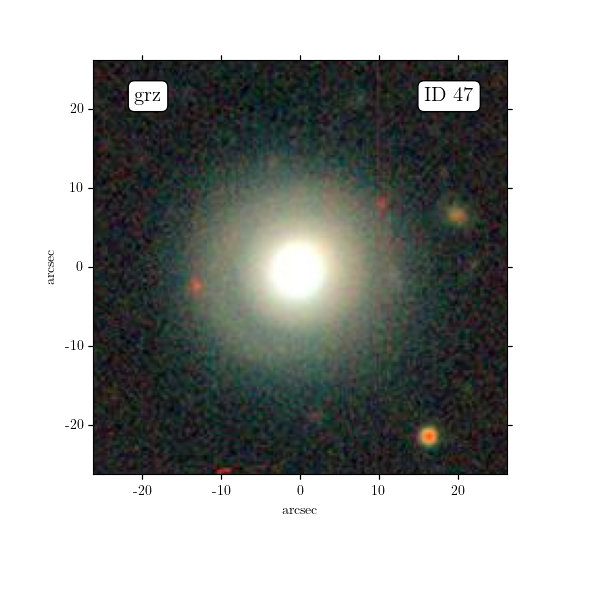}\par 
    \includegraphics[width=6.0cm]{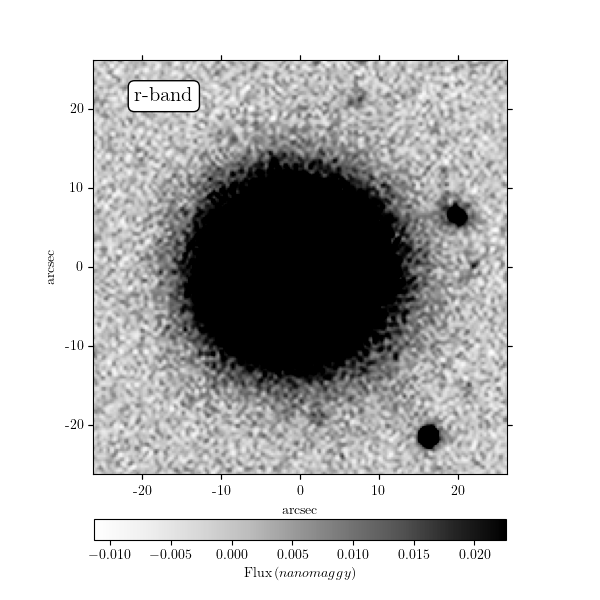}\par 
    \includegraphics[width=6.0cm]{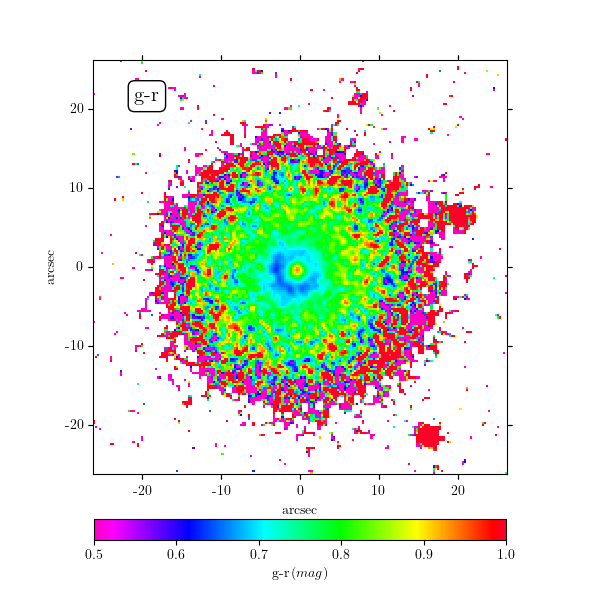}\par
 \end{multicols}
  \begin{multicols}{3}
    \includegraphics[width=6.0cm]{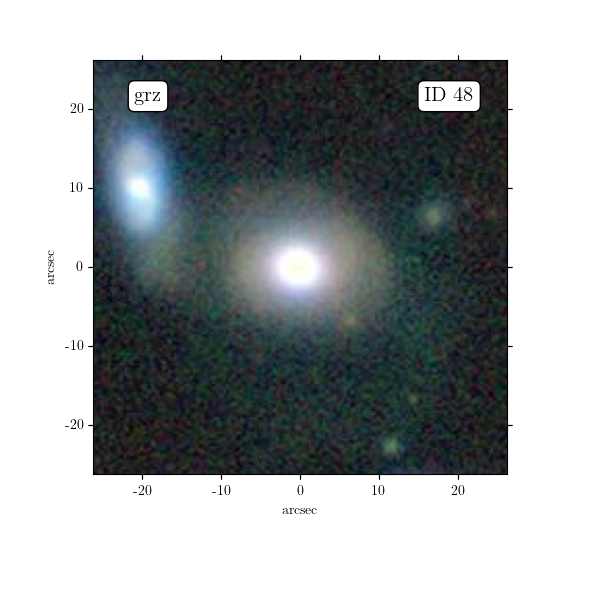}\par 
    \includegraphics[width=6.0cm]{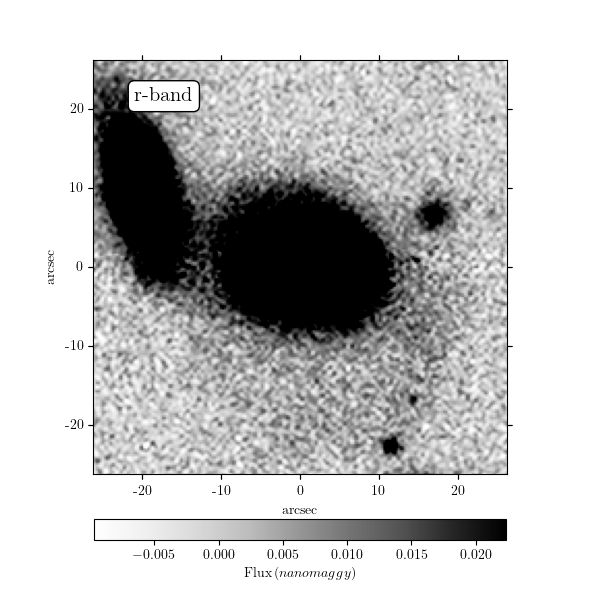}\par 
    \includegraphics[width=6.0cm]{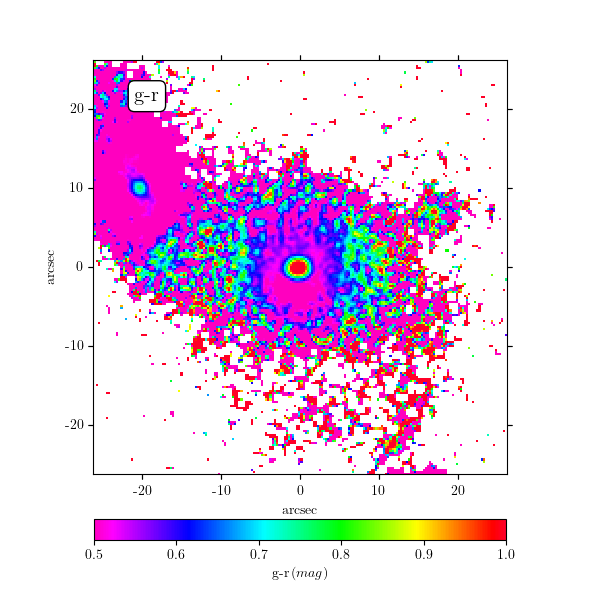}\par
 \end{multicols}
  \end{figure*}

  \begin{figure*}
%\ContinuedFloat 
 \begin{multicols}{3}
    \includegraphics[width=6.0cm]{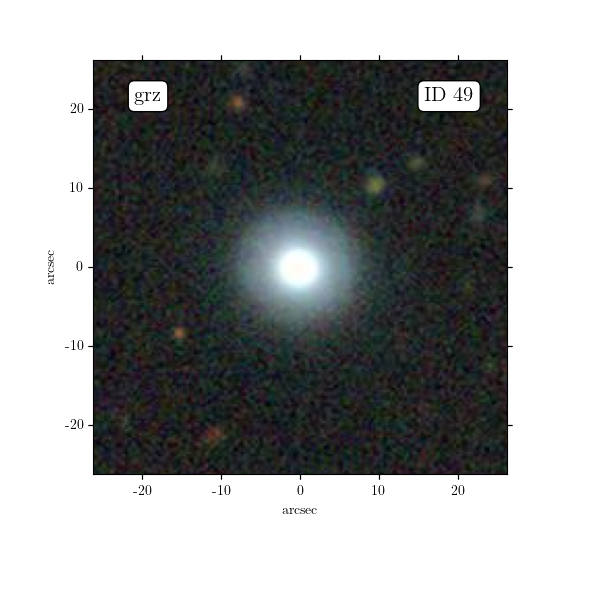}\par 
    \includegraphics[width=6.0cm]{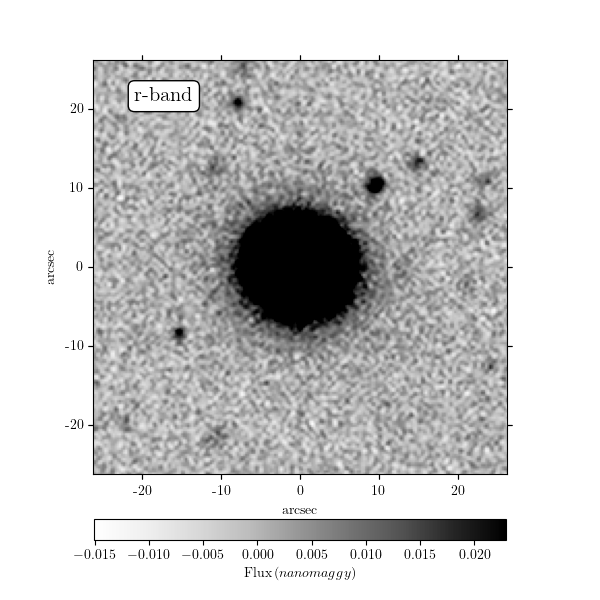}\par 
    \includegraphics[width=6.0cm]{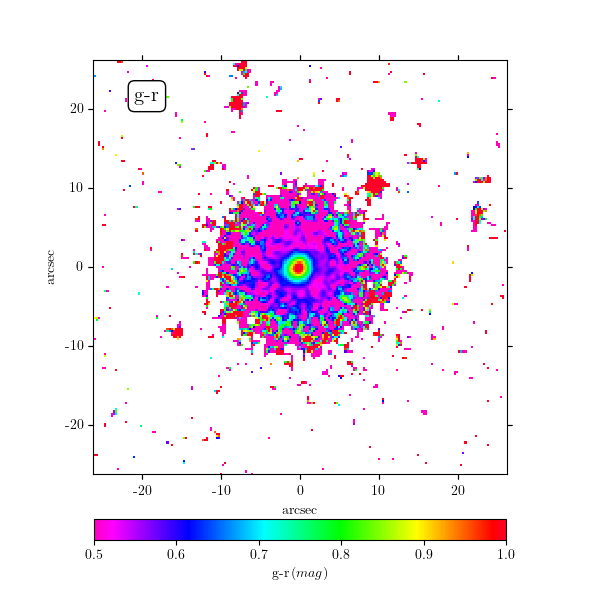}\par
 \end{multicols}
 \begin{multicols}{3}
    \includegraphics[width=6.0cm]{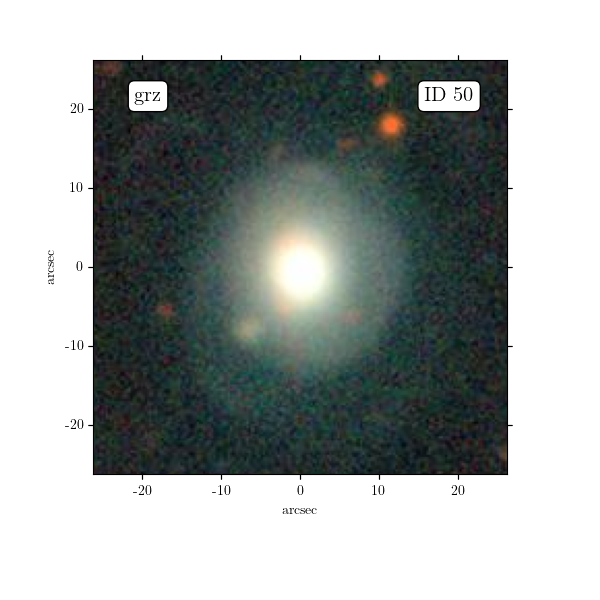}\par 
    \includegraphics[width=6.0cm]{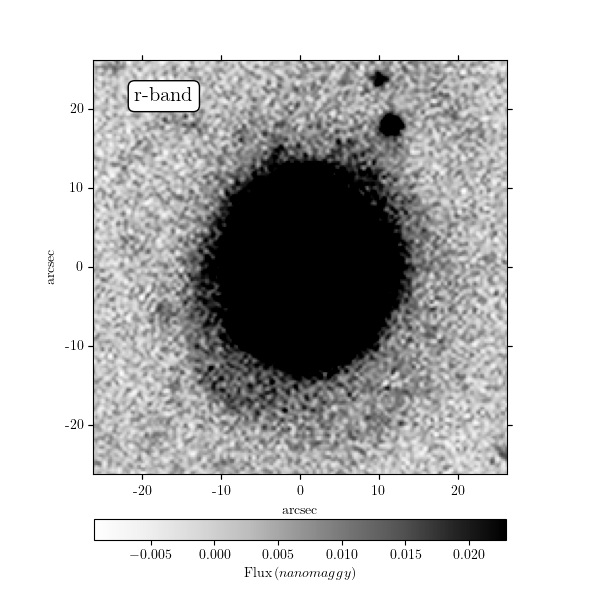}\par 
    \includegraphics[width=6.0cm]{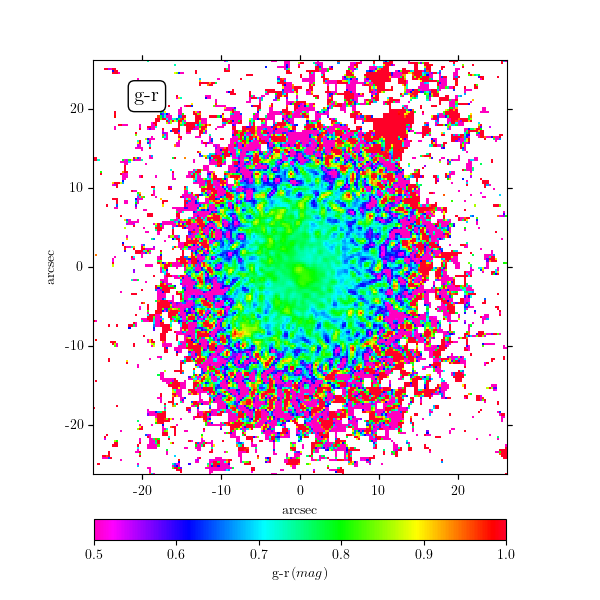}\par
 \end{multicols}
 \begin{multicols}{3}
    \includegraphics[width=6.0cm]{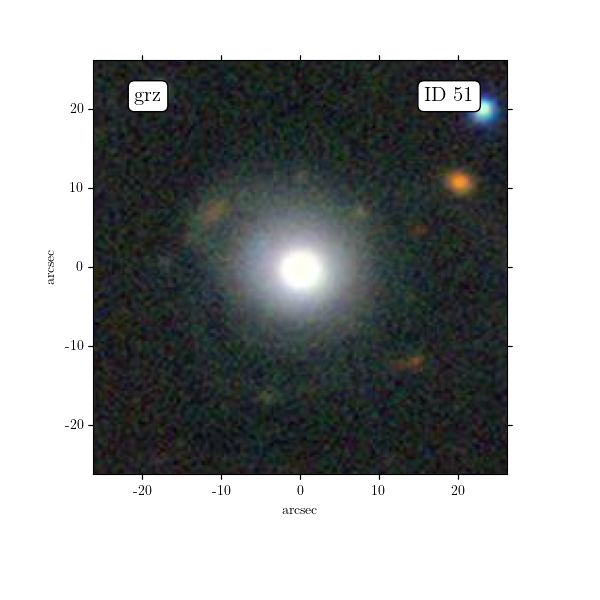}\par 
    \includegraphics[width=6.0cm]{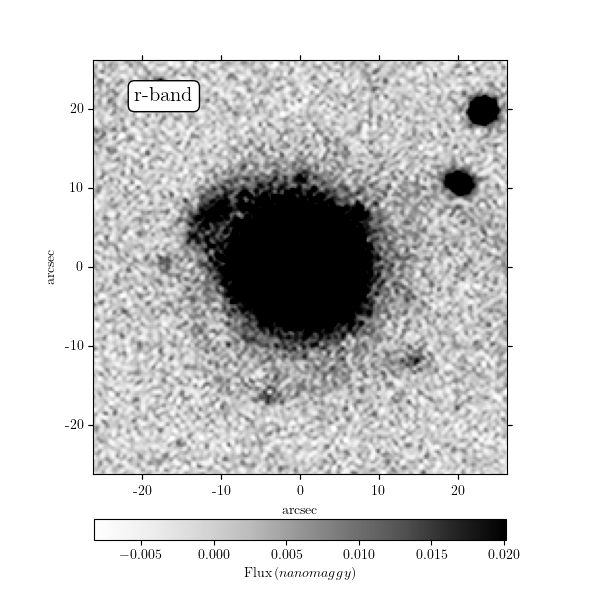}\par 
    \includegraphics[width=6.0cm]{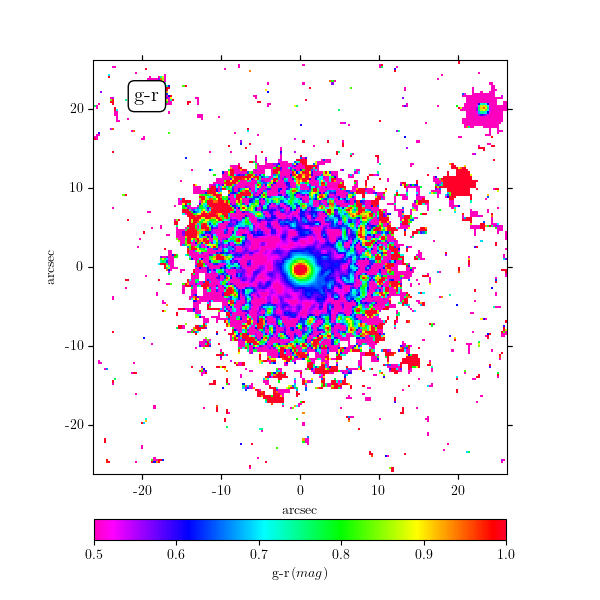}\par
 \end{multicols}
  \begin{multicols}{3}
    \includegraphics[width=6.0cm]{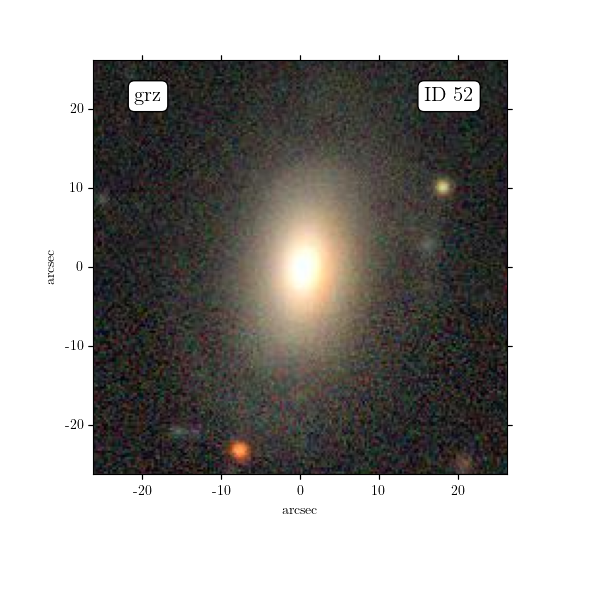}\par 
    \includegraphics[width=6.0cm]{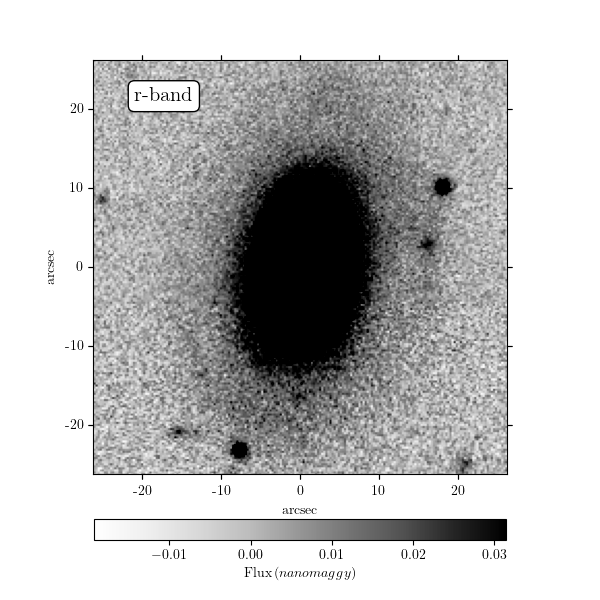}\par 
    \includegraphics[width=6.0cm]{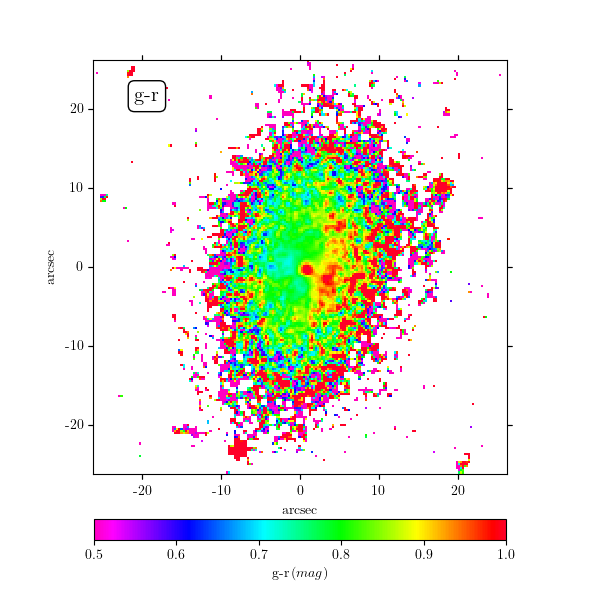}\par
 \end{multicols}
  \end{figure*}

  \begin{figure*}
%\ContinuedFloat 
 \begin{multicols}{3}
    \includegraphics[width=6.0cm]{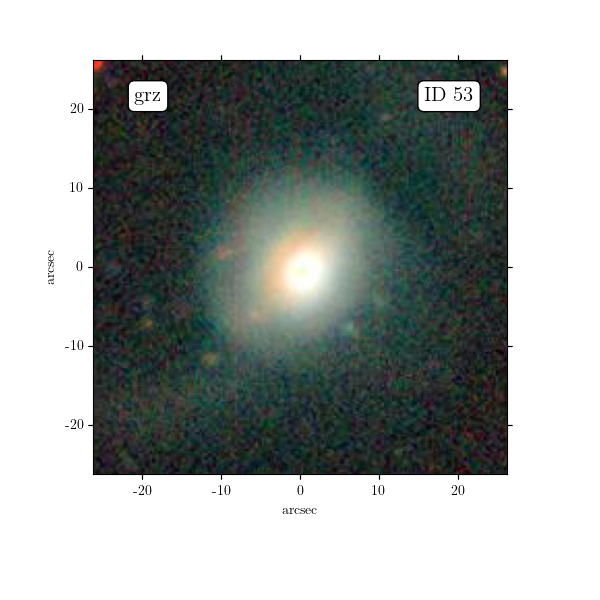}\par 
    \includegraphics[width=6.0cm]{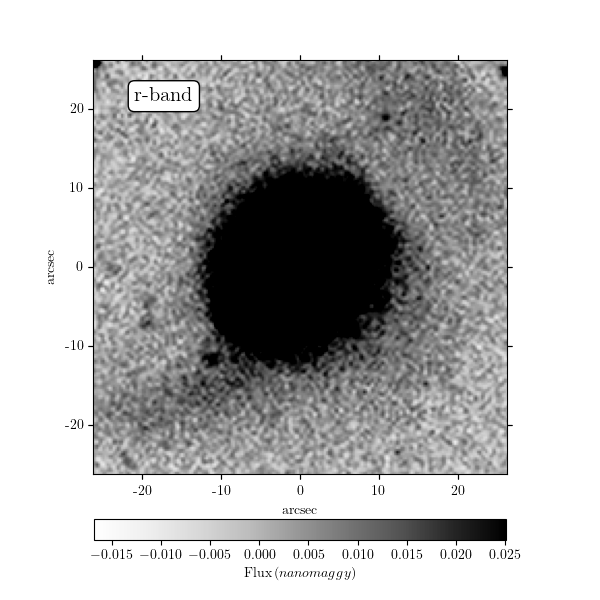}\par 
    \includegraphics[width=6.0cm]{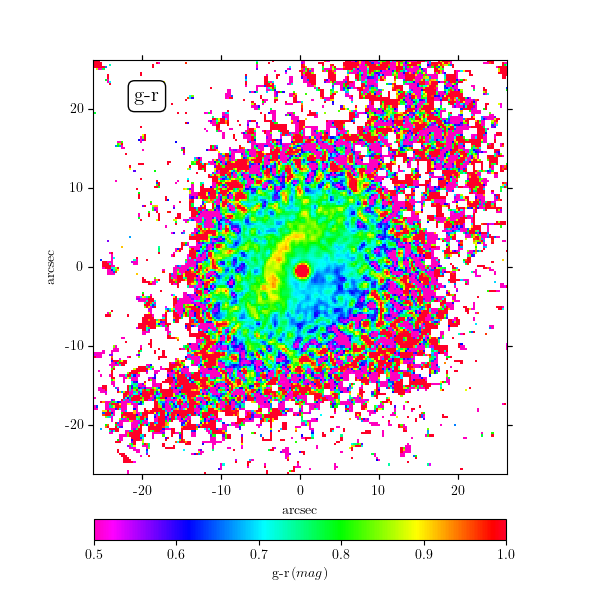}\par
 \end{multicols}
 \begin{multicols}{3}
    \includegraphics[width=6.0cm]{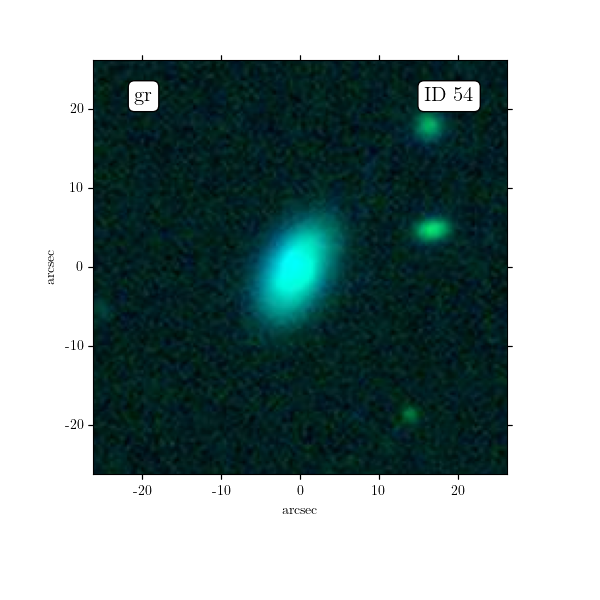}\par 
    \includegraphics[width=6.0cm]{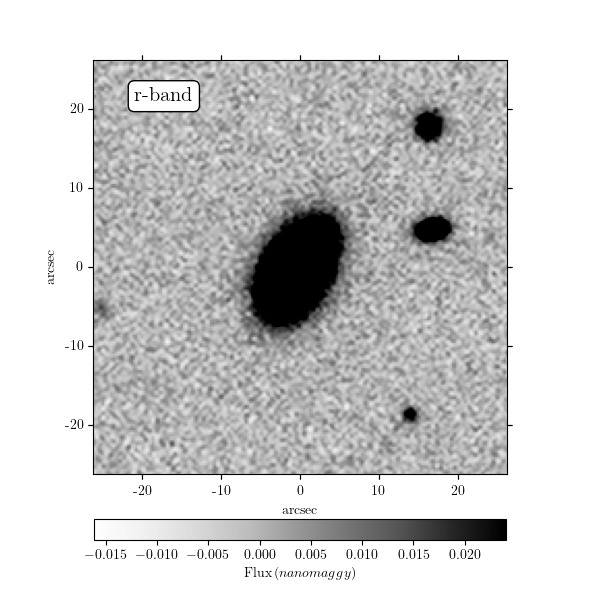}\par 
    \includegraphics[width=6.0cm]{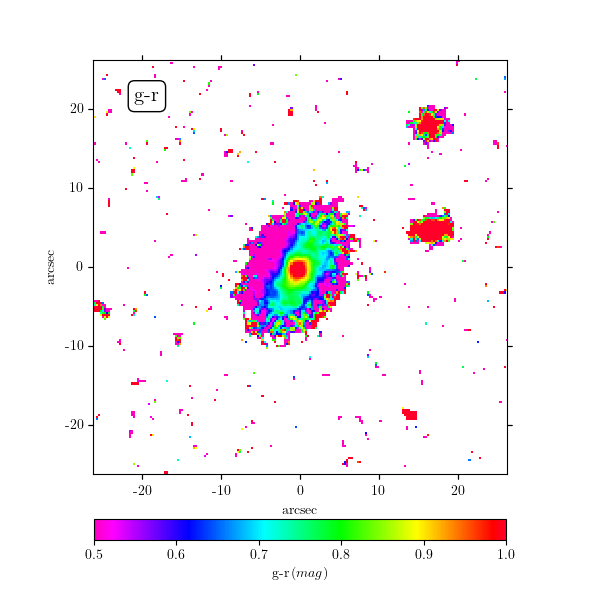}\par
 \end{multicols}
 \begin{multicols}{3}
    \includegraphics[width=6.0cm]{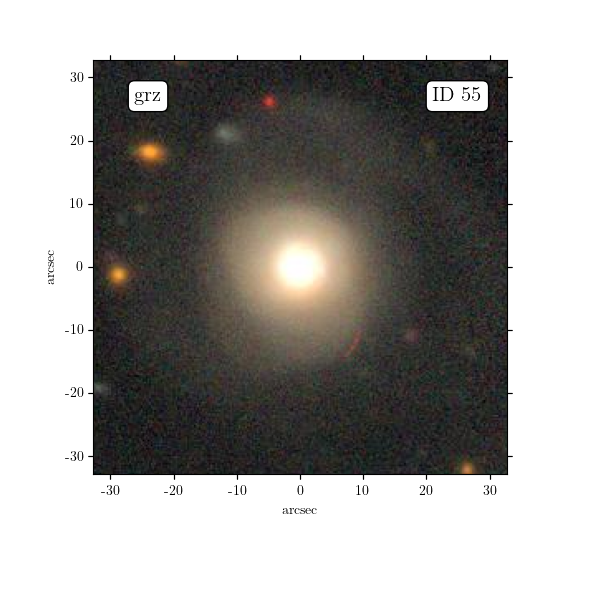}\par 
    \includegraphics[width=6.0cm]{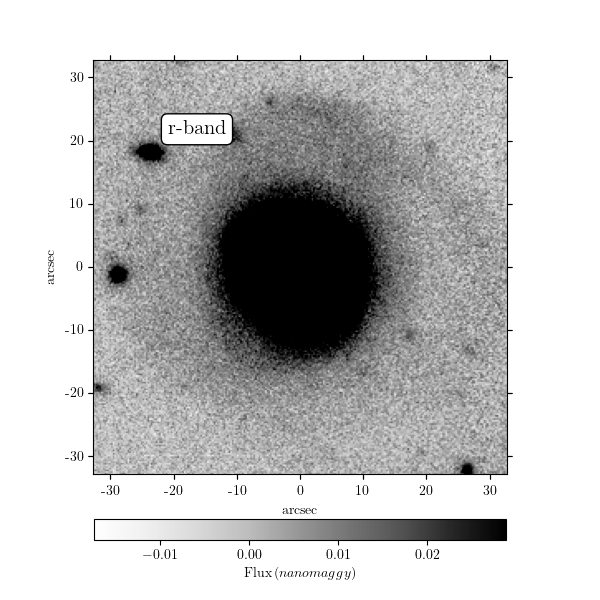}\par 
    \includegraphics[width=6.0cm]{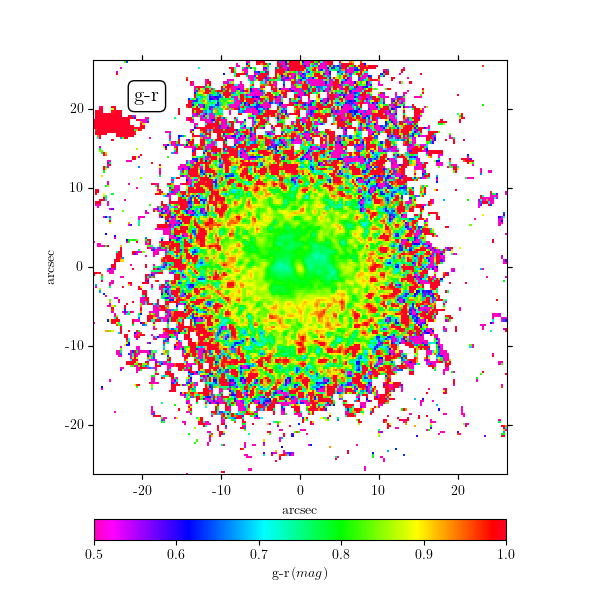}\par
 \end{multicols}
%\end{figure*}
\caption{Deep imaging of 55 blue ETGs. Left panel: Color-composite image of 55 star-forming blue ETGs created using $g$ - (blue), $r$ - (green), and $z$ -(red) band images from the legacy survey. Middle panel: Grayscale color-inverted $r$-band image of the galaxies with the scaling set to highlight faint features in the  galaxy outskirts. Right panel: $g-r$ color map of the galaxies.}
 \label{figure:fig2}
\end{figure*}

\end{document}